\documentclass[pre,aps,10pt,twocolumn,showpacs,floatfix,superscriptaddress]{revtex4-2}
\usepackage{graphicx}
\usepackage{amsmath,amssymb}
\usepackage{mathtools}
\usepackage{dcolumn}
\usepackage{bm}
\usepackage{subfigure}
\usepackage{epsfig}
\usepackage{afterpage}
\usepackage[colorinlistoftodos]{todonotes}
\usepackage[colorlinks=true, allcolors=blue]{hyperref}

\newcommand{\ra}{\rangle}
\newcommand{\la}{\langle}
\newcommand{\bea}{\begin{eqnarray}}
\newcommand{\eea}{\end{eqnarray}}
\newcommand{\beq}{\begin{equation}}
\newcommand{\eeq}{\end{equation}}
\newcommand{\be}{\begin{equation}}
\newcommand{\ee}{\end{equation}}
\newcommand{\beqa}{\begin{eqnarray}}
\newcommand{\eeqa}{\end{eqnarray}}

\newcommand{\eref}[1]{Eq.~(\ref{#1})}%
\newcommand{\Eref}[1]{Equation~(\ref{#1})}%
\newcommand{\fref}[1]{Fig.~\ref{#1}} %
\newcommand{\rahul}{}

\begin{document}

\title{Hard core run and tumble particles on a one dimensional lattice}
\author{Rahul Dandekar} 
\affiliation{The Institute of Mathematical Sciences,
	C.I.T. Campus, Taramani, Chennai-600113, India} 
\affiliation{Homi Bhabha National Institute, Training School Complex, Anushakti Nagar, Mumbai-400094, India}
\author{Subhadip Chakraborti} 
\affiliation{International Centre for Theoretical Sciences, TIFR, Bangalore
	560089, India} 
\author{R.Rajesh} 
\affiliation{The Institute of Mathematical Sciences,
	C.I.T. Campus, Taramani, Chennai-600113, India} 
\affiliation{Homi Bhabha National Institute, Training School Complex, Anushakti Nagar, Mumbai-400094, India}
\begin{abstract}
We study the large scale behavior of a collection of hard core run and tumble particles on a one dimensional lattice with periodic boundary conditions. Each particle has persistent motion in one direction decided by an associated spin variable until the direction of spin is reversed. We map the run and tumble model to a mass transfer model with fluctuating directed bonds. We calculate the steady state single{\rahul-}site mass distribution in the mass model within a mean field approximation for larger spin-flip rates and by analyzing an appropriate coalescence{\rahul-}fragmentation model for small spin-flip rates. We also calculate the hydrodynamic coefficients of diffusivity and conductivity for both large and small spin-flip rates and show that the Einstein relation is violated in both regimes. We also show how the  non-gradient nature of the process can be taken into account in a systematic manner to calculate the hydrodynamic coefficients.
\end{abstract}
\maketitle

\section{Introduction}\label{sec:intro}

Motion in natural contexts, such as bacterial movement,  flocks of birds, etc., consists of units which achieve motility by converting their chemical energy to mechanical energy. The continuous supply of energy  drives the system out of equilibrium and also results in some remarkable collective phenomena, such as clustering and pattern formation~\cite{DombrowskiPRL2004,PalacciScience2013,SurreyScience2001,KemkemerEPJE2000}, and giant number fluctuations~\cite{NarayanScience2007,PeruaniPRL2012}. These systems, widely known as active matter, have been studied through {\rahul various} simple non-equilibrium models. Examples include  Vicsek models~\cite{VicsekPRL1995,ChatePRL2004,vicsekreview} with alignment interaction to study flocks of birds, active Brownian particles~\cite{FilyPRL2012,RednerPRL2013} and run and tumble particles (RTPs)~\cite{tailleur2008statistical,elgeti2015run} to describe the interacting micro-organisms in a liquid medium, and 
active lattice gases~\cite{EvansPRE2006,TailleurPRL2013,WhitelamJCP2018,slowman2016jamming,mallmin2019exact,kourbane2018exact,ChakrabortyCondmat2020}; for reviews see Refs.~\cite{CatesReview2012,MarchettiReview2013}. Phenomenological active hydrodynamics has been developed to characterize flocking phenomena~\cite{TonerPRL1995,TonerPRE1998} and to explore the motion of swarms of bacteria in a liquid medium~\cite{ramaswamy2017active}. There have also been attempts to formulate a thermodynamic structure for models of active matter by characterizing equilibrium-like intensive thermodynamic variables, such as temperature~\cite{LevisEPL2015}, chemical potential~\cite{CatesPRL2013,chakraborti2016additivity,chakraborti2019additivity} or pressure~\cite{KafriNature2015}. However, a general theoretical understanding of the hydrodynamics and large-scale behavior of the steady-states is still lacking. In this paper, with a view to understanding the unique features of active matter systems, we study the steady-state behaviour and calculate transport coefficients in a paradigmatic microscopic model of active matter, interacting RTPs on a one dimensional lattice. 

{\rahul Run and tumble particles (RTPs) are defined as particles which move (`run') with a constant average velocity, while the direction of this velocity changes intermittently (the `tumble'). They differ from random walkers in the persistence of the direction of their motion between such tumbling events.} In the past decade, there have been many studies of the motion of RTPs as models for bacteria, and as interesting non-equilibrium models of interacting particles in their own right. 
Initial studies focused on a macroscopic approach by constructing a hydrodynamic theory for RTPs in one dimension. In Ref.~\cite{tailleur2008statistical}, considering a density dependent mean run speed and  a fixed tumble rate, it was argued that the system exhibits self trapping, also known as motility-induced phase separation which happens in the absence of any attractive microscopic interaction, unlike a passive phase separation. Numerical simulation of RTPs {\rahul with hard-core interaction} showed that the particles cluster more for small tumble rates, though no true condensation occurs. By examining the absorption and evaporation of dimers, the scaling behavior of the steady state cluster distributions for small tumble rates could be obtained~\cite{soto2014run}. RTPs with multiple occupancy of lattice sites have also been studied numerically~\cite{sepulveda2016coarsening}. When the number of allowed particles on a site is larger than two, there is evidence of a condensation transition. A comparative study of RTPs with other models such as active Brownian particles and active Ornstein-Uhlenbeck particles can be found in Refs.~\cite{CatesEPL2013,DolaiCondmat2020}.

More recent approaches have focused on {\rahul exact studies of} systems with only one or two RTPs. For a single RTP, the large deviation function for the displacement shows a first order dynamical transition {\rahul with} a `condensed' phase for large displacements {\rahul where} the large deviation function is dominated by a single run~\cite{gradenigo2019first}. Studies of {\rahul a single RTP} in bounded domains and confining potentials~\cite{malakar2018steady,dhar2019run,basu2020exact} have shown that the steady-state distribution, under some conditions, shows an unusual structure with peaks away from the center of the domain. Similar results have been obtained in two dimensions~\cite{basu2018active,majumdar2020toward}. For a system of two RTPs, the steady state distribution of inter-particle distance has three contributions: a uniform contribution, an exponentially decaying correlation, and a `bound state' in which two RTPs with opposite spins sit on adjacent sites of the lattice~\cite{slowman2016jamming,slowman2017exact}. This analysis was extended in Ref.~\cite{mallmin2019exact} where the Markov matrix for the time evolution of one or two RTPs was diagonalised to show that the system undergoes a dynamical transition at certain values of the tumble rate. When additional thermal noise is added to the  RTPs, the steady state gap distribution becomes exponential~\cite{das2020gap}. The survival probability of two annihilating RTPs in one dimension can also be calculated exactly, {\rahul and a new length scale, the} `Milne length' {\rahul can be identified which characterizes} correlations between the particles~\cite{doussal2019non}.

In this paper, we look at the case of many RTPs {\rahul with a hard-core interaction} on a one dimensional lattice. {\rahul The hard-core interaction implies that each lattice site cannot be occupied by more than one RTP at a time.} We work in the frame of the gaps between particles, which allows use of the framework of mass transport models,{\rahul which has proven useful in other 1D models to calculate various thermodynamic and hydrodynamic quantities, such as the chemical potential~\cite{das2015additivity} and transport coefficients~\cite{das2017einstein,ChakrabortyCondmat2020}}. We go beyond previous studies on 1D RTPs by deriving the cluster and gap distributions for moderate and large tumble rates in the Independent Interval Approximation, and also in deducing the time-dependent behavior for small tumble rates. For moderate tumble rates, {\rahul our approach gives} results which are well supported by numerical simulations. For very small tumble rates, we develop an alternate approach, also in the mass transport picture, based on a mapping to a diffusion-limited coalescence and fragmentation process~\cite{burschka1989transition,ben1990statics}. We show that the results from this model for the steady-state and the relaxation to the steady-state are in excellent agreement with simulations. In addition to the steady state gap distribution, we also calculate the diffusivity and conductivity for large tumble rates, and in the limit of very small tumble rates. We show that the Einstein relation is violated in both regimes. 

The remainder of the paper is organized  as follows. In Sec.~\ref{sec:model} we define the model in the particle picture as well as the  corresponding mass transport model in the gap picture. In Sec.~\ref{sec:mf}, we calculate the mass distribution within a mean field approximation and  compare it to results from Monte Carlo simulation for moderate tumble rates, showing excellent agreement. In Sec.~\ref{sec:smallD}, we determine, for  small tumble rates, the steady state mass distribution as well as relaxation to the steady state, by developing and analyzing a model of diffusion, fragmentation, and  coalescence. In Secs.~\ref{sec:hydro} and \ref{sec:hydro2}, we calculate hydrodynamic quantities such as the diffusion constant and conductivity  for moderate and low tumble rates respectively, and compare the analytical results with results from Monte Carlo simulations. Section~\ref{sec:conclusion} contains a summary of the paper and discussion of the results. 

\section{\label{sec:model} Model} 

Consider $N$ RTPs on a ring of $L$ sites, where each site can be occupied by at most one particle. Each particle is characterized by a spin $S$ that can take the values  $+1$ (pointing right) or $-1$ (pointing left). A particle hops at rate $1$ to the neighboring lattice site in the direction of its spin, provided the target site is empty. Thus, particles with $S=1$ hop to the right and those with $S=-1$ hop to the left. In addition, each particle reverses the direction of its spin at rate $\eta$. Clearly, when $\eta \rightarrow \infty$, the value of the spin is random, and the model reduces to the well-studied symmetric exclusion process \cite{derrida2011microscopic}.

\begin{figure}
	\centering
	\includegraphics[width=\columnwidth]{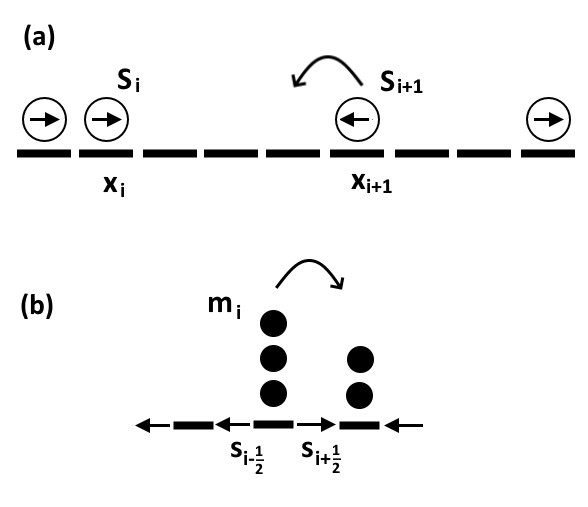}
	\caption{\label{rtpgap}An example of the mapping between (a) the RTP model and (b) the mass transport model. In the mapping, mass at site $i$ equals the number of empty sites between the $i^{th}$  and $(i+1)^{th}$ RTP, and the spins transform as $s_{i-1/2}=-S_i$. In the mass model, a unit mass hops in the direction of the spins on the neighboring bonds. A particle hopping to the left in the RTP model (shown by a bent arrow in (a)) corresponds to a unit mass being transferred to the right in the mass model (shown by a bent arrow in (b)).} 
\end{figure}

It is convenient to study the dynamics of the gaps between particles and define a corresponding mass transport model \cite{Subhadip_thesis}. In this picture, the gaps between RTPs are mapped onto masses on a corresponding lattice, and the spins live on the bonds on this lattice. We now describe the mapping more precisely. Let $x_i$ be the position of the $i^{th}$ RTP on the original lattice, and let $S_i$ denote its spin as shown in Fig.~\ref{rtpgap}(a). In the mass transport model, the $i^{th}$ site has a mass $m_i = x_{i+1} - x_i - 1$, the number of empty sites between the $i^{th}$ and $(i+1)^{th}$ RTPs. Spin $s_{i-\frac{1}{2}}$ in the mass model lives on the bonds between sites $(i-1)$ and $i$, and is equal to $-S_i$, the negative of the spin of the $i^{th}$ RTP. An example of this mapping is shown in Fig.~\ref{rtpgap}. It is clear that the  mass model has $N$ sites with total mass $L-N$. The mass density $\rho_{mass}$, which we {\rahul henceforth} denote by $\rho$ is related to the density  in the particle picture, denoted $\rho_{RTP}$, as 
\beq
\rho \equiv \rho_{mass} = \frac{1}{\rho_{RTP}}-1.
\eeq

The dynamics of the mass model can be derived as follows. Particle $(i+1)$ moving one unit to the left in the RTP picture maps to a unit mass transfer from site $i$ to the right across the bond $(i+\frac{1}{2})$ (where $s_{i+1/2}=1$) in the mass model, as  shown in Fig.~\ref{rtpgap}, {\rahul and similarly, mass transfers to the left across a bond correspond to an RTP moving to the right.} The value of $s_{i+\frac{1}{2}}$ dictates the direction of mass transfer between sites $i$ and $i+1$. If $s_{i+\frac{1}{2}}=+1$, mass from $i$ can move to site $i+1$, but not from $i+1$ to $i$. If $s_{i+\frac{1}{2}}=-1$, the reverse is the case.  Thus, the dynamics of the mass model may be summarized as follows: from a non-zero  mass at site $i$, unit mass is transferred  with rate one to the right and rate one to the left, provided the spin on the bond that it crosses is in the direction of transfer, 
while the spins $s_{i+\frac{1}{2}}$ themselves flip at rate $\eta$ independently of the masses and the other spins.

\section{\label{sec:mf} Steady-state solution within a mean field approximation} 

In this section, we derive the steady state mass distribution, or equivalently the distribution of gaps between particles in the RTP picture, within a mean field approximation. Let $P(\{m_i\},\{s_{i+\frac{1}{2}}\},t)$ denote the probability of finding the system in a certain configuration of masses $\{m_i\}$ and spins $\{s_{i+\frac{1}{2}}\}$ at time $t$. $P(\{m_i\},\{s_{i+\frac{1}{2}}\},t)$ may be expressed as 
\be
P(\{m_i\},\{s_{i+\frac{1}{2}}\}, t) = P(\{m_i\}|\{s_{i+\frac{1}{2}}\}, t) P(\{s_{i+\frac{1}{2}}\}, t),
\label{eq:conditional}
\ee 
where the notation $P(x|y)$ denotes the conditional probability of $x$ given $y$. \Eref{eq:conditional} is exact, since it simply re-expresses joint probabilities in terms of conditional ones. Now, {\rahul since} each spin flips independently, we have 
\be
P(\{s_{i+\frac{1}{2}}\}) = \prod_i P(s_{i+\frac{1}{2}}) = \frac{1}{2^N}.
\ee
We will assume that the spins are initially chosen with this steady state probability so that the probabilities for spins are  invariant in time.

We implement a single-site mean field approximation where, for each mass, we {\rahul only keep track of the neighbouring spins}. Thus, we approximate the conditional mass distribution in \eref{eq:conditional} as
\be
P(\{m_i\} |\{s_{i+\frac{1}{2}} \}) \approx \prod_i P(m_i|s_{i-\frac{1}{2}},s_{i+\frac{1}{2}}).
\label{eq:four}
\ee
{\rahul Thus, there are} four conditional distributions $P(m_i|s_{i-\frac{1}{2}},s_{i+\frac{1}{2}})$, which we denote as $P^{++}_m$, $P^{-+}_m$, $P^{+-}_m$ and $P^{--}_m$. The first (second) superscript refers to the neighboring spin on the left (right) bond, with $+$ referring to $+1$ (or rightward) and $-$ referring to $-1$ (or leftward).  Among these, only three are independent, since due to left-right symmetry
\be
P^{++}_m = P^{--}_m, ~m=0, 1, 2, \ldots. \label{eq:symmetry}
\ee
This is because  both probability distributions correspond to a site with one of its neighboring spins pointing inwards and the other outwards. 
Also, due to translational invariance,  the probability distributions do not depend on the site index.

For ease of notation, we introduce {\rahul the quantities} $\alpha$, $\beta$, $\gamma$, $\delta$:
\beqa
\alpha &\equiv& P^{-+}_0,\nonumber \\
\beta &\equiv& P^{++}_0,  \label{eq:alpha}\\
\gamma &\equiv& P^{--}_0, \nonumber\\
\delta &\equiv& P^{+-}_0, \nonumber
\eeqa
as they will appear repeatedly in the calculations. These quantities will be determined in terms of the mass density $\rho$ and spin flip rate $\eta$. From Eq.~\eqref{eq:symmetry}, we see that in the steady state, $\beta=\gamma$, and we will use the symbol $\beta$ for both $P^{++}_0$ and $P^{--}_0$ in this section. In Sec.~\ref{sec:hydro},  in the presence of a field or a density gradient, we will see that this symmetry is broken.

First, let us consider the temporal evolution of $P^{-+}_m$. For this combination of neighboring spins, the site cannot receive mass, as both spins point away from it. The master equation for $P^{-+}_m$ is
\be
\frac{dP^{-+}_m}{dt} = 2 \eta (P^{++}_m - P^{-+}_m) + 2 P^{-+}_{m+1} -2 (1-\delta_{m,0}) P^{-+}_{m}, \label{eq:meanfieldmp}
\ee
where the first  term on the right hand side describes spin flips and the last two terms describe transfer of unit mass to {\rahul and from} neighboring sites. In the steady state, the time derivative may be set to zero and we obtain
\be
P^{-+}_{m+1} = -\eta \left(P^{++}_m - P^{-+}_m \right) +  (1-\delta_{m,0}) P^{-+}_{m},
\label{eq:meanfield2}
\ee

\Eref{eq:meanfield2} can be solved using the generating function method. Let
\be
\widetilde{P}^{-+}(z) = \sum_{m=0}^\infty P^{-+}_m  z^m,
\label{eq:genfunct}
\ee
with similar definitions for the other two distributions: $\widetilde{P}^{++}(z)$ and $\widetilde{P}^{+-}(z)$.
Multiplying \eref{eq:meanfield2} by $z^m$ and summing over $m$, we obtain
\be
\widetilde{P}^{-+}(z) =  \frac{(1-z) \alpha - z \eta \widetilde{P}^{++}(z)}{1- z(\eta+1)}, \label{pmpz}
\ee
where $\alpha$ is as defined in Eq.~(\ref{eq:alpha}).
Note that $\widetilde{P}^{-+}(z)$ depends upon the generating function $ \widetilde{P}^{++}(z)$.

In Appendix A we similarly solve for the generating functions $P^{++}(z)$ and $P^{+-}(z)$. Finally, we obtain

\begin{widetext}
	\bea
	\delta &=& \frac{2 \eta\beta}{2\eta + 2 - \alpha -\beta}, \label{eq:gamma} \\
	\widetilde{P}^{++}(z) &=& \frac{2 \left[(\alpha +\beta-2 )(z-1) +2 \eta\right] \left[ (\alpha +\beta) \eta z + \beta( z-1) \right]}{h(z)},
	\nonumber \\
	\widetilde{P}^{+-}(z) &= &\frac{2 \left[(\alpha +\beta) \eta z + \beta( z-1) \right] }{h(z)},
	\label{pppz}\\
	\widetilde{P}^{-+}(z) &=& \frac{2 \eta  (z-1) \left[3 \alpha ^2 z+ \alpha  ((4 \beta -6)
		z+2)+(\beta -2) \beta  z\right]+4 \eta ^2 z (\alpha
		+\beta )+\alpha  (z-1)^2 (\alpha +\beta -2) \left[z
		(\alpha +\beta -2)+2\right] }{h(z)}, \nonumber
	\eea
	where the denominator $h(z)$ is given by
	\begin{align}
	h(z) = \left[ \left(\alpha  +\beta   -2 \right) (1+\eta)   z +2 -\alpha -\beta +2 \eta \right] \times \left[ \left(\alpha +\beta  -2 \right) z^2 -\left( \alpha +\beta -4 \eta -4\right) z -2 \right].
	\end{align}
	We note that  the denominator $h(z)$ is common to all three generating functions, and $\alpha$ and $\beta$ remain undetermined.
\end{widetext}

The behavior of the probability distributions for large $m$ is determined by the poles of the generating functions, which in turn are determined by the  three zeros of $h(z)$:
\bea
z_1&=&\frac{2-\alpha -\beta +2 \eta }{(\eta +1) (2-\alpha
	-\beta)},\nonumber \\
z_2&=&\frac{\alpha +\beta -4 \eta -4-\sqrt{(\alpha +\beta -4 \eta )^2+32 \eta}}
{2 (\alpha +\beta -2)}, \label{eq:roots}\\
z_3&=&\frac{\alpha +\beta -4 \eta -4-\sqrt{(\alpha +\beta -4 \eta )^2+32 \eta}}
{2 (\alpha +\beta -2)}. \nonumber
\eea
Among these, $z_1,z_2> 1$ and $z_3<1$. A root whose magnitude is less than unity implies an exponentially diverging $P(m)$ for large $m$, which is unphysical. Hence, $z_3$ cannot be a valid pole. This can be true only if the numerator of all the three generating functions vanish at $z_3$, thus, removing the pole at $z_3$~\cite{rajesh2000conserved,rajesh2000exact}. This leads to the constraint (for all three generating functions) that
\be
\alpha = \frac{2 \beta \sqrt{\beta+ 2 \eta + \eta^2}-\beta^2}{\beta+2 \eta}, \label{ab}
\ee
leaving only $\beta$ to be determined.

To determine $\beta$, we use the fact the total mass is a conserved quantity. The generating function $\widetilde{P}(z) = \sum_{m=0}^\infty P_m  z^m$, where $P_m$ is the probability of a randomly chosen site having mass $m$, is given by
\be
\widetilde{P}(z) = \frac{1}{4} \widetilde{P}^{+-}(z) + \frac{1}{4} \widetilde{P}^{-+}(z) + \frac{1}{2}\widetilde{P}^{++}(z). \label{eq:Pm}
\ee
The density $\rho$ is then given by
\begin{align}
&\rho = z \frac{d P(z)}{dz}\bigg\lvert_{z=1}, \nonumber\\
&= \frac{(\alpha+\beta)(3 \alpha+\beta) +8(2-2 \alpha -\beta) }{8 \eta (\alpha+\beta)}
+\frac{2 - \alpha - \beta }{ \alpha+\beta}.\label{rb}
\end{align}
Equations~\eqref{ab} and \eqref{rb} may be solved to obtain $\alpha$ and $\beta$ in terms of $\rho$ and $\eta$. Thus, we have obtained the full solution to the single site mass distribution within the  mean field approximation.

As there remain two poles of magnitude larger than one, the mass distribution $P(m)$ will be a sum of two exponentials. For large $m$, the pole closest to origin, which turns out to be $z_1$, will have the dominant contribution. All three conditional distributions will show the same asymptotic decay,
\begin{equation}
P(m) \sim e^{-m/m^*} \mbox{ for } m \gg 1,
\end{equation}
where
\begin{equation}
m^* = -\frac{1}{\ln z_1}, \label{mflambda}
\end{equation}
with $z_1$ as in Eq.~(\ref{eq:roots}). This completes the calculation of the single-site distributions within a mean-field framework.

For large $\eta$, the RTP model approaches a symmetric simple exclusion process, since the particle motion decouple from the spin fluctuations.  For large $\eta$, the Eqs.~\eqref{ab} and \eqref{rb}  may be solved as a series expansion to give 
\beqa
\alpha &=& \frac{1}{1+\rho} + \frac{\rho (2+3\rho)}{2 (1+\rho)^3 \eta} + O(\eta^{-2}), \nonumber\\
\beta &=& \frac{1}{1+\rho} + \frac{\rho^2}{2 (1+\rho)^3 \eta} + O(\eta^{-2}), \label{eq:mftinfty}\\
\delta &=& \frac{1}{1+\rho} - \frac{\rho(2+\rho)}{2 (1+\rho)^3 \eta} + O(\eta^{-2}), \nonumber\\
\frac{1}{m^*} &=& -\ln{\left(\frac{\rho}{1+\rho}\right)} -\frac{1}{2 (1+\rho)^2\eta} + O(\eta^{-2}).
\eeqa
Also, the probability of a site being empty $P(0) = \frac{1}{1+\rho} + O(\eta^{-2})$. The leading terms, corresponding to $\eta=\infty$, are consistent with the results for the symmetric simple exclusion process. 

For small $\eta$, the probabilities have the series expansion
\beqa
\alpha &=& 1-8 \rho  (2 \rho +1) \eta^2 + O(\eta^3), \\
\beta &=& 1 - 8 \rho \eta + 56 \rho(\rho+2\rho^2) \eta^2 + O(\eta^3),\\
\delta &=& \frac{1}{1 + 4 \rho}+\frac{16 \rho (\rho+1) \eta}{(4 \rho +1)^2} + O(\eta^2), \\
\frac{1}{m^*} &=& \log{\left(\frac{1}{4 \rho }+1\right)} + \frac{(8 \rho +5) \eta}{4\rho +1} + O(\eta^2).
\eeqa
In the limit $\eta \rightarrow 0$, we see that $m^*$ tends to a finite number larger than zero, thus predicting an exponential decay.  However, simulations show that the average number of particles per occupied site diverges as $\eta \rightarrow 0$, leading to clustering. We explain this deviation from mean field theory for small $\eta$ in Sec.~\ref{sec:smallD}. 

\begin{figure}
	\centering
	\includegraphics[width=\columnwidth]{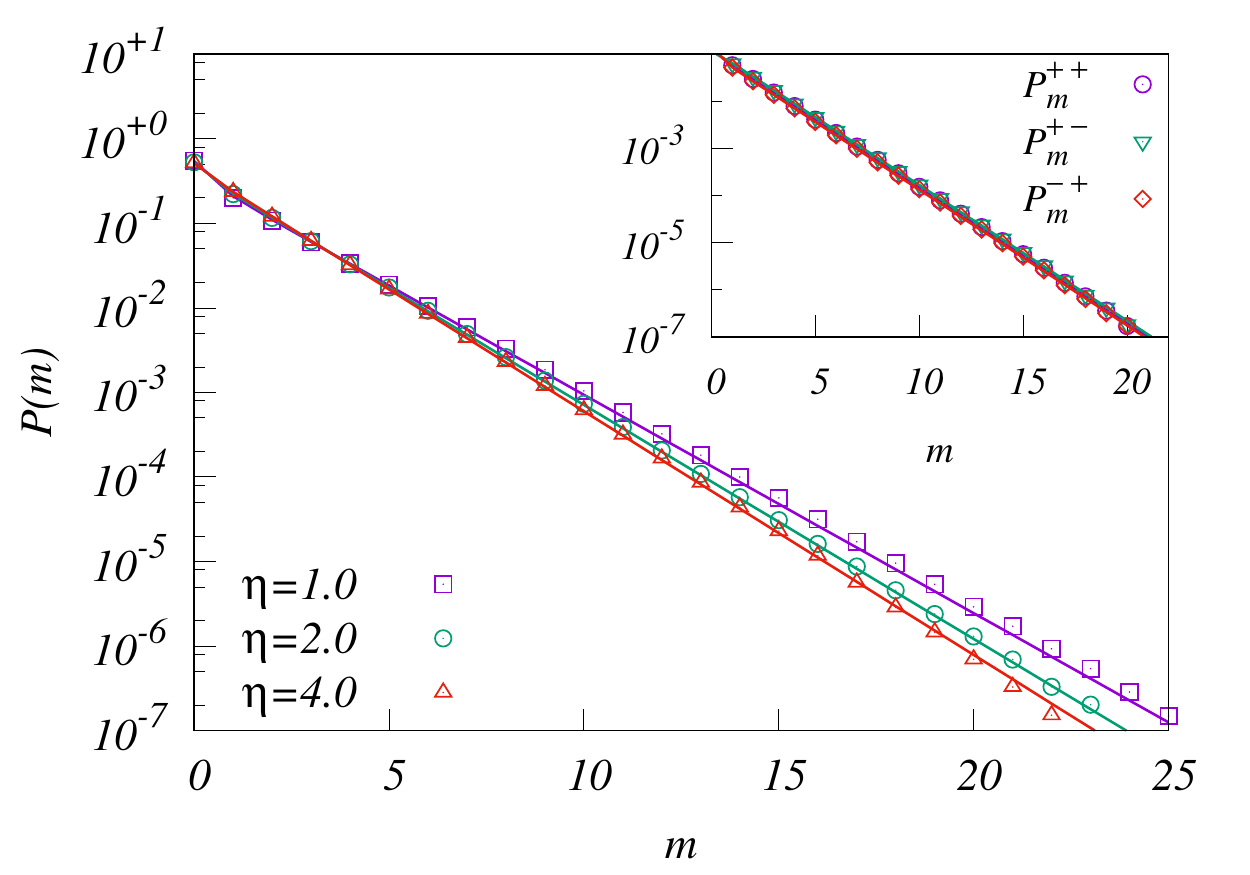}
	\caption{The single site mass distribution $P(m)$, obtained from Monte Carlo simulations, is compared with the predictions from the mean field theory (shown as solid lines) for moderate values of the spin flip rate $\eta$. The data are for systems with density $\rho=1$ and $L=500$. The inset shows, for $\eta = 4.0$, the comparison of the simulations and the mean field results for the distributions $P^{++}=P^{--}$, $P^{+-}$ and $P^{-+}$.} 
	\label{mfsimhigh}
\end{figure}

\begin{figure}
	\centering
	\includegraphics[width=\columnwidth]{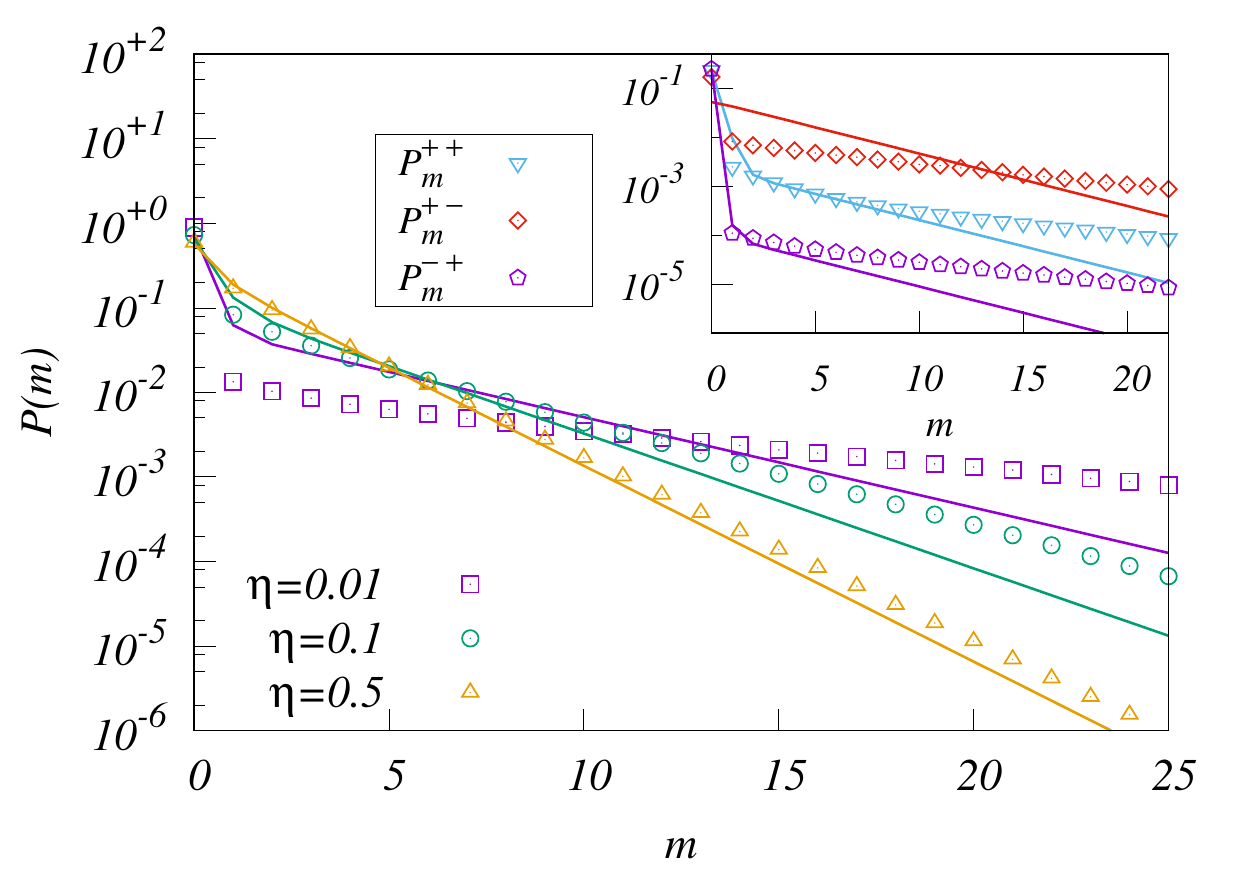}
	\caption{The single site mass distribution $P(m)$, obtained from Monte Carlo simulations are compared with the predictions from the mean field theory (shown in solid lines) for moderate and small values of  the spin flip rate $\eta$. The data are for systems with density $\rho=1$ and $L=500$. The inset shows, for $\eta = 0.01$, the comparison of the simulations and the mean field results for the distributions $P^{++}=P^{--}$, $P^{+-}$ and $P^{-+}$.}
	\label{mfsimlow}
\end{figure}

Figures~\ref{mfsimhigh} and \ref{mfsimlow} compare the predictions of the mean-field single-site mass distributions with results from simulations for $\rho=1$. It can be seen that for $\eta \geq 1$ there is excellent agreement with the simulation results, for all distributions $P^{++}=P^{--}$, $P^{-+}$, and $P^{+-}$ (see Fig.~\ref{mfsimhigh}). For $\eta<1$, we see that the distributions obtained from the simulations decay much more slowly at large $m$ than the mean-field prediction, showing that the mean-field approximation fails for small $\eta$.

We now determine the particle cluster distributions in the RTP picture within the mean field approximation for the mass model. The gaps in the RTP picture correspond to masses in the mass model,  and conversely an empty site in the mass model  corresponds to two neighboring particles in the RTP picture. 
Let $p^{RTP}(n)$ denote the probability in the RTP picture that a given empty site has a cluster of exactly $n$ particles to its right ($n=0$ corresponds to there being no cluster immediately to the right of the site). In the mass model, this corresponds to the probability of a \emph{given non-empty site} having $n$ empty sites to its right, with the $(n+1)^{th}$ site being non-empty.
This probability, in the mean field approximation, may be written as a product of single site mass distributions at different sites. Recall that $P(m_i|s_{i-\frac{1}{2}}, s_{i+\frac{1}{2}})$ is the probability of site $i$ having mass $m_i$, given the two neighboring spins [see Eq.~(\ref{eq:four})]. Let $P (\bullet | s_{i-\frac{1}{2}},s_{i+\frac{1}{2}})$ denotes the probability that there is a non-zero mass at site $i$. Then, 
$p^{RTP}(n)$ can be written as
\begin{align}
&p^{RTP}(n) = \frac{1}{2^{n+2}} \sum_{s_{-\frac{1}{2}}} \ldots \sum_{s_{n+\frac{1}{2}}} \nonumber \\ 
& \left[\prod_{k=1}^{n+1}P(0 | s_{k-\frac{1}{2}}, s_{k+\frac{1}{2}})\right]  P (\bullet | s_{n+\frac{1}{2}}, s_{n+\frac{3}{2}}).
\label{eq:empty}
\end{align}
The sum over the spins are more easily evaluated using  the transfer matrices 
\bea
T = 
\begin{pmatrix} 
	P^{++}_0 & P^{+-}_0 \\
	P^{-+}_0 & P^{--}_0
\end{pmatrix}
= 
\begin{pmatrix} 
	\beta & \alpha \\
	\delta & \beta 
\end{pmatrix},
\eea
and 
\bea
\widetilde{T} = 
\begin{pmatrix} 
	1 - \beta & 1- \alpha \\
	1 - \delta & 1- \beta 
\end{pmatrix},
\eea 
where $\alpha, \beta$ and $\delta$, defined in Eq.~(\ref{eq:alpha}),  are functions of $\rho$ and $\eta$. Eq.~(\ref{eq:empty}) then simplifies to
\beq
p^{RTP}(n) = \frac{1} {2^{n+2}}\sum_{i=1}^2 \sum_{j=1}^2 (T^{n} \widetilde{T})_{i,j}.
\label{eq:occprob}
\eeq

We now check that in the limit $\eta \to \infty$, we recover the results for the symmetric exclusion process. 
In this limit, from Eq.~\eqref{eq:mftinfty}, we know that $\alpha = \beta = \delta = (1+\rho)^{-1} = \rho_{RTP}$, and hence $T = \rho_{RTP} J$, while $\widetilde{T} = (1- \rho_{RTP} )J$ , where  $J$ is the $2\times 2$ matrix with all entries one. Clearly,  $J^n = 2^{n-1} J$.  It is straightforward to simplify Eq.~\eqref{eq:occprob} to 
\be
p^{RTP}(n) \stackrel{\eta \to \infty}{=} \rho_{RTP}^n (1-\rho_{RTP}),
\ee
in agreement with results for the symmetric exclusion process on a ring which has a product measure in the steady-state.

\begin{figure}
	\centering
	\includegraphics[width=\columnwidth]{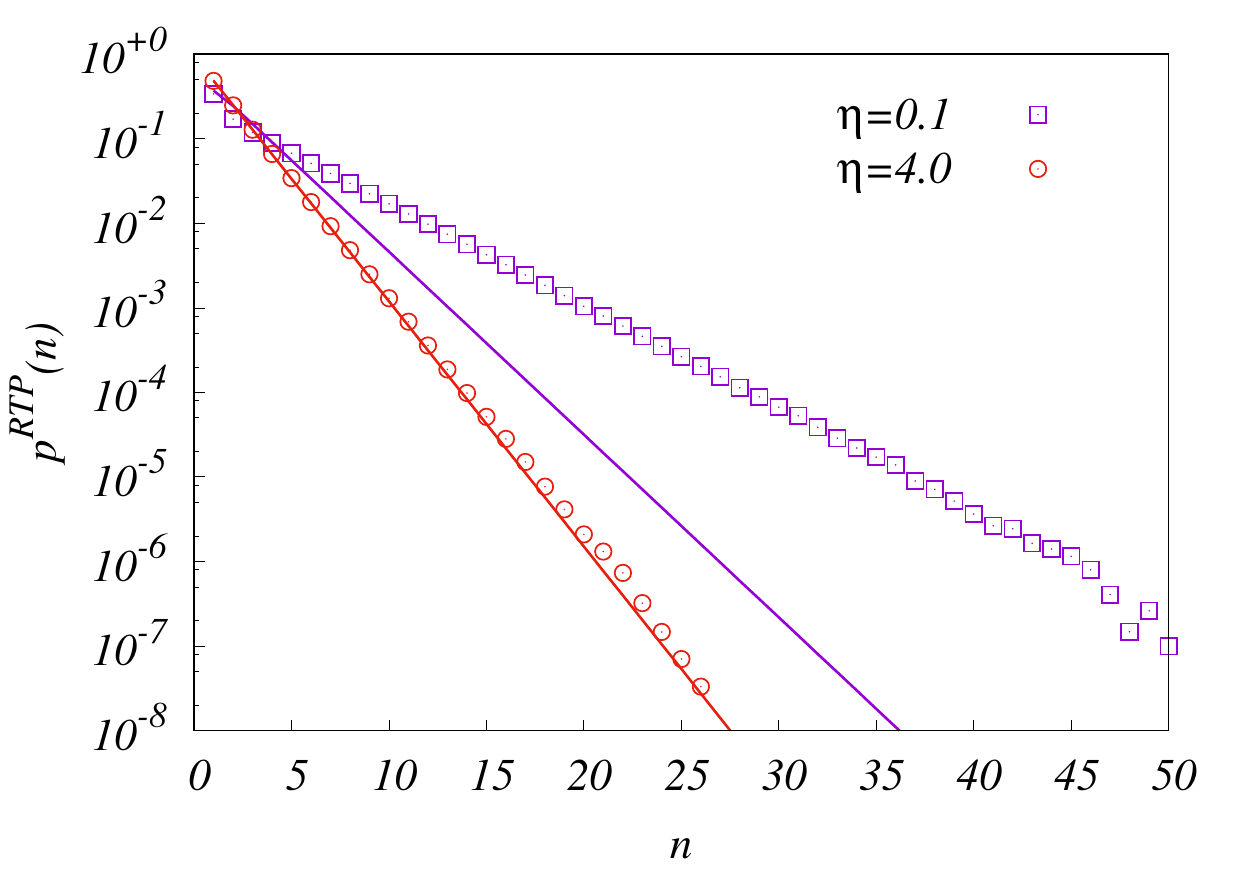}
	\caption{The probability of finding a cluster of size $n$ in the RTP picture, $p^{RTP}(n)$, for $\eta = 0.1$ and $4$ and $\rho=1$ or equivalently $\rho_{RTP} = 0.5$. The solid lines correspond to the results from mean field theory [see Eq.~(\ref{eq:occprob})]. The mean field predictions match with results from simulations  for large $\eta$. The data are for system size $L=500$ in the mass model picture.} 
	\label{particledist}
\end{figure}

In Fig.~\ref{particledist}, the cluster distribution in the RTP picture, obtained from Monte Carlo simulations, is compared with the mean field prediction in Eq.~(\ref{eq:occprob}) for different values of $\eta$. The distribution $p^{RTP}(n)$ is exponential for large $n$. We see that the mean field expression is able to describe the distribution accurately for larger $\eta$. For values of $\eta$ less than $1$, the mean field result under predicts the probabilities, thus failing to capture tendencies towards clustering. 

We now estimate the regime of validity of the mean field approximation. Consider the RTP picture. The mean distance between two particles is $\rho_{RTP}^{-1}$. If no spin flips occur, then the collision times are proportional to the inter-particle distance. For mean field theory to hold, there must be multiple spin flips during this time. Thus, we obtain the criteria for the validity of mean field theory to be $\eta^{-1} \ll  \rho^{-1}_{RTP}$, or effectively $\eta \gg (1+\rho)^{-1}$. The results from simulations [see Figs.~\ref{mfsimhigh}, \ref{mfsimlow}, and \ref{particledist}] are consistent with this criterion, where we find a good match between results from simulations and mean field predictions for $\eta \gtrsim 1$ for $\rho = 1$.

We now provide an alternative description of the mass model that is valid for small $\eta$, based on a mapping to  a single species birth 
and coalescence process.

\section{\label{sec:smallD}Coalescence-Fragmentation model for $\eta\ll 1$} 

In Sec.~\ref{sec:mf}, we showed that the numerical results for the mass distribution for $\eta \gtrsim 1$ are well described by the mean field results, while the mean field analysis fails for small values of $\eta$. In this section, we provide an alternate description of the dynamics that allows us to capture the mass distribution for the case  $\eta\ll 1$. We work, as above, in the mass transfer picture. For small $\eta$, when no site contains large masses, there is a separation of the time scales associated with spins and hopping. In this limit, we can  treat the mass transfers in an adiabatic approximation, wherein we assume the spins to be fixed when the masses move. 

\subsection{Relaxation to the steady state}

For simplicity, we assume that the initially the mass is distributed uniformly on the lattice. The system {\rahul first} evolves to a state in which only sites whose neighboring spins on the left and right are $+$ and $-$ respectively (denoted as $+-$ sites) have non-zero mass. These sites have both spins pointing into the site, and thus the mass at that site cannot move out. On the other hand, the masses at $++$, $--$, and $-+$ sites hop out to neighboring sites. Each $+-$ site has a basin of attraction, which consists of all the sites that are connected to it by spins pointing towards the site. Thus, for $\eta \ll 1$, the initial stage of evolution, which is of very short duration $\sim O(1)$ (the mean size of a basin of attraction is four), consists of  masses  moving to the $+-$ sites and staying there.

We now look at the dynamics on time scales of $O(\eta^{-1})$ and larger. In the adiabatic approximation, the masses are assumed to move only between $+-$ sites, considering occupancy of other sites as transient. A mass can only move out of a $+-$ site if one of its neighboring spins flips, leading to the mass being transferred to the nearest $+-$ site in the direction of the {\rahul flipped spin}. The typical time taken for this transfer is of the order of the typical mass-cluster size at that time. If this time is smaller than $\eta^{-1}$ (the time for a spin flip), the spins involved in this transfer can be assumed to be fixed. Looking at sites with non-zero mass as units $A$, this process can thus be modeled by the single-species coalescence reaction $A + A \rightarrow A$ ~\cite{krapivsky2010kinetic,burschka1989transition,ben1990statics}, with diffusion constant $\eta \la (\Delta d)^2 \ra$, where $\Delta d$ is the average distance over which a mass cluster is transferred after a spin-flip event.  From the known exact solution of $A + A \rightarrow A$, we obtain that the mean density of sites with non-zero mass $n_c(t)$, during this coalescence stage of evolution, decreases as
\be
n_c(t) \approx \frac{1}{\sqrt{2 \pi \eta \la (\Delta d)^2 \ra t}},\quad t \ll \eta^{-2}, \label{eq:nc_t0}
\ee
where the regime of validity  $t \ll \eta^{-2}$ will be shown later in this section. This implies that the typical mass $m(t)$ of an occupied site increases as
$
m(t) \approx \sqrt{2 \pi \eta \la (\Delta d)^2 \ra t}.
$

We now give an argument to calculate $\la (\Delta d)^2 \ra$ in the coalescence regime. A mass transfer is initiated when one of the spins of a $+-$ site flips, say the spin on the right. The mass is then transferred to the right from site to site until it encounters a $-$ arrow. The probability of the first $-$ arrow being a distance $\Delta d$ to the right is $P(\Delta d) = 2^{-\Delta d}$, $\Delta d = 1,2, \ldots$, and thus
\beq
\la (\Delta d)^2 \ra = 6. \label{eq:Deltad}
\eeq
Hence
\be
n_c(t) \approx \frac{1}{\sqrt{12 \pi \eta t}},\quad t \ll \eta^{-2}.
\label{eq:ntdecay} 
\ee

\begin{figure}
	\centering
	\includegraphics[width=\columnwidth]{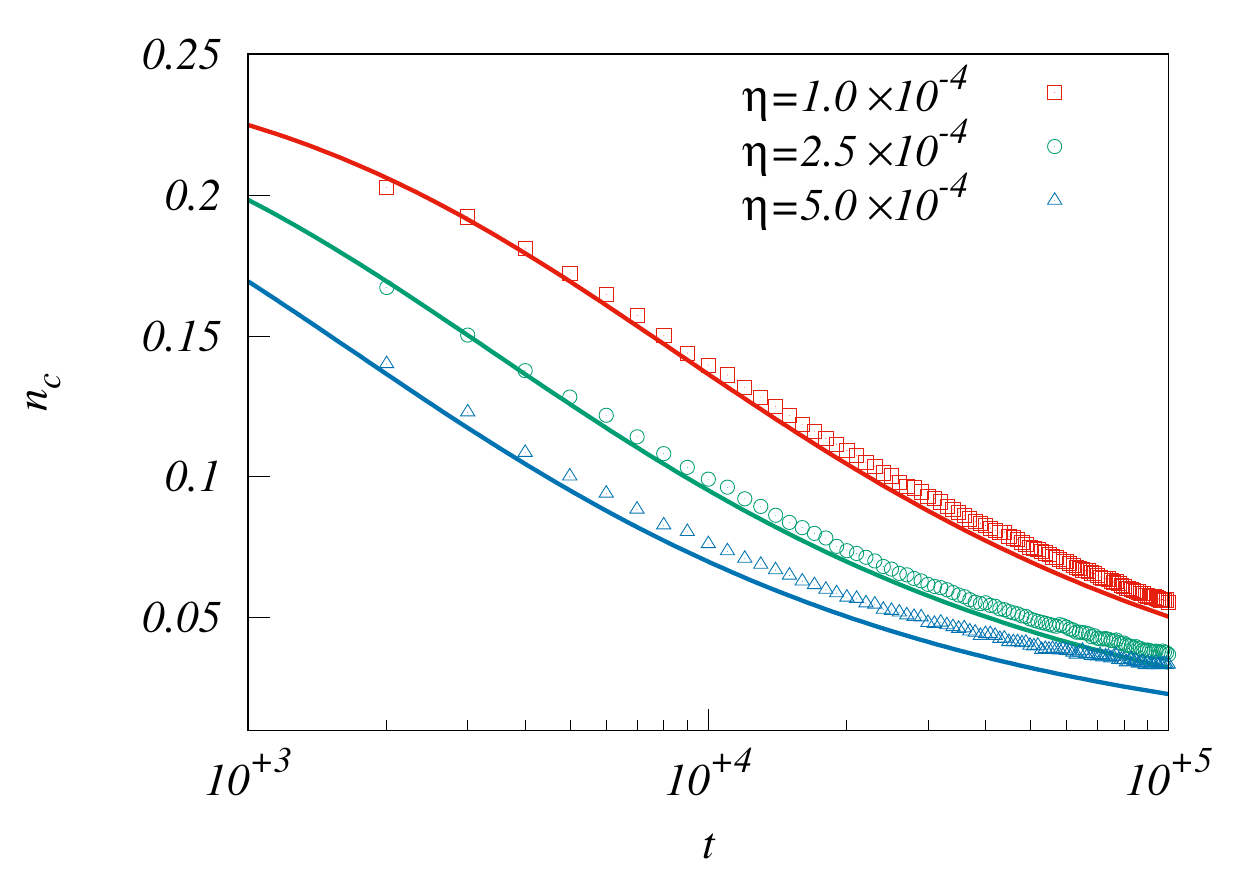}
	\caption{\label{lowDrnt}The variation of the mean density of clusters $n_c(t)$ with time $t$ for three different values of $\eta \ll 1$. The data are for density $\rho=1$, system size $L=1000$ and have been averaged over $200$ histories. The solid lines correspond to the analytical result in Eq.~(\ref{eq:nc_t}), with $a$ and $b$ as given in the text.} 	
\end{figure}

This result  is valid for times $t \gg \eta^{-1}$. In the case of the RTP in the mass transport picture, $A$ sites are $+-$ sites, which are an average distance $4$ apart. For $t \sim \eta^{-1}$, we find better agreement with the interpolation
\be
n_c(t) \approx \frac{b}{\sqrt{12 \pi \eta \left(t + a \eta^{-1}\right)}}, \label{eq:nc_t}
\ee
where $a$ and $b$ are constants that do not depend on $\eta$ (for $\eta \ll 1$). {\rahul Effectively, the equation says that for such an initial condition,} time is shifted to $\tau = t + a\eta^{-1}$. The constants $a$ and $b$ are easily estimated by the observation that for $t \ll \eta^{-1}$, all masses are present only on $+-$ sites, and for initial densities larger than $1/4$, all $+-$ sites are occupied. This implies that $n_c(t) \rightarrow \frac{1}{4}$ as $t \rightarrow 0$. We also know that $n_c(t)$ should approach the form in Eq.~(\ref{eq:nc_t0}) for $t \gg \eta^{-1}$. Hence, we obtain $b=1$ and $a= 4 (3 \pi)^{-1}$. From Fig.~\ref{lowDrnt}, we see that Eq.~(\ref{eq:nc_t}) describes the numerical data for $n_c(t)$ well.  

\begin{figure}
	\centering
	\includegraphics[width=\columnwidth]{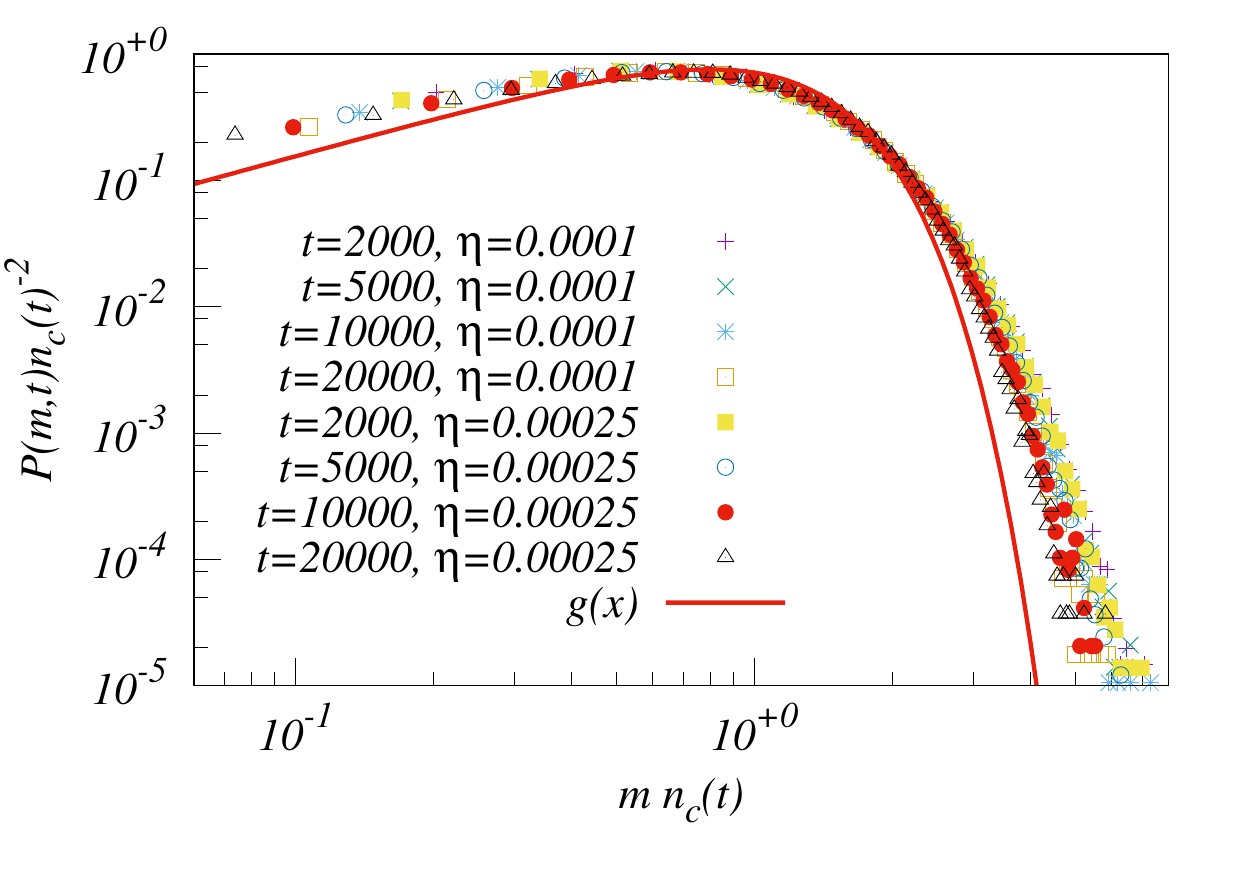}
	\caption{The numerically obtained data for $P(m,t)$, the probability of having a mass $m$ on a site at time $t$, for different times and flip rate $\eta$ collapse onto a {\rahul single} curve when  $P(m,t)$ and mass are scaled as in  Eq.~(\ref{eq:Pmt}). The scaled data is reasonably well described by the scaling function $g(x)$ (shown by solid line) in Eq.~(\ref{eq:scaling}) with $\rho=1$.}
	\label{lowDrscal}
\end{figure}

It is possible to keep track of the time dependent  mass distribution by considering the coalescence process as  
constant-kernel aggregation reaction 
$A_i + A_j \rightarrow A_{i+j}$ in one dimension. From the known solution of this problem~\cite{kang1985fluctuation,spouge1988exact,krishnamurthy2002kang,krapivsky2010kinetic}, the mass distribution, during the coalescence stage of evolution, can be predicted  to be 
\be
P(m,t) \approx n_c(t)^2 g \left(mn_c(t)\right). \label{eq:Pmt}
\ee
For a uniform initial distribution of mass and the case where the diffusion is only to the nearest neighbour sites, the scaling function $g(x)$ is known to be given by
\be
g(x) = \frac{\pi x}{2 \rho^2} \exp\left( \frac{-\pi x^2}{4 \rho^2}\right).
\label{eq:scaling}
\ee
In Fig.~\ref{lowDrscal}, we show the numerically obtained $P(m,t)$ for different times and spin rates $\eta$. When the variables are scaled as in Eq.~(\ref{eq:Pmt}), the data for different $\eta$ and times collapse onto a single curve, showing that the scaling collapse is excellent, except at large values of the argument, where it is possible that the data has not reached the scaling limit. This shows that the adiabatic approximation is able to capture the approach to the steady state very well. It is seen that the curve of the scaling collapse deviates from the known scaling function $g(x)$ in  Eq.~(\ref{eq:scaling}), due to the fact that the evolution in the RTP system does not start with uniformly distributed mass, but with mass only on $+-$ sites.

\subsection{Characterization of the steady state}

The adiabatic approximation of treating the spins as fixed during mass transfer breaks down when typical masses become large and there is no longer a clear separation between the spin flip rates and mass transfer rates.  The evolution can no longer be modeled as a pure coalescence process. An example of interrupted mass transfer is when  a second spin flips while mass transfer is going on.

\begin{figure}
	\centering
	\includegraphics[width=\columnwidth]{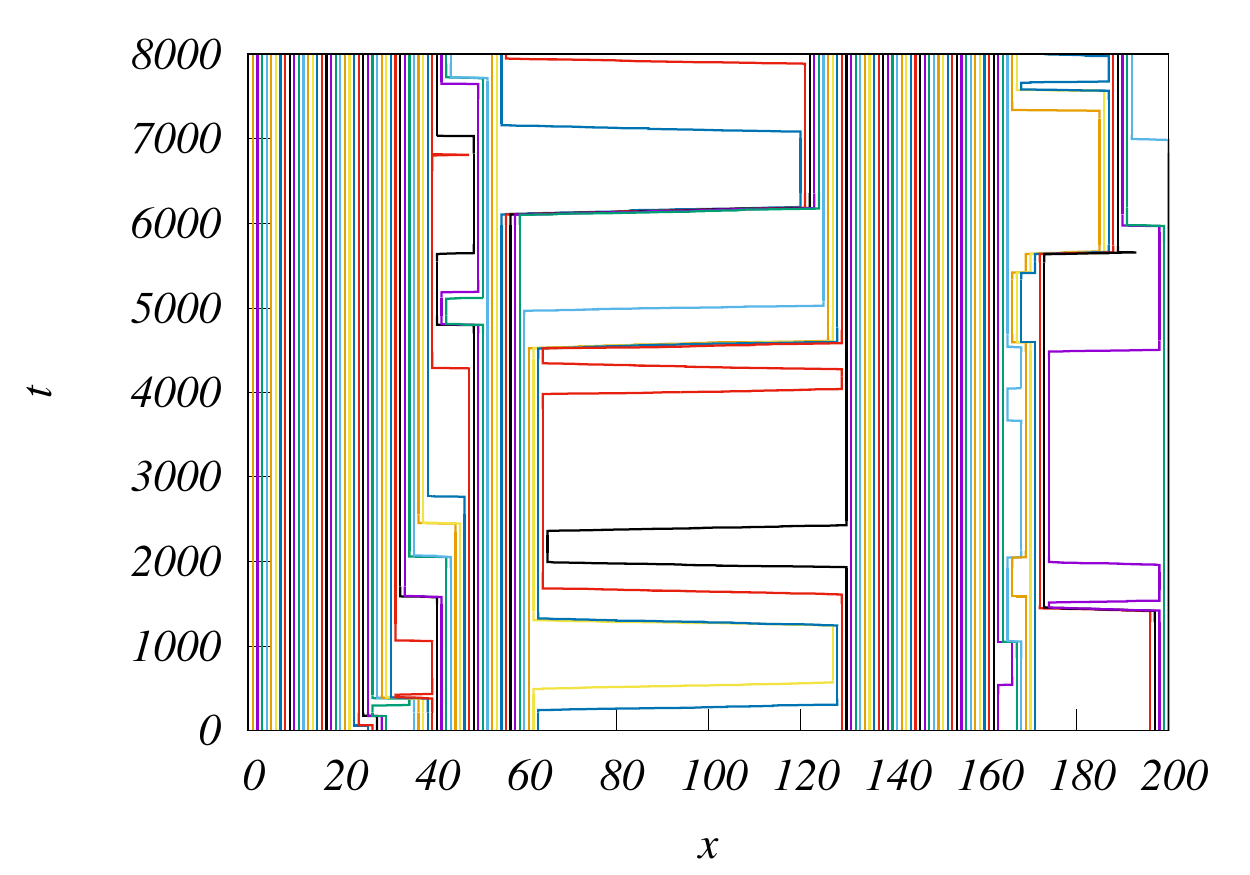}
	\caption{\label{trajectory} { Steady state space-time trajectories of $N=100$ run and tumble particles on a lattice of size $L=200$ for a low spin flip rate $\eta=0.001$. The trajectories of different particles are  represented by different colors. Large clusters of particles are observed, with infrequent movements of particles triggered by spin flips.}}
\end{figure}

Consider a $+-$ site. Suppose one of the neighboring spins flips so that the mass at the site starts transferring in the direction of the spin one by one. If the other neighboring spin flips before the total mass has been transferred, then the mass at the site ends up at both its neighbors.  This leads  the reaction $0A0 \rightarrow A0A$, where $0$ denotes a vacancy. If, on the other hand, the other neighboring spin does not flip, we have a diffusion move $A0 \rightarrow 0A$. At large times $t \gg \eta^{-2}$, there is a balance between diffusion, coalescence and breaking up, and the system reaches a steady-state.
{
As a visual demonstration of the dynamics, Fig.~\ref{trajectory} shows the steady state space-time trajectories of $100$ RTPs, in the RTP picture, for a low spin flip rate $\eta=0.001$. We observe the formation of large clusters of particles and also of vacancies (which can be identified with mass clusters in the mass description). Infrequent movements of particles are seen, corresponding to spin flips, can also be seen, leading over a large time-scale to equilibration of the clusters.
}

We now determine the steady-state value of $n_c$ by considering the coalescence and fragmentation rates. We assume below that in addition to $\eta \ll 1$, the number of clusters $n_c \ll 1$, and only consider terms to leading order in $n_c$.

We first calculate the coalescence rate. For the purposes of this section, we assume that sites are independently occupied or empty with probability $n_c$ and $(1-n_c)$ respectively, except for the necessary caveat that two occupied sites cannot be next to each other. To leading order in $n_c$, this is consistent with the more detailed calculation of the steady-state using empty interval probabilities, given in Appendix C. 

Then, to leading order in $n_c$, the probability that two sites on a lattice separated by a distance $n$ are both occupied is $n_c^2 (1+ \epsilon(n))$, where $\epsilon(n)$ is $O(n_c)$. Now, consider the two occupied $+-$ sites separated by a distance $n$. The sites will coalesce into one if (a) the $-$ arrow on the site on the left or the $+$ arrow on the site on the right flip, which happens at rate $2 \eta$, and (b) all the arrows on the (empty) sites in between the two sites are {\rahul to} pointing the right, in the former case, and to the left in the latter case. If the sites are separated by a distance $n$, there are $n-2$ arrows in between, this probability is $2^{-n+1}$. Hence, the total rate of coalescence is 
\beq
r_c = 2 \eta n_c^2 \sum_{n=2} 2^{-n+2} (1+ \epsilon(n)) = 4 \eta n_c^2 + O(\eta n_c^3).
\eeq

Now, we calculate the fragmentation probability. A fragmentation occurs when, during the course of a mass transfer, a spin between the original site and the target site flips, interrupting the transfer. Now, consider an initial flip of the right spin of an occupied $+-$ site, leading to a transfer to a site a distance $k$ to the right. This requires that the arrows at all intervening sites be pointed to the right, and the $k^{th}$ arrow be pointed to the left, to stop the mass transfer. The rate of this process is 
\beq
\eta \times n_c \times 2^{-k}.
\eeq

Now, consider the case where the $(k-1)^{th}$ spin from the original site flips. For $k=1$ this means the $-$ spin on the original site flips back from $+$ to $-$. Then all the mass simply ends up at a new site a distance $k-1$ from the original, since the original target site is no longer $+-$ but is $--$ now, and thus cannot hold mass. In this case fragmentation does not occur. 

However, if any of the $k-1$ sites between the original and the target flip during the transfer, the mass will end up at two sites. If the left spin on the original site flips during the transfer, some of the mass will start getting transferred to the left, and in this case too, the mass will end up at two sites. A mass transfer over $k$ sites takes a time $m^* + \delta(k)$, where $m^*$ is the average mass on an occupied site, and $\delta(k)$ is a correction of $O(1)$. The probability that one of $k$ spins flips during this time is, to first order,
\beq
\eta \times k \times (m^* + \delta(k)).
\eeq

And hence, the total fragmentation rate is (the factor of 2 in front signifies that the case of the mass transfer being initiated by the $+$ spin on the original site flipping can be treated in exactly the same fashion)
\beqa
r_f &=& 2 \eta^2 n_c (m^* + \delta(k) ) \sum_{k=1} 2^{-k} k \nonumber\\
 &=& 4 \eta^2 n_c m^* + O(\eta^2 n_c m^*).
\eeqa

A steady-state is reached when the coalescence rate is equal to the fragmentation rate, $r_c = r_f$, which gives
\beq
n_c = \eta m^*.
\eeq

We also use the equality $n_c m^* = \rho$, the average density, to eliminate $m^*$ from the above equation. Thus, we finally obtain
\beqa
n_c &=& \sqrt{\eta \rho}, \label{n_cm}\\
m^* &=& \sqrt{\frac{\rho}{\eta}}, \label{eq:mstar}
\eeqa
to leading order.

\begin{figure}
\centering
\includegraphics[width=\columnwidth]{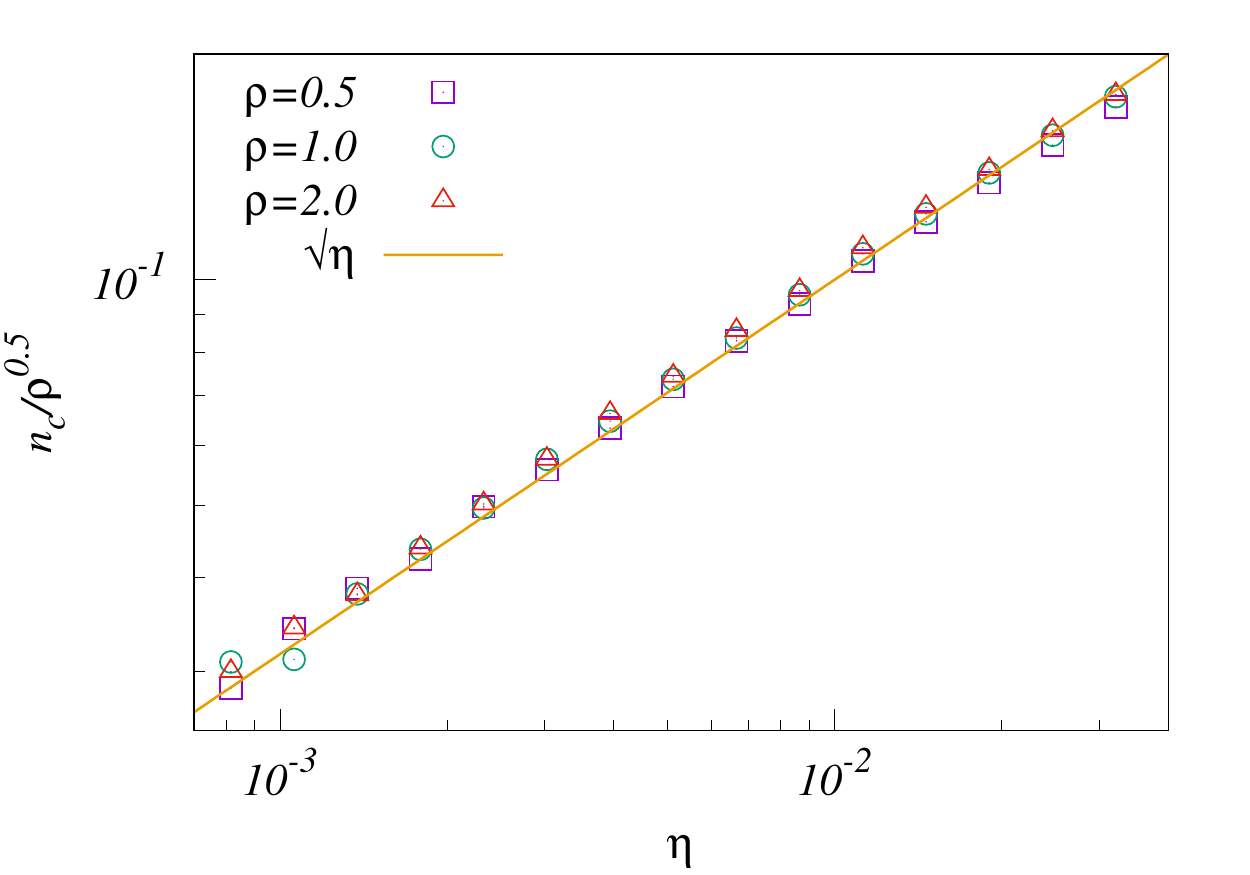}
\caption{\label{lowDrscal3}The variation of the steady state mean density of clusters $n_c$ with spin flip rate $\eta$ for different mass densities $\rho$. The data for different $\rho$ collapse onto a single curve when scaled as in Eq.~(\ref{n_cm}). The scaled data is well described the theoretical prediction (shown as solid line) in  Eq.~(\ref{n_cm}). The data are for system size $L= 500$.}
\end{figure}

We now compare the predictions from the effective aggregation{\rahul-}fragmentation model discussed above with results from simulations of the mass model. Fig. \ref{lowDrscal3} shows the variation of  the steady state mean $n_c$ with different densities $\rho$ and spin flip rates $\eta$. The data for different densities collapse onto a curve when $n_c$ is scaled by $\sqrt{\rho}$, as predicted by Eq.~(\ref{n_cm}). From the figure, it is clear that the dependence on $\eta$ is as predicted by Eq.~(\ref{n_cm}), right upto the constants. 

\begin{figure}
	\centering
	\includegraphics[width=\columnwidth]{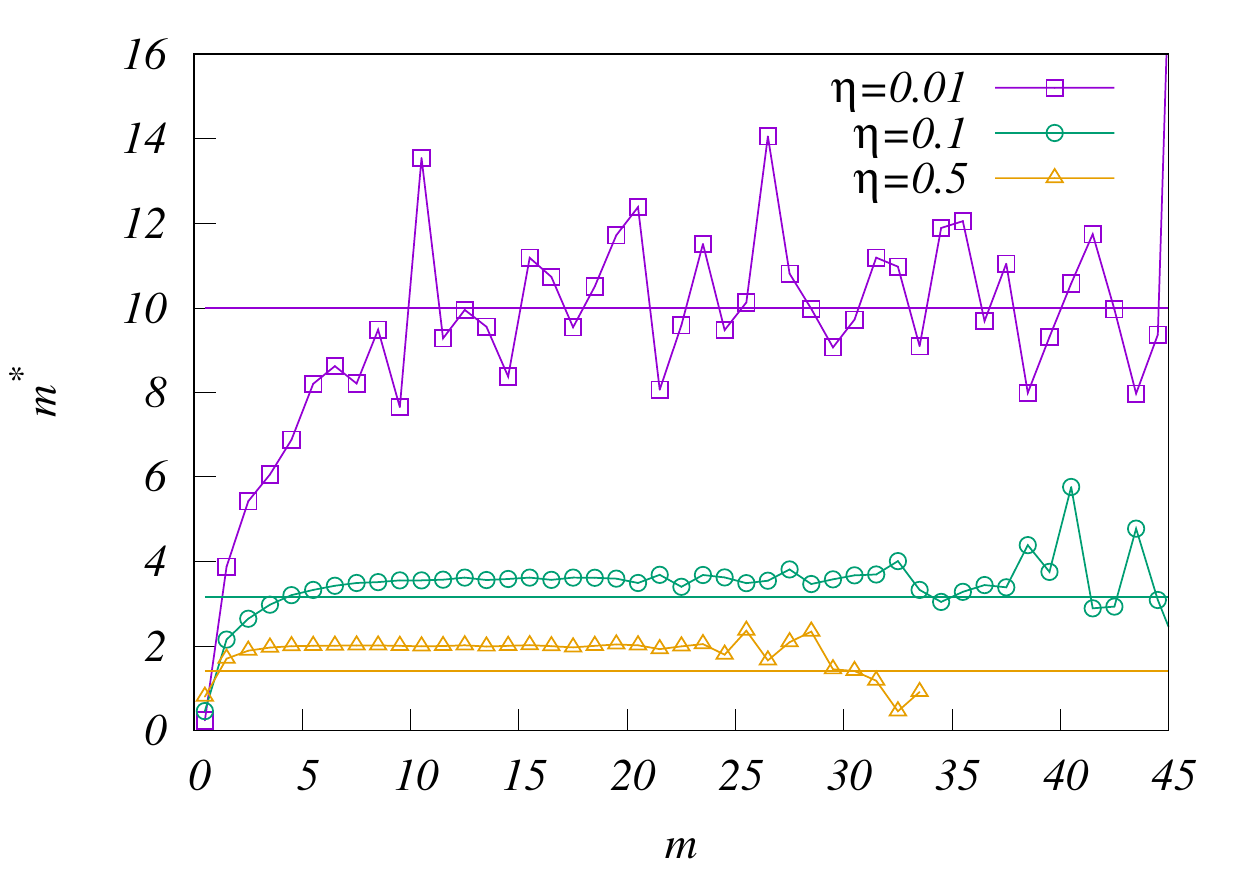}
	\caption{The variation of the typical mass $m^*(m)$ [as defined in Eq.~(\ref{eq:convergence})] with $m$ for three different values of spin flip rate $\eta$. The solid lines are the theoretical predictions for $m^*$ from Eq.~(\ref{eq:mstar}). For small $\eta$, $m^*(m)$ converges to the theoretical value for large $m$. }
	\label{lowDrscal4}
\end{figure}

In Fig.~\ref{lowDrscal4}, we compare the predictions for $m^*$ with estimates from simulations for small $\eta$. To obtain $m^*$ from simulations, we first measure the mass distribution $P(m)$. For each $m$, we obtain $m^*(m)$ as
\be
\frac{1}{m^*(m)}=\frac{1}{2} \ln \frac{P(m-1)}{P(m+1)}.
\label{eq:convergence}
\ee
$m^*(m)$ converges to $m^*$ for large $m$. The variation of $m^*(m)$ with $m$ is shown in Fig.~\ref{lowDrscal4} for three values of $\eta$. It is clear that for small $\eta$, we find an excellent match between the theoretical prediction in Eq.~(\ref{eq:mstar}) and numerical results. 

\begin{figure}
	\centering
	\includegraphics[width=\columnwidth]{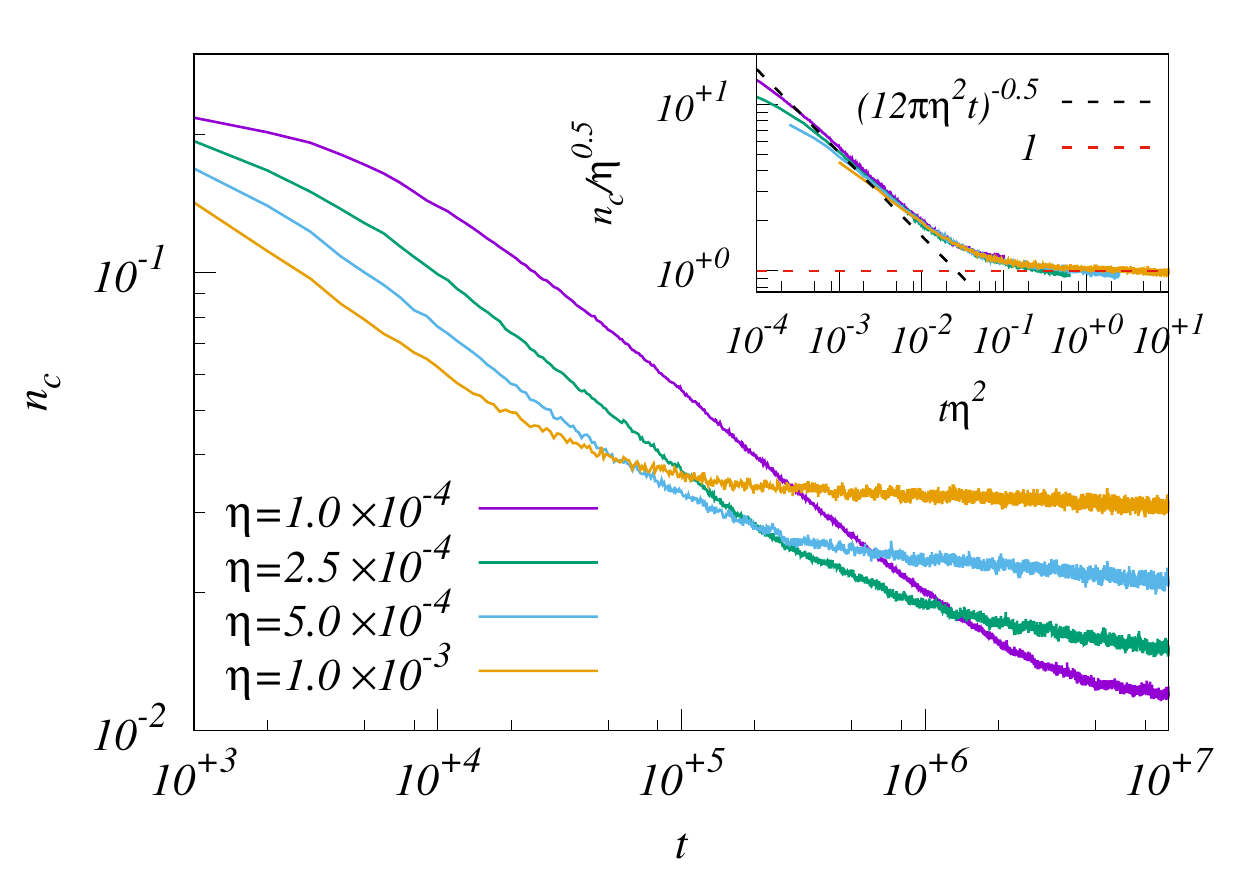}
	\caption{The variation of the mean density of clusters $n_c(t)$ with time $t$ for four different values of $\eta \ll 1$. The data are for density $\rho=1$ and system size $L=1000$ and have been averaged over $200$ histories. Inset: The data for different times collapse onto a single curve when $t$ and $n_c$ are scaled as in Eq.~(\ref{nc_t2}). The dotted lines are the theoretical predictions for the asymptotic behavior of the scaling function (see text after Eq.~(\ref{nc_t2})).  }
	\label{lowDrscalnt}
\end{figure}

Knowing the initial temporal decay $n(t) \approx \frac{1}{\sqrt{12 \pi \eta t}}$ [see Eq.~(\ref{eq:ntdecay})], and the steady state value $n(t) \approx \sqrt{\eta \rho}$, we can develop a scaling form for the behavior of $n(t)$. The crossover time is obtained by equating the early and late time behaviors to give $t^* \sim 1/(\rho \eta^2)$. Thus, we can write,
\be
n_c(t)  = \sqrt{\eta \rho} f(t \rho \eta^{2}), ~t\to \infty, ~\eta \to 0, \label{nc_t2}
\ee
where the scaling  function $f(x) \approx 1$ for $x \rightarrow \infty$, and $f(x) \approx 1/\sqrt{ 12 \pi x}$ for  $x \rightarrow 0$.
\fref{lowDrscalnt} shows the variation of the mean density of clusters $n_c(t)$ with time $t$ for four different values of $\eta \ll 1$. As $\eta$ decreases the extent of the coarsening regime increases, and the final number of clusters in the steady-state decreases, as expected from Eq.~\eqref{n_cm}. The inset in \fref{lowDrscalnt} shows that the data for different times collapse onto a single curve when the different variables are scaled as in Eq.~(\ref{nc_t2}), confirming the scaling hypothesis. In addition, the numerical data are consistent with the predictions for the asymptotic behavior of the scaling function. 

{\rahul We briefly outline how our results translate to the RTP picture. The single-site mass distribution is the distribution of gaps between RTPs, and since only $+-$ sites are occupied (to the lowest order in $\eta$), we conclude that all RTP clusters have particles are bounded by right-moving RTPs on the left and left-moving RTPs on the right. The single-site mass distribution $P(m)$ for $m>0$ is the distribution of the lengths of the gaps between RTP clusters. The empty-interval probabilities $E_n$ for finding an empty interval of length $n$ (see Appendix C) between occupied $+-$ sites give the cluster size distribution of the RTP clusters, through the relation $P_{cluster}(n) = E_{n}+E_{n-2}-2 E_{n-1}$ \cite{ben1990statics}.}

We conclude that the aggregation-fragmentation model complements the mean field approximation in providing an accurate description of the mass model for small values of the spin flip rate $\eta$.

\section{\label{sec:hydro} Hydrodynamics for large $\eta$: mean field approximation} 

In this section, we derive hydrodynamic equations along the lines of recently developed macroscopic fluctuation theory~\cite{BertiniPRL2001}, within the mean-field approximation, which is valid for $\eta \ge 1/(1+ \rho)$. We derive explicit expressions for the diffusion and drift coefficients, $D(\rho)$ and $\chi(\rho)$ [see Eq.~(\ref{eq:hydrodef}) for definitions], and test the Einstein relation
\beq
\frac{1}{\sigma^2(\rho)} = \frac{D(\rho)}{\chi(\rho)},
\label{eq:einstein}
\eeq
where $\sigma^2(\rho)$ is the single-site mass fluctuations. 
To derive the hydrodynamic  equations for the mass model, we consider the current across a bond in presence of a small density gradient $\partial \rho / \partial x$, and in presence of a small field of strength $F \ll 1$, that couples to the mass \cite{BertiniPRL2001,das2017einstein,ChakrabortyCondmat2020}. In the presence of the field, particles hop to the right with rate $1+F$ and to the left with rate $1-F$. The average current across a bond $\langle J \rangle$ is then expressed in terms of the hydrodynamic coefficients $D(\rho)$ and $\chi(\rho)$ as
\beq
\la J \ra = -D(\rho) \frac{\partial \rho}{\partial x} + F \chi(\rho). \label{eq:hydrodef}
\eeq

{\rahul In Appendix B, we derive the relationship between the hydrodynamic coefficients in the original, RTP, picture, and those in the gap picture, which are our main focus for the next two sections.}

The mean current between sites $i$ and $(i+1)$ can be written in terms of the steady state mass distribution. In the presence of the field, these distributions become site dependent. We introduce the following notation. Let the conditional probability of finding a mass $m$ on site $i$, given the site has a $-$ bond to the left and a $+$ bond to the right, be denoted as $\mathcal{P}^{-+}_{m,i}$. Similar definitions hold for $\mathcal{P}^{++}_{m,i}$, $\mathcal{P}^{--}_{m,i}$ and $\mathcal{P}^{+-}_{m,i}$. These distributions are, in the presence of a field, different from the distributions $P^{+-}, P^{-+}, P^{++},$ and $P^{--}$ defined in Sec.~\ref{sec:mf} for the homogeneous case. We also note that the symmetry $\mathcal{P}^{++}_{m,i}= \mathcal{P}^{--}_{m,i}$ is no longer valid in the presence of a field or density gradient.

The mean current between sites $i$ and $i+1$ has contributions from particles hopping from $i$ to $i+1$ (denoted by  $\la J^{+}_{i,i+1} \ra$) and particles hopping from $i+1$ to $i$ (denoted by $\la J^{-}_{i,i+1} \ra$).  Clearly,
\beqa
\la J^{+}_{i,i+1} \ra &=& \frac{1+F}{4}\left[ \sum_{m=1}^\infty \mathcal{P}_{m, i}^{++} + \sum_{m=1}^\infty \mathcal{P}_{m, i}^{-+} \right],\nonumber\\
&=& \frac{1+F}{4}\left( 2-\mathcal{P}^{-+}_{0, i}-\mathcal{P}^{++}_{0, i} \right).
\eeqa
Similarly, 
\beq
\la J^{-}_{i,i+1} \ra = -\frac{1-F}{4} \left(2 - \mathcal{P}^{-+}_{0, i+1} - \mathcal{P}^{--}_{0, i+1} \right).
\eeq
The average current across the bond is the sum of the two currents:
\beqa
\la J_{i,i+1} \ra &=&\frac{  \mathcal{P}^{-+}_{0, i+1}+ \mathcal{P}^{--}_{0, i+1} - \mathcal{P}^{-+}_{0, i}- \mathcal{P}^{++}_{0, i}}{4} 
\nonumber \\
&+&F \frac{4- \mathcal{P}^{-+}_{0, i+1}-\mathcal{P}^{--}_{0, i+1} - \mathcal{P}^{-+}_{0, i}- \mathcal{P}^{++}_{0, i}}{4}. 
\label{Jbpbm1}
\eeqa

The current depends on the site dependent values of $ \mathcal{P}^{-+}_{0, i}$, $\mathcal{P}^{++}_{0, i}$ and $\mathcal{P}^{--}_{0, i}$. The mass model is not a pure mass transfer process, but has other degrees of freedom, the spins on the bonds that do not relax infinitely fast.  It is not surprising that even a small density gradient or field modifies the steady-state probabilities. Thus, to calculate the current in the presence of a field or gradient, we also need to calculate the modified  steady-state mass distributions. 

The different probability distributions evolve in time according to
\begin{widetext}
	\bea
	\frac{d\mathcal{P}^{-+}_{m,i}}{dt} &= & \eta \left[\mathcal{P}^{++}_{m,i}+\mathcal{P}^{--}_{m,i} - 2\mathcal{P}^{-+}_{m,i} \right] + 2 \mathcal{P}^{-+}_{m+1,i} -2 (1-\delta_{m,0}) \mathcal{P}^{-+}_{m,i},
	\label{eq:mp}\\
	\frac{d\mathcal{P}^{+-}_{m,i}}{dt} &= &  \eta \left[ \mathcal{P}^{++}_{m,i} +\mathcal{P}^{--}_{m,i} - 2 \mathcal{P}^{+-}_{m,i} \right] \!+\!
	\left[ \mathcal{P}^{+-}_{m-1,i} (1 \!- \! \delta_{m,0})-\mathcal{P}^{+-}_{m,i}  \right] 
	\!\!\left[ \!(1\!+\!F) \frac{2- \mathcal{P}^{-+}_{0, i-1} \!-\!\mathcal{P}^{++}_{0, i-1}}{2} \!+\! (1\!-\!F) \frac{2- \mathcal{P}^{-+}_{0, i+1} -\mathcal{P}^{--}_{0, i+1}}{2} \right], \label{eq:pm}\\
	\frac{d\mathcal{P}^{++}_{m,i}}{dt}&=&  \eta \left[ \mathcal{P}^{+-}_{m,i} +\mathcal{P}^{-+}_{m,i} - 2 \mathcal{P}^{++}_{m,i} \right]+
	\frac{1+F}{2}(2- \mathcal{P}^{-+}_{0, i-1} -\mathcal{P}^{++}_{0, i-1}) \left[\mathcal{P}^{++}_{m-1,i} (1-\delta_{m,0})-\mathcal{P}^{++}_{m,i} \right] \nonumber \\
	&&+(1+F) \left[ \mathcal{P}^{++}_{m+1,i} - (1-\delta_{m,0}) \mathcal{P}^{++}_{m,i} \right], \label{eq:pp}\\
	\frac{d\mathcal{P}^{--}_{m,i}}{dt}&=&  \eta \left[ \mathcal{P}^{+-}_{m,i} +\mathcal{P}^{-+}_{m,i} - 2 \mathcal{P}^{--}_{m,i} \right]+
	\frac{1-F}{2}(2- \mathcal{P}^{-+}_{0, i+1} -\mathcal{P}^{--}_{0, i+1}) \left[\mathcal{P}^{--}_{m-1,i} (1-\delta_{m,0})-\mathcal{P}^{+--}_{m,i} \right] \nonumber \\
	&&+(1-F) \left[ \mathcal{P}^{--}_{m+1,i} -(1-\delta_{m,0}) \mathcal{P}^{--}_{m,i} \right]. \label{eq:mm}
	\eea
\end{widetext}
In the following we solve for the probabilities $\mathcal{P}_{m,i}$ as a series expansion in the density gradient $\rho'(x)=\partial\rho/\partial x$ in Sec.~\ref{sec:gradient} and field $F$ in Sec.~\ref{sec:field}.

\subsection{Current in the presence of a density gradient \label{sec:gradient}}

In this section, we solve for the steady state mass distribution in the presence of a non-zero density gradient $\rho'$, but field $F=0$. 
We look for perturbative solutions of the kind
\be
\mathcal{P}^{ab}_{m,i}= P^{ab}_m (\rho_i) + \rho' Q^{ab}_{m,i} + O(\rho'^2), ~~~a,b=+,-,
\ee
where $P^{ij}_m$ is the steady state solution in the absence of a field, as obtained in Sec.~\ref{sec:mf}, and $Q^{ab}_{m,i}$ is the 
first order correction. 

Some of the $Q^{ab}_{m,i}$ may be determined from symmetry considerations. If the sign of $\rho'$ is reversed, it is clear that $\mathcal{P}^{++}_{m,i} (\rho') = \mathcal{P}^{--}_{m,i} (-\rho')$. In addition, $\mathcal{P}^{+-}_{m,i} (\rho') = 
\mathcal{P}^{+-}_{m,i} (-\rho')$, and $\mathcal{P}^{-+}_{m,i} (\rho') = 
\mathcal{P}^{-+}_{m,i} (-\rho')$. This implies that
\bea
Q^{--}_{m,i} &=& -Q^{++}_{m,i}, \label{eq:symmetry1}\\
Q^{+-}_{m,i} &=& Q^{-+}_{m,i}=0. \label{eq:symmetry0}
\eea
For convenience, we also introduce the  notation
$
\beta_{1,i}=Q^{++}_{0,i}
$, such that
\be
\mathcal{P}^{++}_{0,i} (\rho') = \beta(\rho_i)+  \rho' \beta_{1,i},
\label{eq:beta1def}
\ee
consistent with the notation in Eq.~(\ref{eq:alpha}).
Note that the leading order term depends on the site $i$ through the site-dependent density. For notational convenience, henceforth, we will not explicitly denote this dependence.

To determine $\beta_{1,i}$, consider Eq.~(\ref{eq:pp}) for $\mathcal{P}^{++}_{m,i}$. 
Expanding the different quantities to order $\rho'$, and using Eq.~(\ref{eq:symmetry0}), it is straightforward to show that
the term proportional to $\rho'$ satisfies
\bea
&&0=  Q^{++}_{m+1,i} - Q^{++}_{m,i} \left[2 \eta + 1-\delta_{m,0} +\frac{2- \alpha -\beta}{2}\right] \nonumber \\
&& +\frac{2- \alpha-\beta}{2} (1-\delta_{m,0}) Q^{++}_{m-1,i} \nonumber\\
&& +
\frac{1}{2}\left[\frac{d (\alpha+\beta)}{d \rho} -\beta_1 \right] \left[P^{++}_{m-1} (1-\delta_{m,0})-P^{++}_{m}
\right]. \label{eq:pp1}
\eea
Consider the generating function
\be
\widetilde{Q}^{++}_i(z) = \sum_{m=0}^{\infty} Q^{++}_{m,i} z^m,
\ee
with $\widetilde{P}^{++}(z) = \sum_{m=0}^{\infty} P^{++}_{m} z^m$ as defined earlier in Eq.~(\ref{eq:genfunct}). Multiplying Eq.~(\ref{eq:pp1}) by $z^m$ and summing over $m$, we obtain
\be
\widetilde{Q}^{++}_i(z)  = 
\frac{ z (1-z)   \left[\frac{d (\alpha+\beta)}{d \rho} -\beta_1 \right]   \widetilde{P}^{++}(z) +2 (1-z) \beta_1}
{(2-\alpha-\beta) z^2 -\left( 4-\alpha -\beta+4 \eta \right) z +2}.
\label{eq:pp2}
\ee

Here, $\beta_1$ is still undetermined. We determine it using the root cancellation method that was used in the mean field approximation. Equation~(\ref{eq:pp2}) has two poles. Of these, the pole 
\beq
z_c = \frac{4+4 \eta-\alpha-\beta \!- \! \sqrt{\left(4+4 \eta-\alpha-\beta\right)^2 \!-\!8 (2-\!\alpha\!-\beta)}}{2 (2-\alpha-\beta)} \label{eq:zc}
\eeq
is less than $1$. This pole will contribute to an exponentially diverging probability unless $z_c$ is a zero of the numerator. This implies that
\be
\beta_1=\frac{ z_c  \widetilde{P}^{++}(z_c) }{z_c    \widetilde{P}^{++}(z_c)-2}   \frac{d }{d \rho} (\alpha+\beta).
\label{eq:beta1}
\ee

\begin{figure}
	\centering
	\includegraphics[width=\columnwidth]{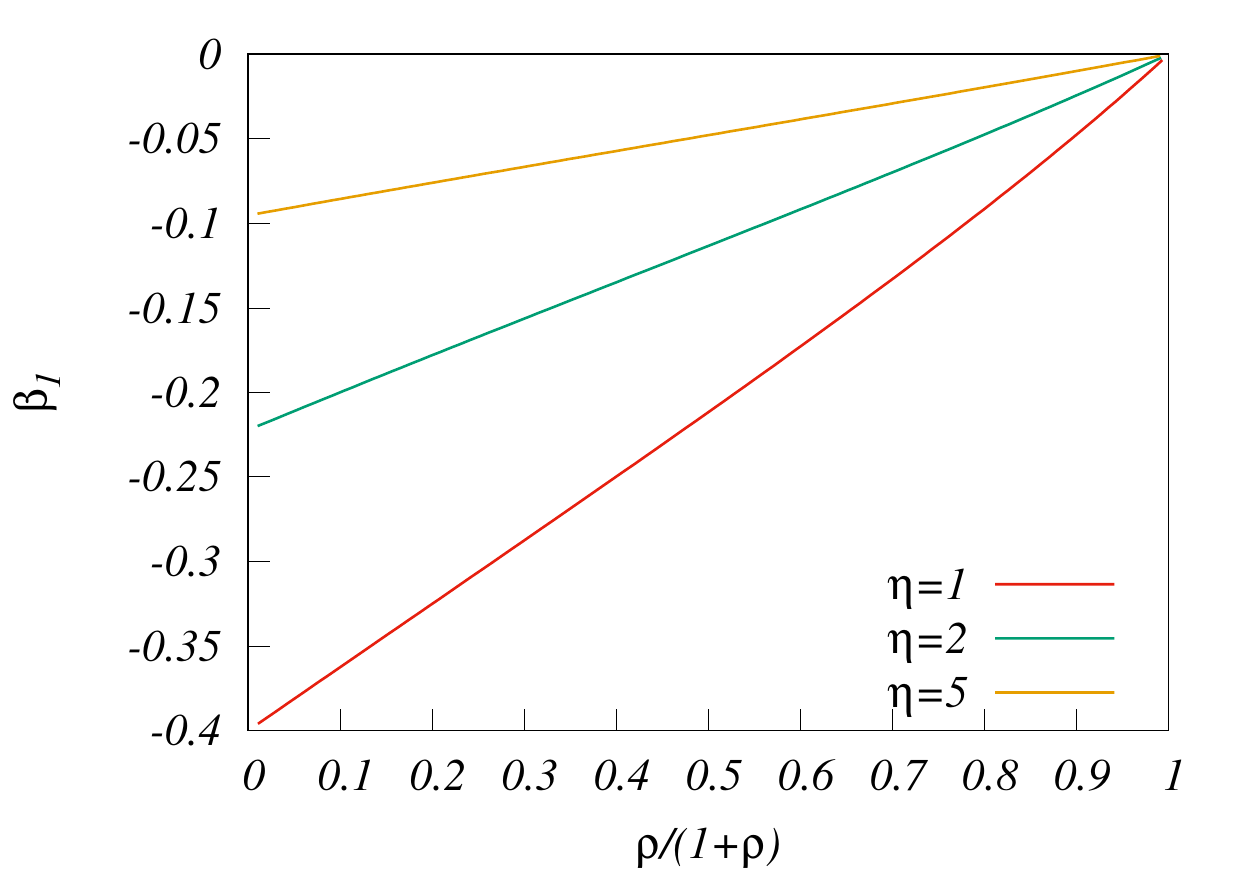}
	\caption{The variation of $\beta_1$, the correction to the steady state due to the presence of a density gradient [see Eq.~(\ref{eq:beta1def}) for definition] with density  $\rho/(1+\rho) = 1-\rho_{RTP}$ for (from top to bottom) $\eta = 1, 2$ and $5$. We see that the contribution of $\beta_1$ to $D(\rho)$ is significant for moderate $\eta$ and low $\rho$, but decreases with increasing $\eta$ and $\rho$.}
	\label{b1}
\end{figure}

In Fig.~\ref{b1}, we show the variation of  $\beta_1$ with density $\rho/(1+\rho) = 1-\rho_{RTP}$ for different spin flip rate $\eta$. It can be seen that for most values of  $\rho$, the correction is $O(1)$, and hence not negligible. It is also seen that $|\beta_1|$ decreases with $\eta$ for fixed density. This is reasonable since in the $\eta \rightarrow \infty$ limit the RTP approaches a simple exclusion process, for which $\beta_1 = 0$. $|\beta_1|$ is also smaller for larger $\rho$, as the probability of a site being empty decreases in this limit, and hence both $\beta_1$ and $\beta$ also decrease.

We can now calculate the current in Eq.~(\ref{Jbpbm1}) when the field $F=0$. Expanding to order $\rho'$, we obtain
\be
\la J \ra = \frac{1}{4} \left[\frac{d}{d\rho} (\alpha+\beta) - 2 \beta_1\right] \rho'.
\ee
Comparing with Eq.~(\ref{eq:hydrodef}), we immediately read {\rahul off} the  diffusion constant,
\be
D(\rho)= -\frac{1}{4} \left[\frac{d}{d\rho} (\alpha+\beta) - 2 \beta_1\right].
\ee
Note that $D(\rho)$ depends on the correction $\beta_1$ to the steady state due to the presence of a non-zero density gradient. Substituting for $\beta_1$ from Eq.~(\ref{eq:beta1}), we obtain
\be
D(\rho) =\frac{1}{4}\left[ \frac{z_c P^{++}_0(z_c)+2}{z_c P^{++}_0(z_c)-2}\right] \frac{d}{d\rho} (\alpha+\beta).
\label{eq:mfD}
\ee

\begin{figure}
	\centering
	\subfigure{
		\includegraphics[width=0.9\columnwidth]{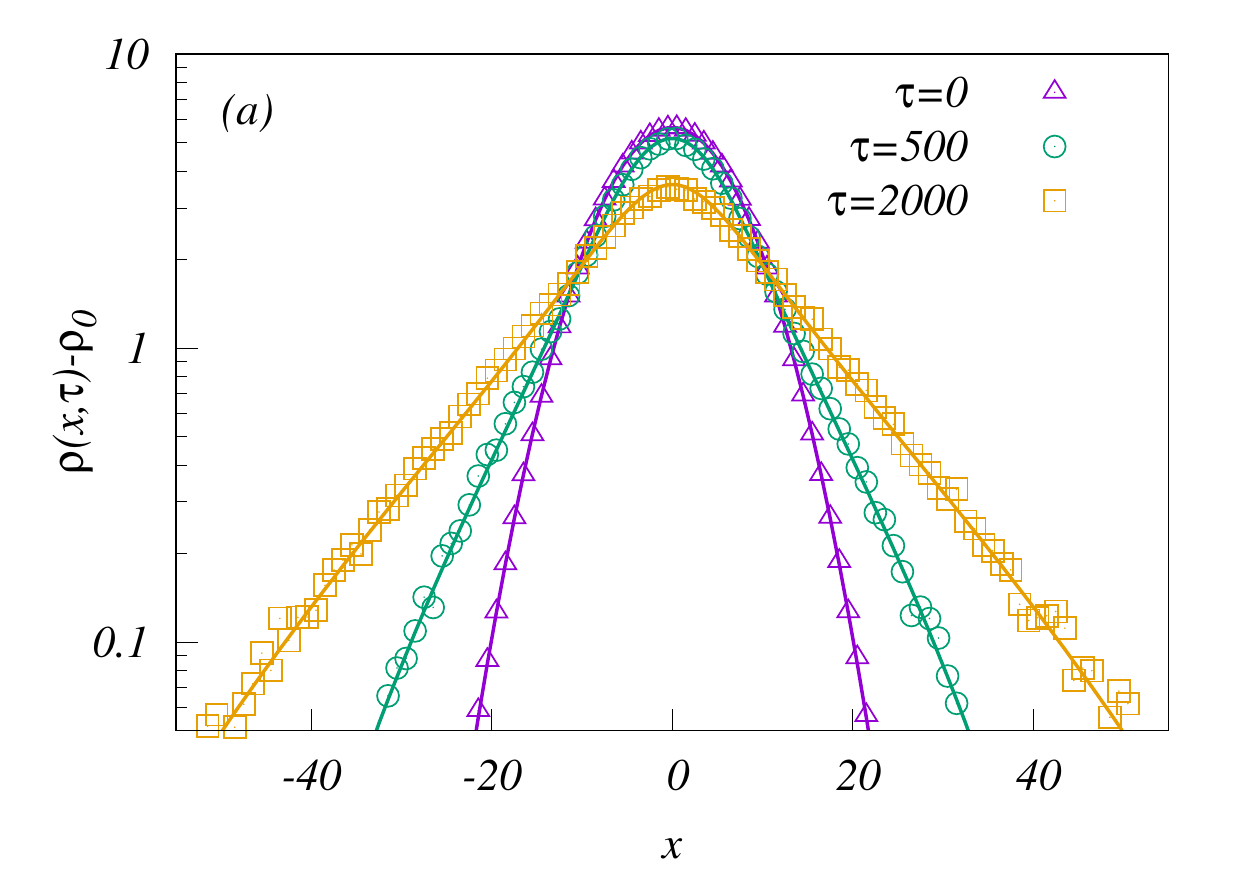}
	}
	\subfigure{
		\includegraphics[width=0.9\columnwidth]{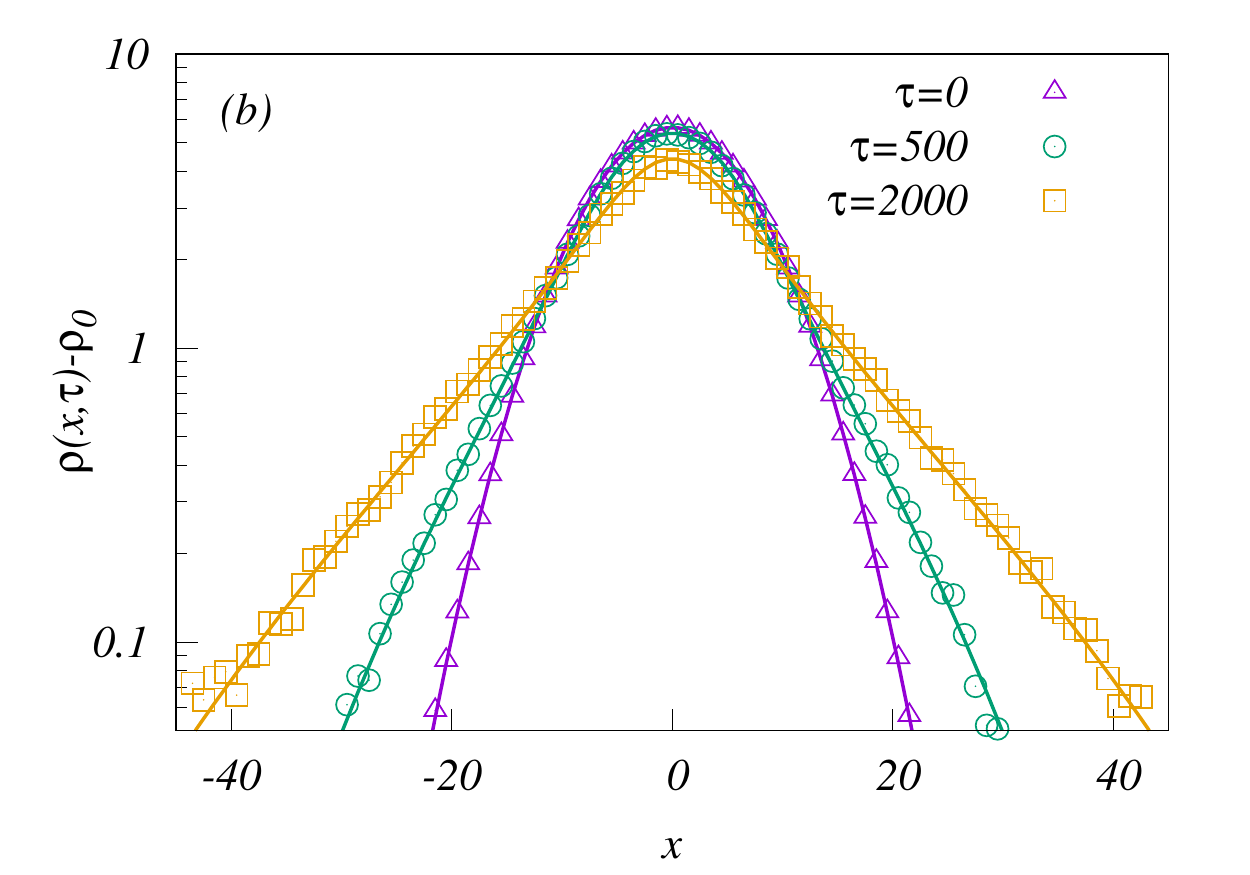}
	}
	\caption{Numerical results for the spreading of an initial Gaussian density perturbation, as described in Eq.~(\ref{eq:densitypert}), with time $\tau$ is compared with the  results from the analytical  mean-field expression for $D(\rho)$ [see Eq.~(\ref{eq:mfD})] for (a) $\eta = 1$ and (b) $\eta=2$.}
	\label{spreadingmf}
\end{figure}

In simulations, we test the above prediction for $D(\rho)$ by measuring the spreading of an initial small density inhomogeneity,
\beq
\rho(x) = \rho_0 + \rho_1 \frac{\exp{\left(\frac{-x^2}{\Delta^2}\right)}}{\sqrt{\pi \Delta^2}}.
\label{eq:densitypert}
\eeq
The initial width is chosen to be $\Delta = 10$, and the spread is averaged over many realizations in the Monte Carlo simulations. The analytical prediction is obtained by numerically integrating  Eqs.~(\ref{eq:hydrodef}) and (\ref{eq:mfD}) using the Euler method. In Fig.~\ref{spreadingmf}, we compare the  numerical results for the spreading of an initial density profile with those from the analytically calculated $D(\rho)$ for two different values of $\eta$. As can be seen from the figure, the two are in excellent agreement. We stress that the term proportional to $\beta_1$ turns out to be essential for reproducing the numerical, and comparing with the naive prediction for $D(\rho)$ does not produce similar agreement (plots not shown).

\subsection{Current in the presence of a field \label{sec:field}}

We now consider the case of a non-zero field $F$ but in the absence of a density gradient, i.e., $\rho'=0$. Since the system on a ring is translationally invariant, it is easy to see that $\mathcal{P}^{ab}_{m,i} = \mathcal{P}^{ab}_m$ is independent of the index $i$. 
We expand the probabilities to first order in $F$:
\beq
\mathcal{P}^{ab}_{m} = P^{ab}_m + F R^{ab}_m + O(F^2), ~~~ a,b = +, -,
\eeq
where $P^{ab}_m$ is the steady state mean field solution obtained in Sec.~\ref{sec:mf} for the homogeneous case, and $R^{ab}_m$ denotes the first order correction. 

For non-zero $F$, the system has certain symmetries. For sites with one spin pointing inwards and one spin pointing outwards, the probabilities are invariant under $F \rightarrow -F$, $s\rightarrow -s$. 
This implies that $\mathcal{P}^{--}_m(F) = \mathcal{P}^{++}_m(-F)$.
For sites with both spins pointing inwards or both pointing outwards, the probabilities are invariant under $F \rightarrow -F$.
This implies that $\mathcal{P}^{+-}_m(F) = \mathcal{P}^{+-}_m(-F)$, and $\mathcal{P}^{-+}_m(F) = \mathcal{P}^{-+}_m(-F)$, i.e, they are even functions of $F$. These symmetries thus imply that
\bea
R^{++}_m &=& -R^{--}_m, \label{eq:symmetry2}\\
R^{+-}_m &=& R^{-+}_m =0, \label{eq:symmetry3} ~~m=0, 1, 2, \ldots
\eea

We now solve for the only unknown first order correction, $R^{++}_m$. It is convenient to introduce the notation
\beq
R^{++}_0 = - R^{--}_0 = \beta_1^F.
\label{eq:beta1Fdef}
\eeq
Consider Eq.~(\ref{eq:pp}) with $\rho'=0$ in the steady state. Using the symmetries in Eqs.~(\ref{eq:symmetry2}) and (\ref{eq:symmetry3}), the $\mathcal{O}(F)$ term in Eq.~(\ref{eq:pp}), after simplification, satisfies
\bea
&&R_{m-1}^{++} [1-\delta_{m,0}] \left[2 \!-\!\alpha \!-\!\beta \right]   - R_m^{++} \left[ 4+4 \eta\!-\!\alpha \!-\!\beta \!-\!2 \delta_{m,0} \right] \nonumber \\
&&+2  R^{++}_{m+1}
+ 2  P^{++}_{m+1}  
- P_m^{++} \left[4-\alpha -\beta -2 \delta_{m,0}-\beta_1^F \right] \nonumber \\
&&+P_{m-1}^{++} (1-\delta_{m,0}) \left[2-\alpha -\beta  -\beta_1^F \right] =0.
\label{eq:ppF}
\eea
Equation~(\ref{eq:ppF}) can be solved by the method of generating functions. Let 
$\widetilde{R}^{++}(z) = \sum_{m=0} R^{++}_m z^m$. Multiplying Eq.~(\ref{eq:ppF}) by $z^m$ and summing over $m$, we obtain
\beq
\frac{\widetilde{R}^{++}(z)}{1-z} = \frac{ 2(\beta+\beta^F_1) +\widetilde{P}^{++}(z)\left[ (2\!-\!\alpha\!-\!\beta)z - \!2\! -\!z \beta^F_1 \right]
}
{(2-\alpha-\beta) z^2 -\left( 4-\alpha -\beta+4 \eta \right) z +2}.
\eeq
$\widetilde{R}^{++}(z)$ has two poles, one of which is less than zero. This pole  equals $z_c$, as given in Eq.~(\ref{eq:zc}). At this value of $z_c$, the numerator should also vanish, allowing $\beta_1^F$ to be determined: 
\beq
\beta^F_1 = \frac{\left[2-(2-\alpha-\beta) z_c \right]  \widetilde{P}^{++}(z_c)-2 \beta}
{2-z_c \widetilde{P}^{++}(z_c)}.
\eeq

Knowing $\beta_1^F$, we can now calculate the drift coefficient $\chi(\rho)$.
From Eqs.~(\ref{eq:hydrodef}) and (\ref{Jbpbm1}), we obtain
\beq
\chi(\rho) = \frac{2- \alpha - \beta - \beta_1^F}{2}. \label{eq:mfchi}
\eeq
We see that the non-gradient nature of the process also affects $\chi(\rho)$ through $\beta^F_1$.

\begin{figure}
	\centering
	\includegraphics[width=\columnwidth]{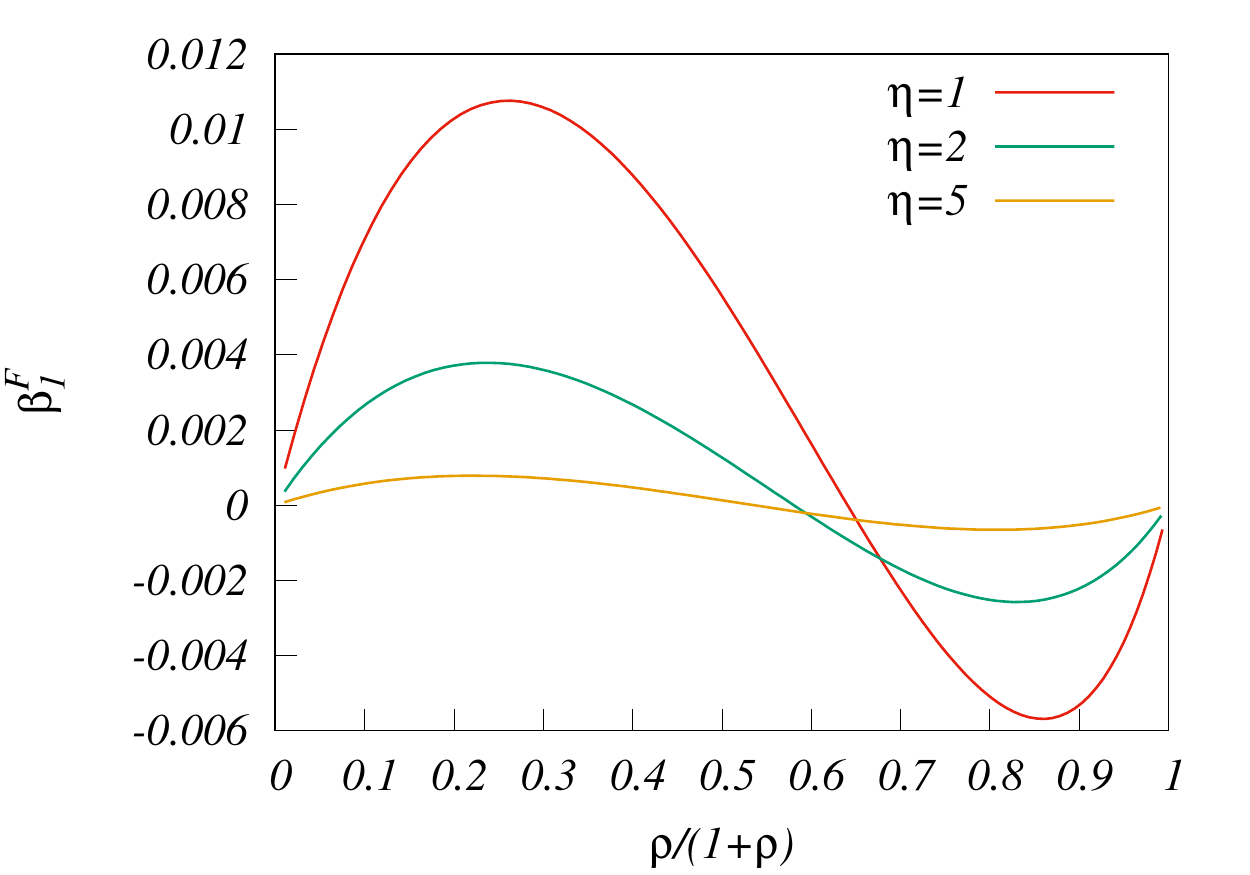}
	\caption{The variation of $\beta_1^F$, the correction to the steady state probability in the presence of a field F [see Eq.~(\ref{eq:beta1Fdef}) for definition] with density $\rho/(1+\rho) = 1-\rho_{RTP}$ for $\eta = 1, 2$ and $5$. We see that the contribution of $\beta^F_1$ to $\chi$ is quite small, and decreases in magnitude with increasing $\eta$.}
	\label{b1F}
\end{figure}

The variation of $\beta_1^F$ with density $\rho/(1+\rho) = 1-\rho_{RTP}$ for different values of $\eta$ is shown in Fig.~\ref{b1F}. It is clear that, in the regime $\eta>1$, where we expect the mean-field theory to be valid, $\beta_1^F$ gives only a small contribution to $\chi(\rho)$. In addition, $|\beta_1^F|$ decreases with increasing $\eta$, consistent with the fact that it equals zero for $\eta=\infty$.

\begin{figure}
	\centering
	\includegraphics[width=\columnwidth]{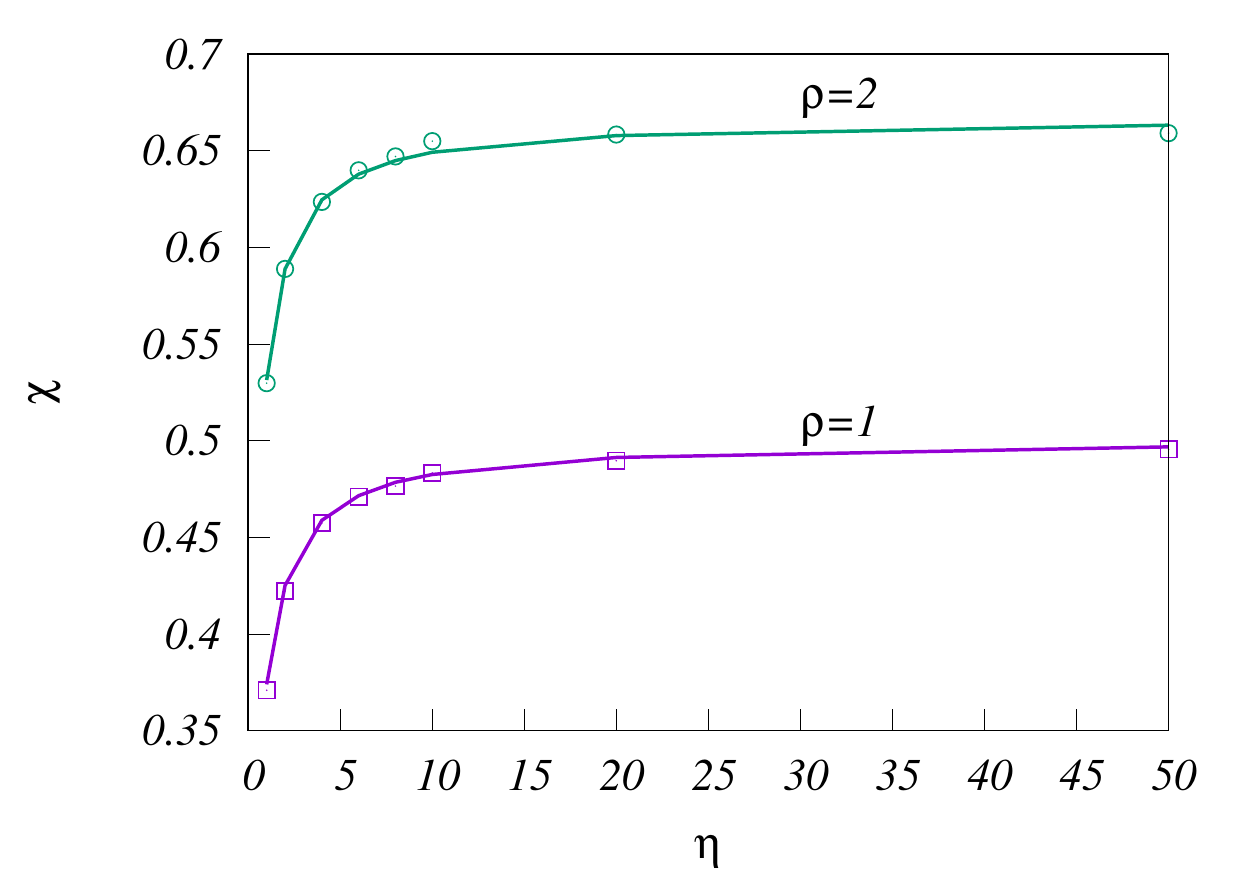}
	\caption{The variation of conductivity $\chi(\rho)$ with spin flip rate $\eta$ for density $\rho = 1, 2$, as measured in simulations. The solid lines are the theoretical predictions as given in  Eq.~(\ref{eq:mfchi}).}
	\label{chiFlarge}
\end{figure}

We now compare the predictions for $\chi(\rho)$ with results from simulations. We measured the conductivity $\chi(\rho)$ in simulations of the system on a periodic ring, with a field $F$ biasing the mass transfers. The current was measured for various values of $\eta$ and $\rho$, and $J/F$ for $F = 1, 2, 3$ was extrapolated to $F=0$ to {\rahul obtain} the linear conductivity. In Fig.~\ref{chiFlarge}, we compare the theoretical result for $\chi(\rho)$ with simulation data for different values of $\eta$ and $\rho$. It is clear that  the mean field hydrodynamic theory is in excellent agreement with simulations.

\subsection{Mass fluctuations}

To test the Einstein relation [see Eq.~(\ref{eq:einstein})], we need to calculate the mass fluctuations $\sigma^2$ defined as
\be
\sigma^2 = \frac{\la m_\ell^2 \ra - \la m_\ell \ra^2}{\ell}, ~~\ell \to \infty,
\ee
where $m_\ell$ denotes the total mass in a subsystem of $\ell$ sites. Although we work in the mean-field approximation, we 
cannot treat the masses as completely independent, as different sites are correlated {\rahul through} the  interconnecting spins. If $P(m_1,m_2,\dots,m_\ell)$ denotes the joint probability distribution function of $\ell$ consecutive sites numbered $1, 2, \ldots, \ell$, then
\beqa
P(m_1,\dots, m_\ell) = \frac{1}{2^{\ell+1}} \sum_{k,n} \left[\prod_{i=1}^{\ell} \left(
\begin{matrix} 
	P^{++}_{m_i} & P^{-+}_{m_i} \\
	P^{+-}_{m_i} & P^{--}_{m_i}
\end{matrix}
\right) \right]_{kn}.
\label{eq:sigmasq}
\eeqa

Now consider the generating function $P_\ell (z)$  for the total mass of the subsystem
\beq
P_\ell(z) = \langle  z^{m_1 + m_2 + \dots+m_\ell} \rangle.
\eeq
Multiplying Eq.~(\ref{eq:sigmasq}) by $z^{m_1+\ldots+m_\ell}$ and summing over the masses, we obtain
\beqa
P_\ell(z) = \frac{1}{2^{\ell+1}} \sum_{i,j} \left[ \left(
\begin{matrix} 
	P^{++}(z) & P^{-+}(z) \\
	P^{+-}(z) & P^{--}(z)
\end{matrix}
\right)^l \right]_{ij}.
\label{eq:genfunctsigma}
\eeqa
For large subsystem sizes $\ell$,  the generating function $P_\ell (z)$ is dominated by the larger eigenvalue, $\Lambda(z)$, of the matrix on the right hand side of Eq.~(\ref{eq:genfunctsigma}). Diagonalising, we obtain
\beq
\Lambda(z) = P^{++}(z) +\sqrt{P^{-+} (z) P^{+-} (z)}.
\eeq
Then,
\beq
\frac{\ln P_\ell(z)}{\ell} = \ln \Lambda(z) - \ln{2} + O\left(\frac{1}{\ell}\right).
\eeq
The mass fluctuations are given by
\be
\sigma^2 = \left(z \frac{d}{dz} \right)^2 \frac{\ln P_\ell(z)}{\ell}. \label{eq:sigma2mf}
\ee

\begin{figure}
	\centering
	\includegraphics[width=\columnwidth]{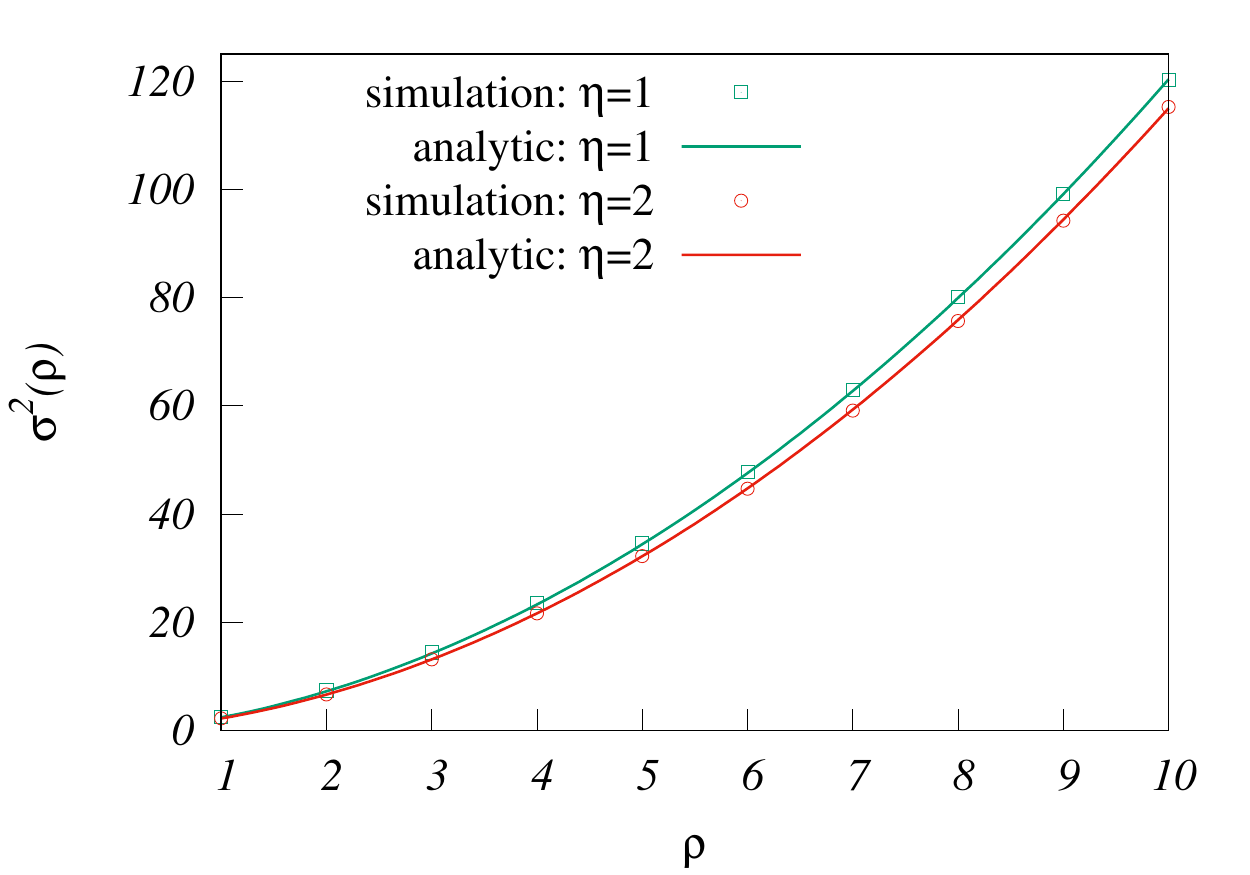}
	\caption{The variation of the numerically obtained subsystem (in this case, single site) mass fluctuations with density $\rho$ for different values of $\eta$. The solid lines are the theoretical predictions as given in Eq.~(\ref{eq:sigma2mf}). }
	\label{fig:sigmasq}
\end{figure}

In Fig.~\ref{fig:sigmasq}, we compare the theoretical result for the subsystem fluctuations with simulations for different $\eta$ and $\rho$. In the simulations, we measure single site mass fluctuations. They are in excellent agreement, further confirming the validity of mean-field theory for moderate and large $\eta$. 

Knowing the subsystem mass fluctuations $\sigma^2(\rho)$, we can proceed to test the Einstein relation.

\subsection{Einstein relation}

\begin{figure}
	\centering
	\includegraphics[width=\columnwidth]{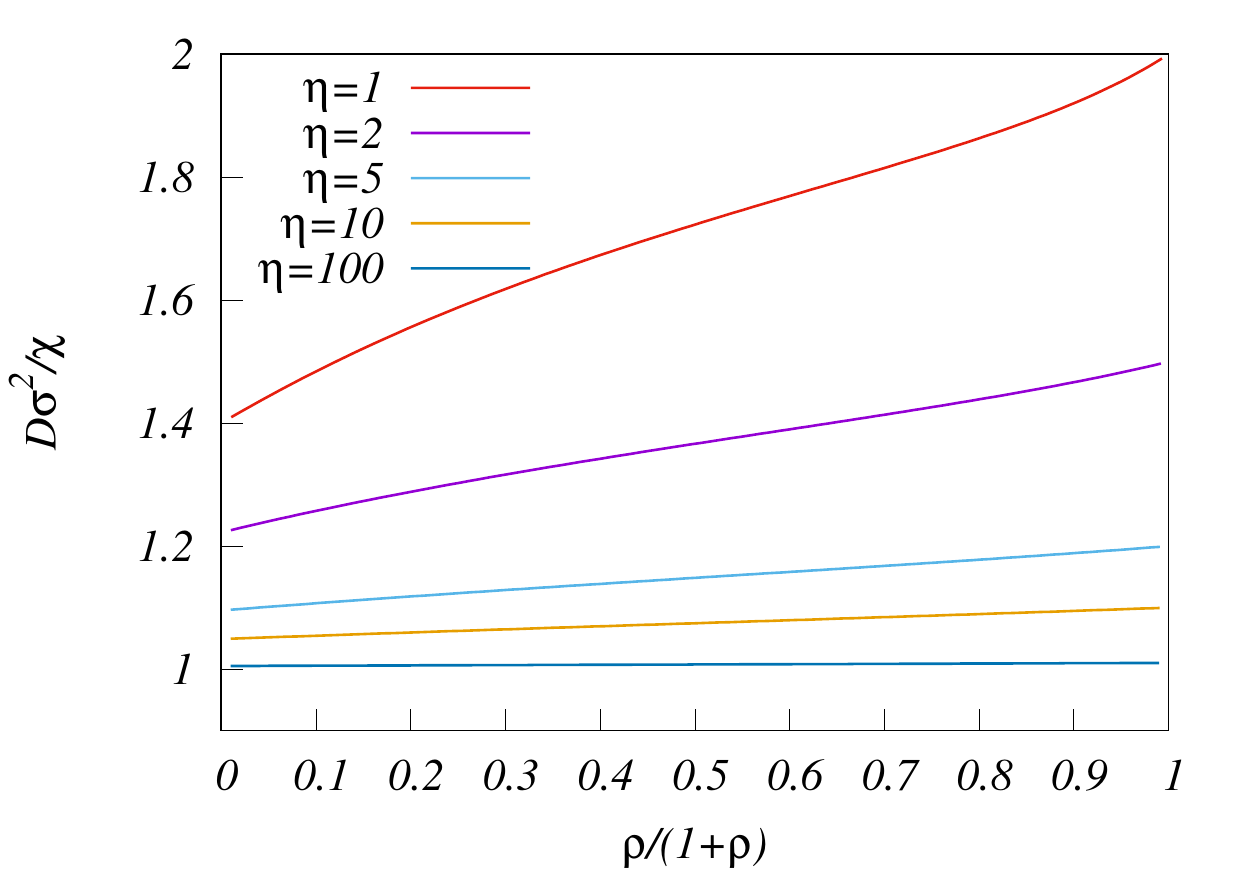}
	\caption{The variation of the ratio $D(\beta) \sigma^2(\beta)/\chi(\beta)${\rahul, calculated within} mean field theory, with density $\rho/(1+\rho) = 1-\rho_{RTP}$ for different values of the spin flip rate $\eta$. If the ratio equals one, then the Einstein relation in Eq.~(\ref{eq:einstein}) is satisfied. The ratio $D(\beta) \sigma^2(\beta)/\chi(\beta)$ approaches $1$ as $\eta \rightarrow \infty$ (the simple exclusion process limit) and for low densities.}
	\label{eintestmf}
\end{figure}

In this subsection, we test the Einstein relation $\frac{1}{\sigma^2(\rho)} = \frac{D(\rho)}{\chi(\rho)}$ given in Eq.~(\ref{eq:einstein}).  To do so, we note that our analytic expressions for the transport coefficients reproduce the numerical results quite accurately, see Figs.~\ref{spreadingmf},  \ref{chiFlarge} and \ref{fig:sigmasq}. Hence, we check for the Einstein relation using these analytic expressions.
Fig. \ref{eintestmf} shows the variation of 
$\frac{D(\rho) \sigma^2(\rho)}{\chi(\rho)}$ with $\rho$ for various values of $\eta$. If the Einstein relation is valid, this ratio should equal one.
It is clear that the  Einstein relation is not valid for moderate to large $\eta$. However, the ratio  $\frac{D(\rho) \sigma^2(\rho)}{\chi(\rho)}$ tends to one as $\eta\to \infty$.

We see that the Einstein relation fails for all finite $\eta$, even in the regime where the mean-field approximation shows excellent agreement with simulations. There are two reasons for this. Firstly, since the current in the mass transfer process depends on the spin configuration, the process is no longer a gradient process. Secondly, the persistence of the spin configuration over a time $\eta^{-1}$ means that the noise in the macroscopic current at two different times is no longer uncorrelated, and hence the conductivity $\chi(\rho)$ no longer represents the strength of the fluctuating part of the current. Thus, care has to be taken when writing a fluctuating hydrodynamics for the RTP, and in analyzing its relation to the conductivity. In the next section, we shall see that these problems are even more severe for small $\eta$, since the persistence times get longer.

\section{\label{sec:hydro2} Hydrodynamics for the Coalescence model} 

We  showed in  Sec.~\ref{sec:smallD} that in the limit of $\eta \ll 1$, the steady state mean density of clusters as well as the approach to the steady state is well approximated by an effective aggregation-fragmentation model. We now calculate the hydrodynamic coefficients $D(\rho)$ and $\chi(\rho)$ [see Eq.~(\ref{eq:hydrodef}) for a definition] by determining the rates of mass transfers across a given bond in the presence of a density gradient $\rho'=\partial \rho/\partial x$ and a field $F$. In the presence of a field,  the rate of mass transfer towards the right is increased by a factor $(1+F)$ and the rate towards the left is decreased by a factor $(1-F)$. We show in Appendix D using the empty interval method that, unlike in the previous section, the presence of a field or density gradient does not change the steady-state measure. This implies that we can use the equilibrium formulae for $n_c(\rho)$ and $m^*(\rho)$ in calculating the current under a small density gradient or field.

\subsection{Calculation of current: preliminaries}

To derive the hydrodynamic coefficients $D(\rho)$ and $\chi(\rho)$, we now calculate the mass current across the bond $(-1,0)$ to first order in $F$ and $\rho'$. As a preliminary observation, we note that the mass current across a given bond might originate in a spin flip at a site far away from the bond, if all the intervening arrows between the site and the bond point towards the bond. We will first consider mass transfers across $(-1,0)$ to the right, initiated by the flip of one of the spins around the site $-k$. {\rahul Symmetrically, one can similarly} calculate the current to the left across $(-1,0)$ from mass transfer from site $(k-1)$, finally summing over $k$ to calculate the total current.

Now, if the site $-k$ is occupied, then $s_{-k+\frac{1}{2}}=-1$ and $s_{-k-\frac{1}{2}}=1$, since we assume that only $+-$ sites are occupied. The probability that a given site is occupied is $n_c(\rho)$. If $s_{-k+\frac{1}{2}}$ flips to $1$, which happens at rate $\eta$, then mass transfer is initiated  to the right. 
The probability of the nearest $-1$ spin to the right being a distance $j$ away is thus
\beq
p(j) = 2^{-j}.
\eeq
Similarly, we denote the probability that the nearest $-1$ spin to the right being a distance $\ge j$ away as
\beq
p_{\ge}(j) = 2^{-j+1} = 2 p(j).
\eeq

The time for a mass transfer over a distance $j \sim O(1)$ is, $m^* + O(1)$ on average, since $m^*(\rho)$ is the average mass on a site, since the rate of mass transfer over each bond is $1$. In the presence of a field, the mass transfer time to the right is $(1-F ) m^*$, and to the left is $(1+F) m^*$, to leading order.

\subsection{Current in a constant density gradient}

We consider the current across the bond $(-1,0)$ due to a mass at site $-k$, in the presence of a small, constant density gradient $\rho'$. The average density at site $j$ is then $\rho_j = \rho_0 + j \rho'$. Our aim is to calculate the leading order term in the diffusion constant. For this purpose, it is sufficient to consider processes where only a single spin flip has occurred.

Consider the mass transfer from site $-k$ across the bond $(-1,0)$ through a process of a single spin flip. The probability that site $-k$ is occupied is given by $n_c(\rho_{-k})$, since $n_c$ depends on position only through the local density. If site $-k$ is occupied, it is of type $+-$, and hence the spin $s_{-k+\frac{1}{2}}=-1$. Mass transfer to the right is initiated by the flipping of the spin $s_{-k+\frac{1}{2}}$. If all the spins between $-k+1$ and $0$ are pointing to the right, the mass is transferred across $(-1,0)$ to the right. Since spin flips happen at rate $\eta$, the current due to site $-k$ is
\beq
J^+_{-k} = \eta n_c(\rho_{-k}) m^*(\rho_{-k}) p(k).
\eeq
The current to the left across $(-1,0)$ due to site $k-1$ is similarly
\beq
J^-_{k-1} = \eta n_c(\rho_{k-1}) m^*(\rho_{k-1}) p(k).
\eeq

Summing over $k$, the total current is
\beqa
J &=& \sum_{k=1}^{\infty} J^+{-k} - J^-_{k-1}, \nonumber\\
 &=& \eta \sum_k p(k) \left(\rho_{-k} - \rho_{k-1} \right), \\
 &=& -\eta \sum_k p(k) (2 k -1) \rho' = -6 \eta \rho', \label{eq:currentgradient}
\eeqa
which gives, to leading order, 
\be
D(\rho) = 6 \eta. \label{eq:coalD1}
\ee
In Eq.~(\ref{eq:currentgradient}) we  used 
\beqa
\sum_{k=1}^{\infty} p_{\ge} (k) &=& 2 \sum_{k=1}^{\infty} p (k) = 2,\\
\sum_{k=1}^{\infty} k p_{\ge} (k) &=& 2 \sum_{k=1}^{\infty} k p (k) = 4.
\eeqa

\begin{figure}
	\centering
	\subfigure{
		\includegraphics[width=0.9\columnwidth]{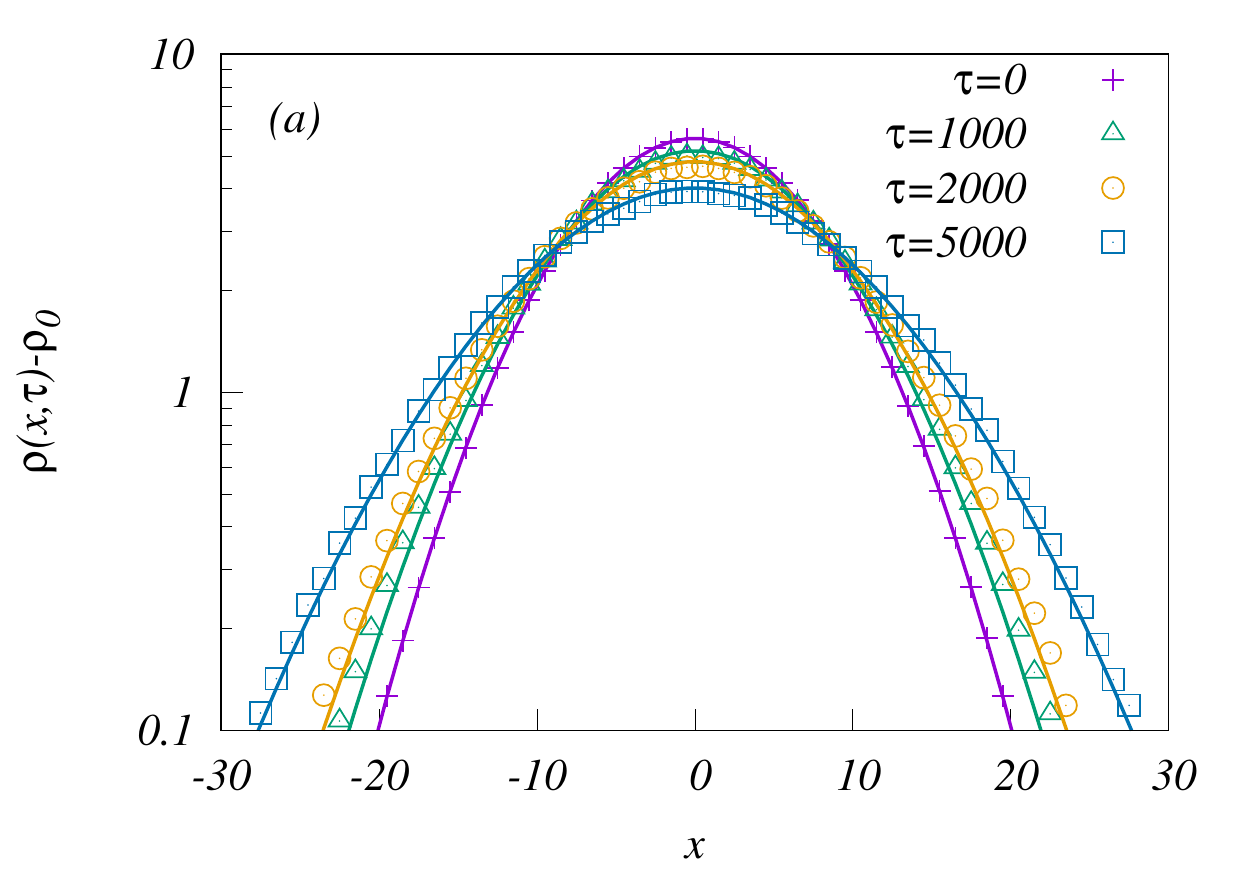}
	}
	\subfigure{
		\includegraphics[width=0.9\columnwidth]{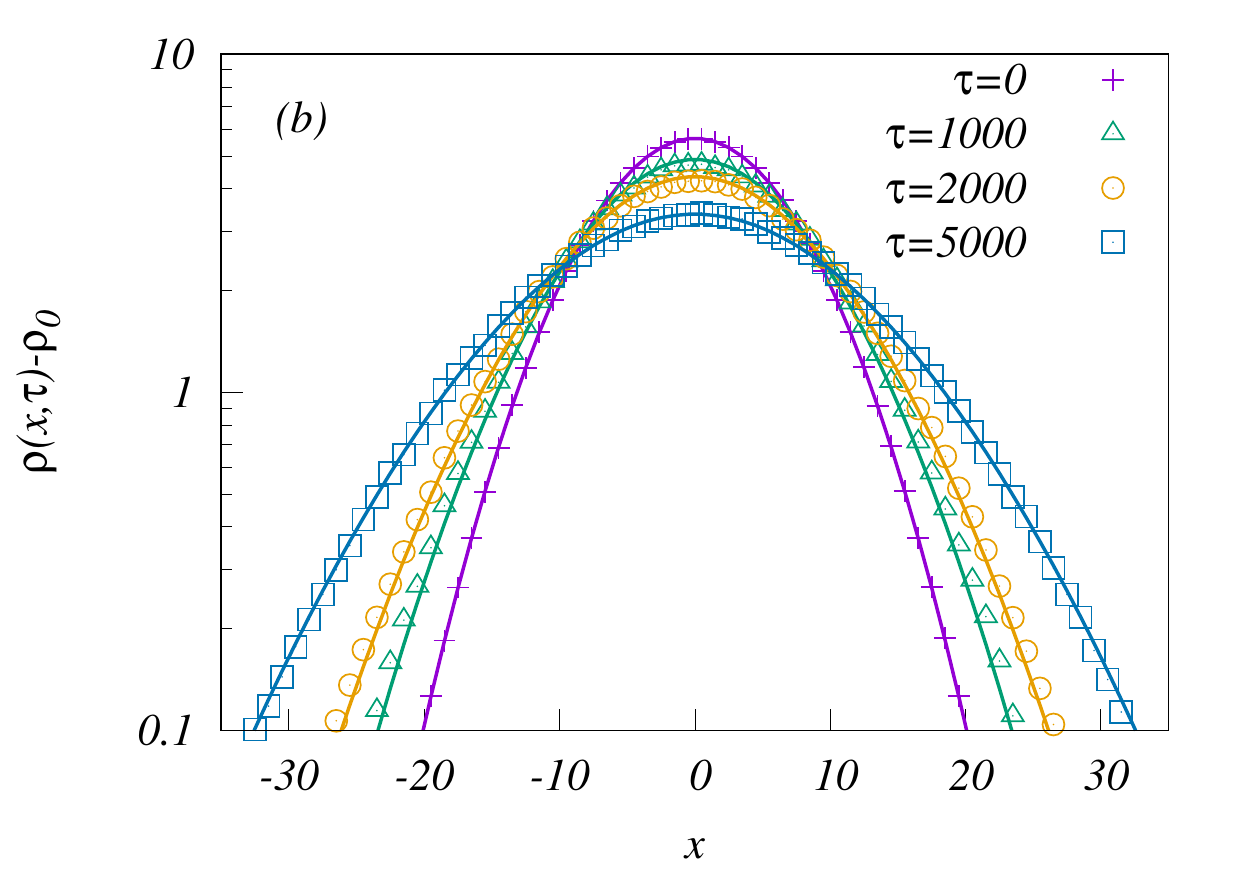}
	}
	\caption{
	Numerical results for the spreading of an initial Gaussian density perturbation, as described in Eq.~(\ref{eq:densitypert}), with time $\tau$ is compared with the  results from the analytical expression for $D(\rho)$ [see Eq.~(\ref{eq:coalD})] for (a) $\eta = 0.001$ and (b) $\eta=0.002$.} 
	\label{diffsimlow}
\end{figure}

The calculation of the next-to-leading-order term is beyond the scope of our calculations, as it requires careful consideration of empty interval probabilities. As in the mean field case, we verify our result for $D(\rho)$ by fitting it to simulations of spreading from an initial density perturbation (see Fig.~\ref{diffsimlow}). Although Eq.~(\ref{eq:coalD1}) gives a reasonable estimate of the spread, we find that postulating a form
\beq
D(\rho) \approx \eta (6 + d \sqrt{\eta \rho}), \label{eq:coalD}
\eeq
allows us to obtain an excellent fit. This is because for the values of $\rho$ and $\eta$ we study, the subleading correction is important for the initial stages of the spread. Our simulations are consistent with the above form, and an excellent fit is obtained with $d = 3$, as shown in Fig.~\ref{diffsimlow}. 

\subsection{Current in a small field $F$}

We now assume that  the field $F$ is nonzero, but {\rahul that there is no} density gradient.  Since the system is on a ring, {\rahul it is translationally invariant}. As before, we consider the current across the bond $(-1,0)$, in a time $T$. The probability that a given bond flips during this time is 
$\eta T$. The probability that $j$ given bonds flip is $(\eta T)^j$ while  the probability that $j$ given bonds do not flip  is, to leading order, 
$1-j \eta T$. The probe time scale $T$ is chosen so that it is greater than the time scale of mass transfer $m^*$, but $T \ll \eta^{-1}$, allowing for an expansion in numbers of bonds flipped during the transfer.

There are five processes which contribute to mass transfer across the bond $(-1,0)$ from the site $-k$, with $k>0$.

(1) \emph{ $s_{-k+\frac{1}{2}}$ flips initially  and no other spin flips}. This leads to a mass $m^*$ being transferred across $(-1,0)$ if all of the intervening spins are pointing to the right. If the mass ends up beyond site $0$, no spins between $-(k+1)$ and $0$ should flip during the mass transfer. If the mass ends up at site $0$, the spin $s_{-\frac{1}{2}}$ should not flip during the probe time-scale $T$, as this will lead to a mass less than $m^*$ being transferred across $(-1,0)$ at the end of the probe time period. The current due to process (1) is
\beqa
J^+_1 &=& n_c \eta m^* \big[ p_{\ge}(k+1) (1- (k+1) \eta m^* (1-F)) \nonumber\\
&+& p(k) (1-\eta m^* k (1-F) - \eta T) \big],
\eeqa
where the first term is the contribution to current from mass transfer to sites $1$ and beyond, and the second describes the contribution to mass to transfer to site $0$.

(2) \emph{$s_{-k+\frac{1}{2}}$ flips initially followed by  $s_{-k-\frac{1}{2}}$ flipping during transfer}. This leads to some mass being transferred to the left of site $-k$. On average, the spin $s_{-k-\frac{1}{2}}$ flips halfway through the transfer, thus the remaining mass is $m^*/2$. Of this remaining mass, due to the field $F$, a fraction $(1+F)/2$ ends up on the right and $(1-F)/2$ on the left of site $-k$. If the spins between $-k$ and $0$ point to the right, the mass transferred across $(-1,0)$ is $m^*/2 + m^*(1+F)/4$. The current is
\beqa
J^+_2 = n_c \eta \left(\frac{3}{4} m^* + \frac{1}{4} m^* F \right) (\eta m^* (1-F) p_{\ge}(k).
\eeqa

(3) \emph{$s_{-k-\frac{1}{2}}$ flips initially followed by  $s_{-k+\frac{1}{2}}$ flipping during transfer}. The initial mass transfer is to the left of $-k$ (and against the field), but during the transfer the flipping of $s_{-k+\frac{1}{2}}$ leads to some mass being transferred across $(-1,0)$. Since the transfer is in the direction of the field, the average mass transferred to the right is $m^*(1+F)/4$. Therefore,
\beqa
J^+_3 = n_c \eta \frac{1}{4} m^* (1+F) (\eta m^* (1+F)) p_{\ge}(k).
\eeqa

(4) \emph{$s_{-k+\frac{1}{2}}$ flips initially followed by a spin between $-k$ and $0$ flipping from right to left during the mass transfer}. Such a flip interrupts the transfer and on average leads to only mass $m^*/2$ being transferred across $(-1,0)$. However, as in (1) above, we have to distinguish between the cases where the transfer is to site $1$ or beyond, and the transfer ending at site $0$. In the latter case, flipping of any spin between $-k$ and $-1$ interrupts the transfer as before, but flipping of spin $s_{-\frac{1}{2}}$ will lead to the whole mass ending up at site $-1$ and no mass transferred across $(-1,0)$ at the end of the probe period $T$.  Hence,
\beqa
J^+_4 &=& n_c \eta \frac{m^*}{2} \big[ p_{\ge}(k+1) k \eta m^* (1-F) \nonumber\\
 &+& p(k) \left( (k-1) \eta m^* (1-F) + \eta m^* (1+F) \right) \big].
\eeqa

(5) \emph{ $s_{-k+\frac{1}{2}}$ flips initially followed by a spin between $-k$ and $0$ flipping from left to right during the mass transfer}. The fifth type of process is in a way the inverse of (4) above: say the initial closest $-1$ spin was $s_{-i+\frac{1}{2}}$, with $0<i<k$, but that spin flips during the time $T$. If the next nearest $-1$ spin is beyond site $0$, then mass will end up across $(-1,0)$. Taking the mass transfer time into account, if this flip happens in $[0,T-m^*(1-F)]$, the whole mass $m^*$ ends up across $(-1,0)$. If it happens in $[T-m^*(1-F),T]$, on average only half the mass ends up across $(-1,0)$ after time $T$. Note that the probability that the nearest spin is at $-i+\frac{1}{2}$ and the next nearest beyond $0$ is $p(k-i) p_{\ge}(i) = p_{\ge}(k)$. The site $i$ can be in one of $(k-1)$ positions. 
If site $i$ was initially occupied, which happens with probability $n_c$, an additional mass $m^*$ will be transferred across $(-1,0)$. However, if $i=k-1$, the site $i$ cannot be occupied, since two occupied sites cannot be directly adjacent. Hence when $i$ is in one of the other $(k-2)$ places, $2 m^*$ is transferred across $(-1,0)$. Thus, the current is
\begin{align}
&J^+_5 = n_c \eta p_{\ge}(k) (k-1) \big[ m^* \eta (T-m^*(1-F)) 
\nonumber\\&
+ \frac{m^*}{2} \eta m^* (1-F) \big] \nonumber\\
&+  n_c \eta p_{\ge}(k) (k-2) \big[ m^* \eta (T-m^*(1-F)) \nonumber\\
&+ \frac{m^*}{2} \eta m^* (1-F) \big] \Theta(k-2).
\end{align}
where the Heaviside Theta function accounts for the fact that the contribution of the second type when $k=1$ is zero.

Summing up the contributions to the current from the above five processes, we obtain the total current from left to right across the bond $(-1,0)$.
To calculate the conductivity $\chi(\rho)$, we {\rahul only need to} consider those terms in $J^+$ that are proportional to $F$. In the absence of a density gradient, the other terms of the current will cancel each other. We denote the part of the current $J$ proportional to $F$ as $J_F$. Using $p_{\ge}(k) = 2 p_{\ge}(k+1) = 2p(k)$, we obtain
\beqa
J^+_F &=& F (\eta\rho)^{3/2} \sum_{k=1}^\infty p(k) \left( \frac{5 k}{2} + \frac{1}{2} \right), \nonumber\\
&=& \frac{11}{2} F (\eta \rho)^{3/2}.
\eeqa
Similarly, the $F$-dependent term of $J_F^-$ is
\beq
J^-_F = \frac{11}{2} F (\eta \rho)^{3/2}.
\eeq
Thus, we obtain
\beq
\chi(\rho) = 11 (\eta \rho)^{3/2}, \quad \eta \ll 1.
\label{eq:conductivity-smalleta}
\eeq

\begin{figure}
	\centering
	\includegraphics[width=\columnwidth]{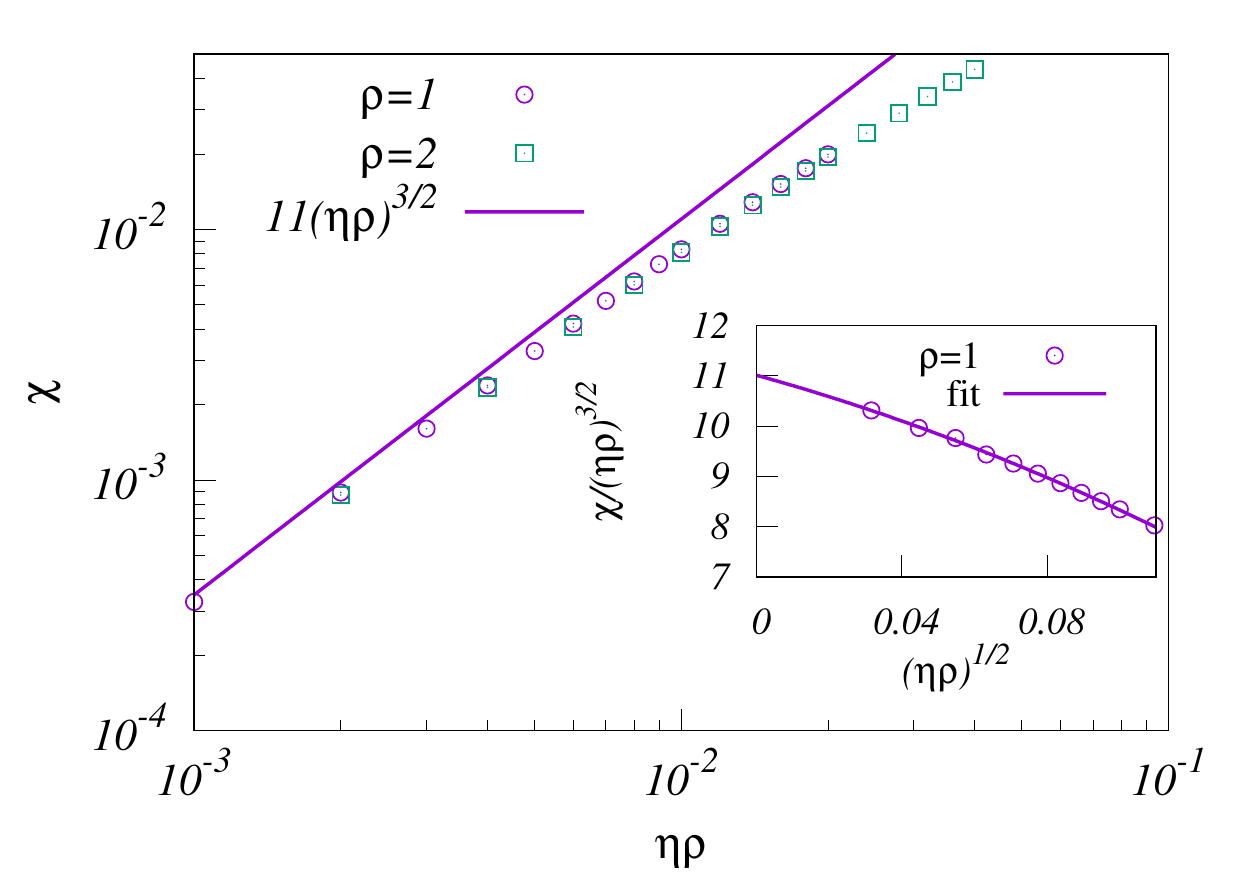}
	\caption{The variation of conductivity $\chi(\rho)$ with $\eta \rho$ for two different values of density $\rho$. The data for different $\rho$ lie on one curve, which for small $\eta \rho$ is compared with the theoretical prediction $11 (\eta \rho)^{3/2}$ (solid line), as given in  Eq.~(\ref{eq:conductivity-smalleta}). Inset: The coefficient of the power law is obtained by plotting $\chi/(\eta \rho)^{3/2}$ against $\sqrt{\eta \rho}$. The best fit, shown by dotted line, is $11.01-20.02 \sqrt{\eta \rho}-68.18 \eta \rho$.}
	\label{chiF}
\end{figure}

In Fig.~\ref{chiF}, we show the variation  of $\chi$ with $\eta$ for two values of density $\rho$. The data for the two densities collapse onto one curve when $\chi$ is plotted against $\eta \rho$, consistent with Eq.~(\ref{eq:conductivity-smalleta}). For small $\eta \rho$, the data is consistent with a power law with exponent $3/2$ as in Eq.~(\ref{eq:conductivity-smalleta}). However, the curve deviates from the power law for larger $\eta \rho$. In the inset, we account for these correction terms in order to measure the prefactor of the power law more accurately to be $\approx 11.01$, consistent with the theoretical prediction in Eq.~(\ref{eq:conductivity-smalleta}). The leading correction to conductivity is numerically estimated to be 
\beq
\chi(\rho) = 11 (\eta \rho)^{3/2}  + \kappa (\eta \rho)^2 + \mathcal{O}\left((\eta \rho)^{5/2}\right),
\eeq
with  $\kappa \approx 20$.

\subsection{Transport coefficients and Einstein relation}

To test the Einstein relation, we need to estimate the mass fluctuations in a subsystem. We will see that the Einstein relation is not obeyed in the low $\eta$ regime, and the discrepancy is of the order of $\eta^{-1}$ and thus not simply a matter of getting the numerical factors correct. So far we have not considered the distribution of mass on an occupied site in the low $\eta$ regime. To proceed with the estimate, we postulate that the mass distribution on occupied sites is $P_{occ}(m) \sim e^{-\sqrt{\frac{\eta}{\rho}} m}$. Since the density of occupied sites is $n_c = \sqrt{\eta \rho}$, we have (after normalization)
\be
P(m) = 
\begin{cases} 
	1 - \sqrt{\eta \rho} & \mbox{ for $m=0$,}\\
	\eta ~ e^{-\sqrt{\frac{\eta}{\rho}} m} &\mbox{ for $m>0$.}
\end{cases}
\ee
If we assume the masses at different sites are independent, this gives 
\beq
\sigma^2(\rho) = \langle m^2 \rangle- \langle m \rangle^2 = 2 \frac{\rho^{3/2}}{\sqrt{\eta}} + O(1),
\eeq
Comparing with
\beq
\frac{D(\rho)}{\chi(\rho)} = \frac{6}{11} \frac{1}{\eta^{1/2} \rho^{3/2}} + O(1),
\eeq
we see that the Einstein relation, given in Eq.~(\ref{eq:einstein}), is not satisfied for the dynamics in the coalescence picture, i.e., for $\eta \ll 1$. The physical reason for this is that mass transfers in the low $\eta$ regime are correlated over a time $\eta^{-1}$, which can be very large. The Einstein relation assumes that $\chi(\rho)$ is also the coefficient of a delta-correlated noise term in the current. However, the long-time {\rahul persistence} makes this assumption false, and hence the Einstein relation is not expected to hold.

\section{\label{sec:conclusion}Conclusion}

In this paper, we studied a system of hard core run and tumble particles on a one dimensional lattice. A particle moves persistently in the direction of its spin, till the spin flips direction.  We study the properties of the gaps between particles by mapping the model to a mass transport model across fluctuating directed bonds that are persistent for a time $\eta^{-1}$. We solve for the mass distribution using a mean field approximation which reproduces well the distribution for moderate and large spin flip rates $\eta$, but fails for small $\eta$. For small $\eta$, we develop an aggregation-fragmentation model which analytically characterizes both the approach to the steady state as well as the steady state, both in excellent agreement with results from simulations. 
For both large and small $\eta$, we also derived the hydrodynamic coefficients of diffusivity and conductivity by calculating the current across a bond in the presence of a small density gradient and a small biasing field respectively. In the moderate and large $\eta$ case, we see that the steady-state changes upon introduction of a small gradient or biasing field, and such a  change in the steady-state contributes significantly to the hydrodynamic coefficients. This is due to the fact that crowding occurs differently behind particles parallel to and anti parallel to the field or gradient, and the direction of particle motion is persistent for a time $\eta^{-1}$. For small $\eta$, we calculate the hydrodynamic coefficients exactly to leading order. 

The Einstein relation between the hydrodynamic coefficients and the subsystem mass fluctuations is not obeyed for $\eta \neq \infty$. This is, again, because of persistence of the motion of individual particles. In particular, unlike in usual lattice gases, the conductivity does not equal the amplitude of the current fluctuations in the steady-state, because the current fluctuations are correlated over a time $\eta^{-1}$. It is the balance between the diffusivity and the current fluctuations in the steady-state that dictates the subsystem mass fluctuations, and not the balance between the diffusivity and the conductivity. Thus, the discrepancy in the Einstein relation in both regimes is proportional to $\eta^{-1}$.

Our calculations explicitly demonstrate the unusual effects of persistence on the hydrodynamics of RTPs, namely the effect of the change in the steady-state in a field, and the mismatch between the conductivity and current fluctuations. {\rahul Our findings have implications for attempts to develop a hydrodynamics of active matter \cite{tailleur2008statistical, MarchettiReview2013}, showing that the nongradient nature of the process has to be carefully accounted for, and that unlike the case of weak persistence ($\eta^{-1} \sim L^{-1}$) \cite{kourbane2018exact}, local equilibrium does not hold in the presence of strongly persistent motion.

There are several promising areas for future study. A straightforward generalisation is to higher dimensions. In dimensions higher than one, the effect of caging is absent, and hence, one would expect the mean field results to be more accurate. Although the mapping to a mass model breaks down in higher dimensions, the mass transport model with directed bonds whose direction fluctuates in time is also an interesting model to study. In the low $\eta$ regime, especially if one introduces interactions between the directed bonds, one expects to find novel clustered and patterned phases. This generalised mass model can easily be studied with the techniques developed in this paper. 

In the RTP model studied here, a site can be occupied by at most one particle. A generalisation of the model is to increase the occupancy of a site to more than one. Clearly, when the maximum occupancy is infinite, the effect of hard core repulsion disappears and the problem reduces of that of independent persistent random walkers. However, when the maximum occupancy is larger than two but finite, it has been reported, based on Monte Carlo simulations, that there is a condensation transition~\cite{sepulveda2016coarsening}.  It would be interesting to apply the techniques developed in this paper to study the hydrodynamics in this model, as well as the cluster size distribution.

The RTP model considered in this paper may be considered as adding spins to particles in the symmetric exclusion process. A tractable generalisation of the symmetric exclusion process where exact results are easy to obtain is the random average process (RAP) \cite{rajesh2001exact}, where particles live on the continuum and make jumps which can be of any distance as long as the order of particles is maintained. An RAP model with persistent spins would be interesting as a more general model to study using the techniques developed here.}

\section{Acknowledgments}
We thank Punyabrata Pradhan for helpful discussions.

\appendix

\section{Calculation of the mean-field solutions for $P^{++}(z)$, $P^{+-}(z)$}

Consider the transfers leading to changes in $P^{+-}_m$. For this configuration of spins, the lattice  site can only receive mass, as both the neighboring spins point inwards. $P^{+-}_m$ evolves in time as
\bea
\frac{dP^{+-}_m}{dt}&=&  \sum_{m'=1}^\infty \left( P^{-\bullet}_{m'} + P^{\bullet+}_{m'} \right)  \left[P^{+-}_{m-1} (1-\delta_{m,0})-P^{+-}_m\right] \nonumber \\
&+& \eta\left( P^{--}_m +P^{++}_m -2P^{+-}_m \right),
\label{eq:master2}
\eea
where the $\bullet$ in the superscript means that the spin could be either $+$ or $-$. Clearly,
\be
P^{-\bullet}_m = \frac{1}{2} P^{--}_m + \frac{1}{2} P^{-+}_m.
\label{eq:conditional1}
\ee
Substituting \eref{eq:conditional1} into \eref{eq:master2} and simplifying, we obtain
\bea
\frac{dP^{+-}_m}{dt}&=&  \left( 2- \alpha -\beta  \right)  \left[ P^{+-}_{m-1} (1-\delta_{m,0})-P^{+-}_m\right] \nonumber \\
&+& 2 \eta \left( P^{++}_m -P^{+-}_m \right).
\label{eq:master2final}
\eea
Setting time derivatives to zero in the steady state and solving for $P^{+-}_m$, we obtain
\be
P^{+-}_m = \frac{2 \eta P^{++}_m+(2- \alpha -\beta) P^{+-}_{m-1}}{2\eta + 2 - \alpha - \beta}.
\ee
Multiplying by $z^m$ and summing over $m$, we obtain the generating function to be
\be
\widetilde{P}^{+-}(z) = \frac{2 \eta \widetilde{P}^{++}(z)}{2\eta + (1-z) (2 - \alpha -\beta)}. \label{ppmz}
\ee
Note that $\widetilde{P}^{+-}(z)$ is also determined in terms of $\widetilde{P}^{++}(z)$. Setting $z=0$ in Eq.~(\ref{ppmz}), we obtain $\delta$ is terms of $\alpha$ and $\beta$ as
\be
\delta = \frac{2 \eta\beta}{2\eta + 2 - \alpha -\beta}.
\ee

We now examine  $P^{++}(m)$. The spin configuration is such that the site  can receive particles from the left neighbor  but also lose particles to the right neighbor.  Like the earlier cases, the time evolution equation for $P^{++}(m)$ may be written. In the steady state, we obtain
\begin{align}
&P^{++}_{m+1}= \frac{2-\alpha-\beta}{2} \left[ P^{++}_m  - P^{++}_{m-1} (1-\delta_{m,0}) \right]\nonumber \\
&+ P^{++}_m (1-\delta_{m,0}) +\eta \left( 2 P^{++}_m - P^{-+}_m - P^{+-}_m  \right).
\end{align}
Multiplying by $z^m$ and summing over $m$, we obtain the generating function to be
\be
\widetilde{P}^{++}(z) = \frac{2 (1-z) \beta +2 \eta z \left[ \widetilde{P}^{+-}(z)+ \widetilde{P}^{-+}(z) \right]}{2 (1-z) -z (1-z) (2 - \alpha -\beta)-4 \eta z}. \label{pppz0}
\ee

Solving for the generating functions $\widetilde{P}^{++}(z)$, $\widetilde{P}^{+-}(z)$ and $\widetilde{P}^{-+}(z)$ from
Eqs.~\eqref{pmpz}, \eqref{ppmz} and \eqref{pppz0}, we finally obtain
\begin{widetext}
	\bea
	\widetilde{P}^{++}(z) &=& \frac{2 \left[(\alpha +\beta-2 )(z-1) +2 \eta\right] \left[ (\alpha +\beta) \eta z + \beta( z-1) \right]}{h(z)},
	\nonumber \\
	\widetilde{P}^{+-}(z) &= &\frac{2 \left[(\alpha +\beta) \eta z + \beta( z-1) \right] }{h(z)},\\
	\widetilde{P}^{-+}(z) &=& \frac{2 \eta  (z-1) \left[3 \alpha ^2 z+ \alpha  ((4 \beta -6)
		z+2)+(\beta -2) \beta  z\right]+4 \eta ^2 z (\alpha
		+\beta )+\alpha  (z-1)^2 (\alpha +\beta -2) \left[z
		(\alpha +\beta -2)+2\right] }{h(z)}, \nonumber
	\eea
\end{widetext}
where the denominator $h(z)$ is given by
\begin{align}
h(z) &= \left[ \left(\alpha  +\beta   -2 \right) (1+\eta)   z +2 -\alpha -\beta +2 \eta \right] \nonumber \\
&\times \left[ \left(\alpha +\beta  -2 \right) z^2 -\left( \alpha +\beta -4 \eta -4\right) z -2 \right].
\end{align}

{\rahul
\section{Relating the hydrodynamic coefficients in the RTP and gap pictures}

As stated in section \ref{sec:model}, the particles in the RTP picture become sites in the gap picture. Thus, a length $\Delta x$ in the gap picture is equivalent to a length $\Delta x_{RTP}$ in the corresponding RTP configuration through
\be
\Delta x = \rho_{RTP} \Delta x_{RTP}
\ee
Flow of gap particles across a bond in the gap picture is equivalent to the movement of the RTP corresponding to the bond in the opposite direction. Thus, current in the gap picture is the negative of the average velocity in the RTP picture:
\be
J = - v_{RTP}
\ee
Now let us consider the current in the RTP picture in the presence of a density gradient. Since the average current is equal to the density multiplied by the average velocity of particles, we have
\bea
J_{RTP} &=& \rho_{RTP} v_{RTP} = - \rho_{RTP} J \nonumber \\
 &=& \rho_{RTP} D(\rho) \frac{d\rho}{dx} \nonumber\\
 &=& \rho_{RTP} D\left(\frac{1}{\rho_{RTP}}-1\right) \rho_{RTP} \frac{d}{dx_{RTP}} \left(\frac{1}{\rho_{RTP} - 1} \right) \nonumber\\
 &=& - D\left(\frac{1}{\rho_{RTP}}-1\right) \frac{d\rho_{RTP}}{dx_{RTP}}
\eea

Thus, the diffusion constants in the two pictures are simply related as
\be
D_{RTP}(\rho_{RTP}) = D_{gap}\left(\frac{1}{\rho_{RTP}}-1\right)
\ee

Similarly, it can be derived that the conductivity in the presence of a field in the RTP picture is related to the conductivity in the gap picture through
\be
\chi_{RTP}(\rho_{RTP}) = -\rho_{RTP} \chi_{gap}\left(\frac{1}{\rho_{RTP}}-1\right) 
\ee
where the negative sign, included for consistency, accounts for the fact that a right-biasing field for the gap particles is a left-biasing field for the RTPs.}

\section{The empty interval approximation for small $\eta$}

Let $E_n$ be the probability that a given set of $n$ contiguous sites is empty. Then the probability that a set of $n$ contiguous sites is empty and has an occupied site on the immediately right is $E_n-E_{n+1} \equiv Q_n$. We now study the time evolution of the probability $E_n$ for $n \gg 1$. There are two contributions: complete mass transfers labeled as (1) and incomplete mass transfers due to fragmentation labeled as (2).

(1) Consider the contribution from complete transfers. An empty cluster of length $n$ is destroyed if there is a transfer over a distance $k$ or greater from an occupied site at the end of an empty cluster of length $n+k-1$. The rate of this happening is given by $\eta ~ p_{>}(k) ~Q_{n+k-1}$, where $p_{>}(k)$ is the probability of mass being transferred a distance larger than $k$. 

A cluster of length $n$ is created if there is a cluster of length $n-k$, followed by an occupied site, followed by an empty cluster of length $k-1$, and there is an event which transfers the mass at the occupied site by a distance  $\ge k$. The probability of this event can be approximated as $\eta p_>(k)  Q_{n-k} Q_{k-1}/n_c$, where $n_c$ is the density of occupied sites in the steady state. Collecting together the creation and destruction processes, we obtain
\beq
\frac{dE_n}{dt}\bigg\lvert_{(1)} = 2 \eta \sum_k p_>(k) \left(Q_{n+k-1} - Q_{n-k} \frac{Q_{k-1}}{n_c} \right),
\eeq
where the subscript $(1)$ denotes the contribution from complete mass transfers.

Since we have assumed $n \gg 1$, we can approximate the above by a continuum description,
\beqa
\frac{dE_n}{dt}\bigg\lvert_{(1)} &\approx& 2 \eta \sum_k p_>(k) \bigg[ (2k-1) \frac{d Q_n}{dn} \nonumber \\
&&+ Q_{n-k} \left( 1- \frac{Q_{k-1}}{n_c} \right) \bigg], \label{En1}\\
&=& 12 \eta \frac{d Q_n}{dn} + 2 \eta \sum_k p_>(k) Q_{n-k} \left[1- \frac{Q_{k-1}}{n_c} \right]. \nonumber 
\eeqa

We note that the quantity $Q_{k-1}$ is of the order of $n_c$ times the probability of finding an interval of length $k$, and hence $1- Q_{k-1}/n_c$ is of $O(n_c)$. We have assumed that we are in the limit $n_c \ll 1$. Now, for large $n$, $Q_n \sim n_c e^{-n_c n}$, $Q_n'$ is $O(n_c^2)$, and thus the two terms are of comparable order, and hence we cannot neglect the second term. 

In order to  evaluate the second term in Eq.~(\ref{En1}), we approximate $Q_{j}$ as follows: $Q_0 = Q_1 = n_c$, since the density of occupied sites is $n_c$ and an occupied site is always followed by an empty site, since two $+-$ sites cannot be contiguous. And for $k>1$, we assume $Q_{j} \approx n_c (1-n_c)^{j-1}$. With this assumption, we can calculate the second term to be
\beqa
&&2 \eta \sum_k p_>(k) Q_{n-k} \left( 1- \frac{Q_{k-1}}{n_c} \right) 
\nonumber \\
&&\approx 2 \eta \sum_k p_>(k) Q_{n-k} (k-2) n_c 
\approx 2 \eta n_c Q_n, \label{En2}
\eeqa
where we  used $Q_{n-k} \approx Q_n - k Q'_n = Q_n + O(n_c^2)$.

(2) We now consider the contributions to the empty interval probabilities due to incomplete mass transfers. Consider a mass transfer event initiated at a $+-$ site due to the flipping of the $-$ spin, and the nearest $-$ spin to the right is at a distance $k$. If, during this transfer, a spin a distance $i < k-1$ to the right flips, the mass ends up at two sites, $i$ and $k$. (If the spin at $k-1$ flips, the whole mass ends up at $k-1$ instead of $k$.) The probability of this event for a given $i$ is $\eta m$ to leading order. This happens also in case the $+$ spin on the original $+-$ site flips as well, where now some of the mass ends up on the right of the original site and some on the left. Note that since some mass does end up a distance $k$ away, the destruction of empty clusters to the right happens whether there is a splitting or not. However, the creation of new empty clusters might not happen.

Consider the change in the creation term due to splitting events. If during a mass transfer over a distance $>k$ from an occupied site inside the cluster, one of the intervening $k$ spins flips ($k-1$ in case of a transfer over exactly $k$ sites), the creation is interrupted. This could also happen in case the spin on the other side of the occupied site also flips. Thus,
\beqa
&&\frac{dE_n}{dt}\bigg\lvert_{(2)} \!\!\!\!\!\!= 2 \eta^2 \sum_k  m^*\! \left[k p(k) \!+\! (k+1) p_>(k+1)\right]  \frac{Q_{n-k} Q_{k-1}}{n_c}
\nonumber\\
&&  =10 \eta^2 m^* (Q_n + O(n_c^2)). \label{En3}
\eeqa

Collecting together the contributions from  Eqs.~\eqref{En1}, \eqref{En2} and \eqref{En3}, we obtain, to leading order,
\beq
\frac{dE_n}{dt} = 12 \eta \frac{d Q_n}{dn} + (2 \eta n_c + 10 \eta^2 m^*) Q_n.
\eeq
Thus, in the steady state,
\beq
-\frac{1}{Q_n} \frac{d Q_n}{dn} = \frac{1}{6}(n_c + 5 \eta m^*).
\eeq
Since the density of clusters is $n_c$, the average distance between clusters is $n_c^{-1}$, and hence, for large $n$, we can write $Q_n \approx C e^{- n_c n}$ to leading order. C is a constant that might depend on $n_c$ and other factors. 
We use the equality $n_c m^* = \rho$, the average density, to eliminate $m^*$ from the above equation. Thus, we finally obtain
\beqa
n_c &=& \sqrt{\eta \rho}, \\
m^* &=& \sqrt{\frac{\rho}{\eta}}, \label{n_cm1}
\eeqa
to leading order.

\section{Invariance of the low $\eta$ steady state to first order in $\rho'$ and $F$}

In this appendix, we show that, for small $\eta$, the empty interval probabilities are invariant upto order $\eta$ in the presence of a field $F$.
Consider the equations  [see \eqref{En1} and \eqref{En3}] for the empty interval probabilities $E_n$, the probability of finding a void of length $n$. Consider the terms in Eqs. \eqref{En1} and \eqref{En3}, but written in the general form
\beq
\frac{d E_n}{dt}\vert{\rm{right}} = \sum_i f_i F_i(E_{k_1},E_{k_2},\dots),
\eeq
where we have written only the equation for the creation and destruction terms at the right boundary of $E_n$. The $f_i$ denote the rates for the various jump processes, which depend on the spin configurations and rates for spin flips, and the $F_i$ are functions of the void probabilities $E_k$. 
Now, in the presence of a constant density gradient, the rates $f_i$ remain unchanged, but the $E_k$ depend on position due to their dependence on the density field $\rho(x)$. It is convenient to specify the position of a void of size $n$ by the position of its midpoint, say at $x$, and denote the probability of this cluster of sites being empty as $E_{n,x}$. In the presence of a density gradient,
\beq
E_{n,x} = E_{n,0} + \rho'(x) {\rahul \frac{dE_{n}}{d\rho} x} + O((\rho')^2).
\eeq
Now, the equation for the rate of change of void probabilities $E_{n}$ in th presence of a density gradient, due to transfer of particles through its right boundary becomes
\beq
\frac{d E_{n,0}}{dt}_{\vert{\rm{right}}} = \sum f_i F_i(E_{k_1,x_1},E_{k_2,x_2},\dots),
\eeq
where for convenience we have taken the original void to be centered at $x=0$, and the void which contribute to $F_i$ to be centered at $x_1, x_2,$ and so on. To first order, this equals
\beqa
\frac{d E_{n,0}}{dt}\vert{\rm{right}} &=& \sum_i f_i F_i(E_{k_1,0},E_{k_2,0},\dots) \nonumber\\
&+& \sum_i f_i \rho'(x) \sum_j x_j \frac{d F_i}{d E_{k_j}} \frac{d E_{k_j}}{d \rho} + O(\rho'^2). \nonumber
\eeqa

By symmetry, the equation for the rate of change of $E_{n,0}$ due to flows across its left boundary is
\beqa
\frac{d E_{n,0}}{dt}\vert{\rm{left}} &=& \sum_i f_i F_i(E_{k_1,-x_1},E_{k_2,-x_2},\dots), \nonumber\\
&=&\sum_i f_i F_i(E_{k_1,0},E_{k_2,0},\dots) \nonumber\\
&-& \sum_i f_i \rho'(x) \sum_j x_j \frac{d F_i}{d E_{k_j}} \frac{d E_{k_j}}{d \rho} + O(\rho'^2). \nonumber
\eeqa
Hence, the equation for the rate of change of $E_{n,0}$ in the presence of a density gradient is
\beq
\frac{d E_{n,0}}{dt} = 2 \sum_i f_i F_i(E_{k_1,0},E_{k_2,0},\dots) + O(\rho'^2),
\eeq
which proves that the steady state does not change in the presence of a density gradient, to first order in $\rho'(x)$.

Now, in the presence of a constant field $F$, assuming the system is on a ring, the probabilities $E_n$ do not depend on position. However, the jump rates to the right and left become functions of $F$. If a jump to the right occurs at rate $p(F)$, the same jump to the left occurs against the field, and hence at a rate $p(-F)$. To first order in $F$, $p(F) + p(-F) = 2 p(0)$. Now, the rate of change of $E_n$ due to a left jump across the right boundary has its corresponding process on the left boundary as a right jump. Hence, denoting the functional dependence of the jump rates $f_i$,
\beqa
\frac{d E_{n}}{dt} &=& \sum_i f_i(F) F_i(E_{k_1},E_{k_2},\dots) \nonumber\\
&+& \sum_i f_i(-F) F_i(E_{k_1},E_{k_2},\dots), \nonumber\\
&=& 2 \sum_i f_i(0) F_i(E_{k_1,0},E_{k_2,0},\dots) + O(F^2).
\eeqa
Hence, the steady state also does not, under a constant field $F$, change to first order in $F$. Thus we can also conclude that $m^{*}$, $P(m)$ and $n_c$ remain unchanged from their steady-state values to first order in $F$ and $\rho'$.


\begin{thebibliography}{57}%
	\makeatletter
	\providecommand \@ifxundefined [1]{%
		\@ifx{#1\undefined}
	}%
	\providecommand \@ifnum [1]{%
		\ifnum #1\expandafter \@firstoftwo
		\else \expandafter \@secondoftwo
		\fi
	}%
	\providecommand \@ifx [1]{%
		\ifx #1\expandafter \@firstoftwo
		\else \expandafter \@secondoftwo
		\fi
	}%
	\providecommand \natexlab [1]{#1}%
	\providecommand \enquote  [1]{``#1''}%
	\providecommand \bibnamefont  [1]{#1}%
	\providecommand \bibfnamefont [1]{#1}%
	\providecommand \citenamefont [1]{#1}%
	\providecommand \href@noop [0]{\@secondoftwo}%
	\providecommand \href [0]{\begingroup \@sanitize@url \@href}%
	\providecommand \@href[1]{\@@startlink{#1}\@@href}%
	\providecommand \@@href[1]{\endgroup#1\@@endlink}%
	\providecommand \@sanitize@url [0]{\catcode `\\12\catcode `\$12\catcode
		`\&12\catcode `\#12\catcode `\^12\catcode `\_12\catcode `\%12\relax}%
	\providecommand \@@startlink[1]{}%
	\providecommand \@@endlink[0]{}%
	\providecommand \url  [0]{\begingroup\@sanitize@url \@url }%
	\providecommand \@url [1]{\endgroup\@href {#1}{\urlprefix }}%
	\providecommand \urlprefix  [0]{URL }%
	\providecommand \Eprint [0]{\href }%
	\providecommand \doibase [0]{https://doi.org/}%
	\providecommand \selectlanguage [0]{\@gobble}%
	\providecommand \bibinfo  [0]{\@secondoftwo}%
	\providecommand \bibfield  [0]{\@secondoftwo}%
	\providecommand \translation [1]{[#1]}%
	\providecommand \BibitemOpen [0]{}%
	\providecommand \bibitemStop [0]{}%
	\providecommand \bibitemNoStop [0]{.\EOS\space}%
	\providecommand \EOS [0]{\spacefactor3000\relax}%
	\providecommand \BibitemShut  [1]{\csname bibitem#1\endcsname}%
	\let\auto@bib@innerbib\@empty
	\bibitem [{\citenamefont {Dombrowski}\ \emph {et~al.}(2004)\citenamefont
		{Dombrowski}, \citenamefont {Cisneros}, \citenamefont {Chatkaew},
		\citenamefont {Goldstein},\ and\ \citenamefont
		{Kessler}}]{DombrowskiPRL2004}%
	\BibitemOpen
	\bibfield  {author} {\bibinfo {author} {\bibfnamefont {C.}~\bibnamefont
			{Dombrowski}}, \bibinfo {author} {\bibfnamefont {L.}~\bibnamefont
			{Cisneros}}, \bibinfo {author} {\bibfnamefont {S.}~\bibnamefont {Chatkaew}},
		\bibinfo {author} {\bibfnamefont {R.~E.}\ \bibnamefont {Goldstein}},\ and\
		\bibinfo {author} {\bibfnamefont {J.~O.}\ \bibnamefont {Kessler}},\
	}\bibfield  {title} {\bibinfo {title} {Self-concentration and large-scale
			coherence in bacterial dynamics},\ }\href
	{https://doi.org/10.1103/PhysRevLett.93.098103} {\bibfield  {journal}
		{\bibinfo  {journal} {Phys. Rev. Lett.}\ }\textbf {\bibinfo {volume} {93}},\
		\bibinfo {pages} {098103} (\bibinfo {year} {2004})}\BibitemShut {NoStop}%
	\bibitem [{\citenamefont {Palacci}\ \emph {et~al.}(2013)\citenamefont
		{Palacci}, \citenamefont {Sacanna}, \citenamefont {Steinberg}, \citenamefont
		{Pine},\ and\ \citenamefont {Chaikin}}]{PalacciScience2013}%
	\BibitemOpen
	\bibfield  {author} {\bibinfo {author} {\bibfnamefont {J.}~\bibnamefont
			{Palacci}}, \bibinfo {author} {\bibfnamefont {S.}~\bibnamefont {Sacanna}},
		\bibinfo {author} {\bibfnamefont {A.~P.}\ \bibnamefont {Steinberg}}, \bibinfo
		{author} {\bibfnamefont {D.~J.}\ \bibnamefont {Pine}},\ and\ \bibinfo
		{author} {\bibfnamefont {P.~M.}\ \bibnamefont {Chaikin}},\ }\bibfield
	{title} {\bibinfo {title} {Living crystals of light-activated colloidal
			surfers},\ }\href {https://doi.org/10.1126/science.1230020} {\bibfield
		{journal} {\bibinfo  {journal} {Science}\ }\textbf {\bibinfo {volume}
			{339}},\ \bibinfo {pages} {936} (\bibinfo {year} {2013})}\BibitemShut
	{NoStop}%
	\bibitem [{\citenamefont {Surrey}\ \emph {et~al.}(2001)\citenamefont {Surrey},
		\citenamefont {N{\'e}d{\'e}lec}, \citenamefont {Leibler},\ and\ \citenamefont
		{Karsenti}}]{SurreyScience2001}%
	\BibitemOpen
	\bibfield  {author} {\bibinfo {author} {\bibfnamefont {T.}~\bibnamefont
			{Surrey}}, \bibinfo {author} {\bibfnamefont {F.}~\bibnamefont
			{N{\'e}d{\'e}lec}}, \bibinfo {author} {\bibfnamefont {S.}~\bibnamefont
			{Leibler}},\ and\ \bibinfo {author} {\bibfnamefont {E.}~\bibnamefont
			{Karsenti}},\ }\bibfield  {title} {\bibinfo {title} {Physical properties
			determining self-organization of motors and microtubules},\ }\href
	{https://doi.org/10.1126/science.1059758} {\bibfield  {journal} {\bibinfo
			{journal} {Science}\ }\textbf {\bibinfo {volume} {292}},\ \bibinfo {pages}
		{1167} (\bibinfo {year} {2001})}\BibitemShut {NoStop}%
	\bibitem [{\citenamefont {R.}\ \emph {et~al.}(2000)\citenamefont {R.},
		\citenamefont {D.}, \citenamefont {D.},\ and\ \citenamefont
		{H.}}]{KemkemerEPJE2000}%
	\BibitemOpen
	\bibfield  {author} {\bibinfo {author} {\bibfnamefont {K.}~\bibnamefont
			{R.}}, \bibinfo {author} {\bibfnamefont {K.}~\bibnamefont {D.}}, \bibinfo
		{author} {\bibfnamefont {K.}~\bibnamefont {D.}},\ and\ \bibinfo {author}
		{\bibfnamefont {G.}~\bibnamefont {H.}},\ }\bibfield  {title} {\bibinfo
		{title} {{Elastic properties of nematoid arrangements formed by amoeboid
				cells}},\ }\href {https://doi.org/https://doi.org/10.1007/s101890050024}
	{\bibfield  {journal} {\bibinfo  {journal} {The European Physical Journal E}\
		}\textbf {\bibinfo {volume} {1}},\ \bibinfo {pages} {215} (\bibinfo {year}
		{2000})}\BibitemShut {NoStop}%
	\bibitem [{\citenamefont {Narayan}\ \emph {et~al.}(2007)\citenamefont
		{Narayan}, \citenamefont {Ramaswamy},\ and\ \citenamefont
		{Menon}}]{NarayanScience2007}%
	\BibitemOpen
	\bibfield  {author} {\bibinfo {author} {\bibfnamefont {V.}~\bibnamefont
			{Narayan}}, \bibinfo {author} {\bibfnamefont {S.}~\bibnamefont {Ramaswamy}},\
		and\ \bibinfo {author} {\bibfnamefont {N.}~\bibnamefont {Menon}},\ }\bibfield
	{title} {\bibinfo {title} {Long-lived giant number fluctuations in a
			swarming granular nematic},\ }\href {https://doi.org/10.1126/science.1140414}
	{\bibfield  {journal} {\bibinfo  {journal} {Science}\ }\textbf {\bibinfo
			{volume} {317}},\ \bibinfo {pages} {105} (\bibinfo {year}
		{2007})}\BibitemShut {NoStop}%
	\bibitem [{\citenamefont {Peruani}\ \emph {et~al.}(2012)\citenamefont
		{Peruani}, \citenamefont {Starru\ss{}}, \citenamefont {Jakovljevic},
		\citenamefont {S\o{}gaard-Andersen}, \citenamefont {Deutsch},\ and\
		\citenamefont {B\"ar}}]{PeruaniPRL2012}%
	\BibitemOpen
	\bibfield  {author} {\bibinfo {author} {\bibfnamefont {F.}~\bibnamefont
			{Peruani}}, \bibinfo {author} {\bibfnamefont {J.}~\bibnamefont
			{Starru\ss{}}}, \bibinfo {author} {\bibfnamefont {V.}~\bibnamefont
			{Jakovljevic}}, \bibinfo {author} {\bibfnamefont {L.}~\bibnamefont
			{S\o{}gaard-Andersen}}, \bibinfo {author} {\bibfnamefont {A.}~\bibnamefont
			{Deutsch}},\ and\ \bibinfo {author} {\bibfnamefont {M.}~\bibnamefont
			{B\"ar}},\ }\bibfield  {title} {\bibinfo {title} {Collective motion and
			nonequilibrium cluster formation in colonies of gliding bacteria},\ }\href
	{https://doi.org/10.1103/PhysRevLett.108.098102} {\bibfield  {journal}
		{\bibinfo  {journal} {Phys. Rev. Lett.}\ }\textbf {\bibinfo {volume} {108}},\
		\bibinfo {pages} {098102} (\bibinfo {year} {2012})}\BibitemShut {NoStop}%
	\bibitem [{\citenamefont {Vicsek}\ \emph {et~al.}(1995)\citenamefont {Vicsek},
		\citenamefont {Czir\'ok}, \citenamefont {Ben-Jacob}, \citenamefont {Cohen},\
		and\ \citenamefont {Shochet}}]{VicsekPRL1995}%
	\BibitemOpen
	\bibfield  {author} {\bibinfo {author} {\bibfnamefont {T.}~\bibnamefont
			{Vicsek}}, \bibinfo {author} {\bibfnamefont {A.}~\bibnamefont {Czir\'ok}},
		\bibinfo {author} {\bibfnamefont {E.}~\bibnamefont {Ben-Jacob}}, \bibinfo
		{author} {\bibfnamefont {I.}~\bibnamefont {Cohen}},\ and\ \bibinfo {author}
		{\bibfnamefont {O.}~\bibnamefont {Shochet}},\ }\bibfield  {title} {\bibinfo
		{title} {Novel type of phase transition in a system of self-driven
			particles},\ }\href {https://doi.org/10.1103/PhysRevLett.75.1226} {\bibfield
		{journal} {\bibinfo  {journal} {Phys. Rev. Lett.}\ }\textbf {\bibinfo
			{volume} {75}},\ \bibinfo {pages} {1226} (\bibinfo {year}
		{1995})}\BibitemShut {NoStop}%
	\bibitem [{\citenamefont {Gr\'egoire}\ and\ \citenamefont
		{Chat\'e}(2004)}]{ChatePRL2004}%
	\BibitemOpen
	\bibfield  {author} {\bibinfo {author} {\bibfnamefont {G.}~\bibnamefont
			{Gr\'egoire}}\ and\ \bibinfo {author} {\bibfnamefont {H.}~\bibnamefont
			{Chat\'e}},\ }\bibfield  {title} {\bibinfo {title} {Onset of collective and
			cohesive motion},\ }\href {https://doi.org/10.1103/PhysRevLett.92.025702}
	{\bibfield  {journal} {\bibinfo  {journal} {Phys. Rev. Lett.}\ }\textbf
		{\bibinfo {volume} {92}},\ \bibinfo {pages} {025702} (\bibinfo {year}
		{2004})}\BibitemShut {NoStop}%
	\bibitem [{\citenamefont {Ginelli}(2016)}]{vicsekreview}%
	\BibitemOpen
	\bibfield  {author} {\bibinfo {author} {\bibfnamefont {F.}~\bibnamefont
			{Ginelli}},\ }\bibfield  {title} {\bibinfo {title} {The physics of the vicsek
			model},\ }\href@noop {} {\bibfield  {journal} {\bibinfo  {journal} {The
				European Physical Journal Special Topics}\ }\textbf {\bibinfo {volume}
			{225}},\ \bibinfo {pages} {2099} (\bibinfo {year} {2016})}\BibitemShut
	{NoStop}%
	\bibitem [{\citenamefont {Fily}\ and\ \citenamefont
		{Marchetti}(2012)}]{FilyPRL2012}%
	\BibitemOpen
	\bibfield  {author} {\bibinfo {author} {\bibfnamefont {Y.}~\bibnamefont
			{Fily}}\ and\ \bibinfo {author} {\bibfnamefont {M.~C.}\ \bibnamefont
			{Marchetti}},\ }\bibfield  {title} {\bibinfo {title} {Athermal phase
			separation of self-propelled particles with no alignment},\ }\href
	{https://doi.org/10.1103/PhysRevLett.108.235702} {\bibfield  {journal}
		{\bibinfo  {journal} {Phys. Rev. Lett.}\ }\textbf {\bibinfo {volume} {108}},\
		\bibinfo {pages} {235702} (\bibinfo {year} {2012})}\BibitemShut {NoStop}%
	\bibitem [{\citenamefont {Redner}\ \emph {et~al.}(2013)\citenamefont {Redner},
		\citenamefont {Hagan},\ and\ \citenamefont {Baskaran}}]{RednerPRL2013}%
	\BibitemOpen
	\bibfield  {author} {\bibinfo {author} {\bibfnamefont {G.~S.}\ \bibnamefont
			{Redner}}, \bibinfo {author} {\bibfnamefont {M.~F.}\ \bibnamefont {Hagan}},\
		and\ \bibinfo {author} {\bibfnamefont {A.}~\bibnamefont {Baskaran}},\
	}\bibfield  {title} {\bibinfo {title} {Structure and dynamics of a
			phase-separating active colloidal fluid},\ }\href
	{https://doi.org/10.1103/PhysRevLett.110.055701} {\bibfield  {journal}
		{\bibinfo  {journal} {Phys. Rev. Lett.}\ }\textbf {\bibinfo {volume} {110}},\
		\bibinfo {pages} {055701} (\bibinfo {year} {2013})}\BibitemShut {NoStop}%
	\bibitem [{\citenamefont {Tailleur}\ and\ \citenamefont
		{Cates}(2008)}]{tailleur2008statistical}%
	\BibitemOpen
	\bibfield  {author} {\bibinfo {author} {\bibfnamefont {J.}~\bibnamefont
			{Tailleur}}\ and\ \bibinfo {author} {\bibfnamefont {M.}~\bibnamefont
			{Cates}},\ }\bibfield  {title} {\bibinfo {title} {Statistical mechanics of
			interacting run-and-tumble bacteria},\ }\href@noop {} {\bibfield  {journal}
		{\bibinfo  {journal} {Physical review letters}\ }\textbf {\bibinfo {volume}
			{100}},\ \bibinfo {pages} {218103} (\bibinfo {year} {2008})}\BibitemShut
	{NoStop}%
	\bibitem [{\citenamefont {Elgeti}\ and\ \citenamefont
		{Gompper}(2015)}]{elgeti2015run}%
	\BibitemOpen
	\bibfield  {author} {\bibinfo {author} {\bibfnamefont {J.}~\bibnamefont
			{Elgeti}}\ and\ \bibinfo {author} {\bibfnamefont {G.}~\bibnamefont
			{Gompper}},\ }\bibfield  {title} {\bibinfo {title} {Run-and-tumble dynamics
			of self-propelled particles in confinement},\ }\href@noop {} {\bibfield
		{journal} {\bibinfo  {journal} {EPL (Europhysics Letters)}\ }\textbf
		{\bibinfo {volume} {109}},\ \bibinfo {pages} {58003} (\bibinfo {year}
		{2015})}\BibitemShut {NoStop}%
	\bibitem [{\citenamefont {Raymond}\ and\ \citenamefont
		{Evans}(2006)}]{EvansPRE2006}%
	\BibitemOpen
	\bibfield  {author} {\bibinfo {author} {\bibfnamefont {J.~R.}\ \bibnamefont
			{Raymond}}\ and\ \bibinfo {author} {\bibfnamefont {M.~R.}\ \bibnamefont
			{Evans}},\ }\bibfield  {title} {\bibinfo {title} {Flocking regimes in a
			simple lattice model},\ }\href {https://doi.org/10.1103/PhysRevE.73.036112}
	{\bibfield  {journal} {\bibinfo  {journal} {Phys. Rev. E}\ }\textbf {\bibinfo
			{volume} {73}},\ \bibinfo {pages} {036112} (\bibinfo {year}
		{2006})}\BibitemShut {NoStop}%
	\bibitem [{\citenamefont {Solon}\ and\ \citenamefont
		{Tailleur}(2013)}]{TailleurPRL2013}%
	\BibitemOpen
	\bibfield  {author} {\bibinfo {author} {\bibfnamefont {A.~P.}\ \bibnamefont
			{Solon}}\ and\ \bibinfo {author} {\bibfnamefont {J.}~\bibnamefont
			{Tailleur}},\ }\bibfield  {title} {\bibinfo {title} {Revisiting the flocking
			transition using active spins},\ }\href
	{https://doi.org/10.1103/PhysRevLett.111.078101} {\bibfield  {journal}
		{\bibinfo  {journal} {Phys. Rev. Lett.}\ }\textbf {\bibinfo {volume} {111}},\
		\bibinfo {pages} {078101} (\bibinfo {year} {2013})}\BibitemShut {NoStop}%
	\bibitem [{\citenamefont {Whitelam}\ \emph {et~al.}(2018)\citenamefont
		{Whitelam}, \citenamefont {Klymko},\ and\ \citenamefont
		{Mandal}}]{WhitelamJCP2018}%
	\BibitemOpen
	\bibfield  {author} {\bibinfo {author} {\bibfnamefont {S.}~\bibnamefont
			{Whitelam}}, \bibinfo {author} {\bibfnamefont {K.}~\bibnamefont {Klymko}},\
		and\ \bibinfo {author} {\bibfnamefont {D.}~\bibnamefont {Mandal}},\
	}\bibfield  {title} {\bibinfo {title} {Phase separation and large deviations
			of lattice active matter},\ }\href {https://doi.org/10.1063/1.5023403}
	{\bibfield  {journal} {\bibinfo  {journal} {The Journal of Chemical Physics}\
		}\textbf {\bibinfo {volume} {148}},\ \bibinfo {pages} {154902} (\bibinfo
		{year} {2018})}\BibitemShut {NoStop}%
	\bibitem [{\citenamefont {Slowman}\ \emph {et~al.}(2016)\citenamefont
		{Slowman}, \citenamefont {Evans},\ and\ \citenamefont
		{Blythe}}]{slowman2016jamming}%
	\BibitemOpen
	\bibfield  {author} {\bibinfo {author} {\bibfnamefont {A.}~\bibnamefont
			{Slowman}}, \bibinfo {author} {\bibfnamefont {M.}~\bibnamefont {Evans}},\
		and\ \bibinfo {author} {\bibfnamefont {R.}~\bibnamefont {Blythe}},\
	}\bibfield  {title} {\bibinfo {title} {Jamming and attraction of interacting
			run-and-tumble random walkers},\ }\href@noop {} {\bibfield  {journal}
		{\bibinfo  {journal} {Physical review letters}\ }\textbf {\bibinfo {volume}
			{116}},\ \bibinfo {pages} {218101} (\bibinfo {year} {2016})}\BibitemShut
	{NoStop}%
	\bibitem [{\citenamefont {Mallmin}\ \emph {et~al.}(2019)\citenamefont
		{Mallmin}, \citenamefont {Blythe},\ and\ \citenamefont
		{Evans}}]{mallmin2019exact}%
	\BibitemOpen
	\bibfield  {author} {\bibinfo {author} {\bibfnamefont {E.}~\bibnamefont
			{Mallmin}}, \bibinfo {author} {\bibfnamefont {R.~A.}\ \bibnamefont
			{Blythe}},\ and\ \bibinfo {author} {\bibfnamefont {M.~R.}\ \bibnamefont
			{Evans}},\ }\bibfield  {title} {\bibinfo {title} {Exact spectral solution of
			two interacting run-and-tumble particles on a ring lattice},\ }\href@noop {}
	{\bibfield  {journal} {\bibinfo  {journal} {Journal of Statistical Mechanics:
				Theory and Experiment}\ }\textbf {\bibinfo {volume} {2019}},\ \bibinfo
		{pages} {013204} (\bibinfo {year} {2019})}\BibitemShut {NoStop}%
	\bibitem [{\citenamefont {Kourbane-Houssene}\ \emph {et~al.}(2018)\citenamefont
		{Kourbane-Houssene}, \citenamefont {Erignoux}, \citenamefont {Bodineau},\
		and\ \citenamefont {Tailleur}}]{kourbane2018exact}%
	\BibitemOpen
	\bibfield  {author} {\bibinfo {author} {\bibfnamefont {M.}~\bibnamefont
			{Kourbane-Houssene}}, \bibinfo {author} {\bibfnamefont {C.}~\bibnamefont
			{Erignoux}}, \bibinfo {author} {\bibfnamefont {T.}~\bibnamefont {Bodineau}},\
		and\ \bibinfo {author} {\bibfnamefont {J.}~\bibnamefont {Tailleur}},\
	}\bibfield  {title} {\bibinfo {title} {Exact hydrodynamic description of
			active lattice gases},\ }\href@noop {} {\bibfield  {journal} {\bibinfo
			{journal} {Physical review letters}\ }\textbf {\bibinfo {volume} {120}},\
		\bibinfo {pages} {268003} (\bibinfo {year} {2018})}\BibitemShut {NoStop}%
	\bibitem [{\citenamefont {Chakraborty}\ \emph {et~al.}(2020)\citenamefont
		{Chakraborty}, \citenamefont {Chakraborti}, \citenamefont {Das},\ and\
		\citenamefont {Pradhan}}]{ChakrabortyCondmat2020}%
	\BibitemOpen
	\bibfield  {author} {\bibinfo {author} {\bibfnamefont {T.}~\bibnamefont
			{Chakraborty}}, \bibinfo {author} {\bibfnamefont {S.}~\bibnamefont
			{Chakraborti}}, \bibinfo {author} {\bibfnamefont {A.}~\bibnamefont {Das}},\
		and\ \bibinfo {author} {\bibfnamefont {P.}~\bibnamefont {Pradhan}},\
	}\bibfield  {title} {\bibinfo {title} {Hydrodynamics, superfluidity, and
			giant number fluctuations in a model of self-propelled particles},\ }\href
	{https://doi.org/10.1103/PhysRevE.101.052611} {\bibfield  {journal} {\bibinfo
			{journal} {Phys. Rev. E}\ }\textbf {\bibinfo {volume} {101}},\ \bibinfo
		{pages} {052611} (\bibinfo {year} {2020})}\BibitemShut {NoStop}%
	\bibitem [{\citenamefont {Cates}(2012)}]{CatesReview2012}%
	\BibitemOpen
	\bibfield  {author} {\bibinfo {author} {\bibfnamefont {M.~E.}\ \bibnamefont
			{Cates}},\ }\bibfield  {title} {\bibinfo {title} {Diffusive transport without
			detailed balance in motile bacteria: does microbiology need statistical
			physics?},\ }\href {https://doi.org/10.1088/0034-4885/75/4/042601} {\bibfield
		{journal} {\bibinfo  {journal} {Reports on Progress in Physics}\ }\textbf
		{\bibinfo {volume} {75}},\ \bibinfo {pages} {042601} (\bibinfo {year}
		{2012})}\BibitemShut {NoStop}%
	\bibitem [{\citenamefont {Marchetti}\ \emph {et~al.}(2013)\citenamefont
		{Marchetti}, \citenamefont {Joanny}, \citenamefont {Ramaswamy}, \citenamefont
		{Liverpool}, \citenamefont {Prost}, \citenamefont {Rao},\ and\ \citenamefont
		{Simha}}]{MarchettiReview2013}%
	\BibitemOpen
	\bibfield  {author} {\bibinfo {author} {\bibfnamefont {M.~C.}\ \bibnamefont
			{Marchetti}}, \bibinfo {author} {\bibfnamefont {J.~F.}\ \bibnamefont
			{Joanny}}, \bibinfo {author} {\bibfnamefont {S.}~\bibnamefont {Ramaswamy}},
		\bibinfo {author} {\bibfnamefont {T.~B.}\ \bibnamefont {Liverpool}}, \bibinfo
		{author} {\bibfnamefont {J.}~\bibnamefont {Prost}}, \bibinfo {author}
		{\bibfnamefont {M.}~\bibnamefont {Rao}},\ and\ \bibinfo {author}
		{\bibfnamefont {R.~A.}\ \bibnamefont {Simha}},\ }\bibfield  {title} {\bibinfo
		{title} {Hydrodynamics of soft active matter},\ }\href
	{https://doi.org/10.1103/RevModPhys.85.1143} {\bibfield  {journal} {\bibinfo
			{journal} {Rev. Mod. Phys.}\ }\textbf {\bibinfo {volume} {85}},\ \bibinfo
		{pages} {1143} (\bibinfo {year} {2013})}\BibitemShut {NoStop}%
	\bibitem [{\citenamefont {Toner}\ and\ \citenamefont
		{Tu}(1995)}]{TonerPRL1995}%
	\BibitemOpen
	\bibfield  {author} {\bibinfo {author} {\bibfnamefont {J.}~\bibnamefont
			{Toner}}\ and\ \bibinfo {author} {\bibfnamefont {Y.}~\bibnamefont {Tu}},\
	}\bibfield  {title} {\bibinfo {title} {Long-range order in a two-dimensional
			dynamical $\mathrm{XY}$ model: How birds fly together},\ }\href
	{https://doi.org/10.1103/PhysRevLett.75.4326} {\bibfield  {journal} {\bibinfo
			{journal} {Phys. Rev. Lett.}\ }\textbf {\bibinfo {volume} {75}},\ \bibinfo
		{pages} {4326} (\bibinfo {year} {1995})}\BibitemShut {NoStop}%
	\bibitem [{\citenamefont {Toner}\ and\ \citenamefont
		{Tu}(1998)}]{TonerPRE1998}%
	\BibitemOpen
	\bibfield  {author} {\bibinfo {author} {\bibfnamefont {J.}~\bibnamefont
			{Toner}}\ and\ \bibinfo {author} {\bibfnamefont {Y.}~\bibnamefont {Tu}},\
	}\bibfield  {title} {\bibinfo {title} {Flocks, herds, and schools: A
			quantitative theory of flocking},\ }\href
	{https://doi.org/10.1103/PhysRevE.58.4828} {\bibfield  {journal} {\bibinfo
			{journal} {Phys. Rev. E}\ }\textbf {\bibinfo {volume} {58}},\ \bibinfo
		{pages} {4828} (\bibinfo {year} {1998})}\BibitemShut {NoStop}%
	\bibitem [{\citenamefont {Ramaswamy}(2017)}]{ramaswamy2017active}%
	\BibitemOpen
	\bibfield  {author} {\bibinfo {author} {\bibfnamefont {S.}~\bibnamefont
			{Ramaswamy}},\ }\bibfield  {title} {\bibinfo {title} {Active matter},\
	}\href@noop {} {\bibfield  {journal} {\bibinfo  {journal} {Journal of
				Statistical Mechanics: Theory and Experiment}\ }\textbf {\bibinfo {volume}
			{2017}},\ \bibinfo {pages} {054002} (\bibinfo {year} {2017})}\BibitemShut
	{NoStop}%
	\bibitem [{\citenamefont {Levis}\ and\ \citenamefont
		{Berthier}(2015)}]{LevisEPL2015}%
	\BibitemOpen
	\bibfield  {author} {\bibinfo {author} {\bibfnamefont {D.}~\bibnamefont
			{Levis}}\ and\ \bibinfo {author} {\bibfnamefont {L.}~\bibnamefont
			{Berthier}},\ }\bibfield  {title} {\bibinfo {title} {From single-particle to
			collective effective temperatures in an active fluid of self-propelled
			particles},\ }\href {https://doi.org/10.1209/0295-5075/111/60006} {\bibfield
		{journal} {\bibinfo  {journal} {{EPL} (Europhysics Letters)}\ }\textbf
		{\bibinfo {volume} {111}},\ \bibinfo {pages} {60006} (\bibinfo {year}
		{2015})}\BibitemShut {NoStop}%
	\bibitem [{\citenamefont {Stenhammar}\ \emph {et~al.}(2013)\citenamefont
		{Stenhammar}, \citenamefont {Tiribocchi}, \citenamefont {Allen},
		\citenamefont {Marenduzzo},\ and\ \citenamefont {Cates}}]{CatesPRL2013}%
	\BibitemOpen
	\bibfield  {author} {\bibinfo {author} {\bibfnamefont {J.}~\bibnamefont
			{Stenhammar}}, \bibinfo {author} {\bibfnamefont {A.}~\bibnamefont
			{Tiribocchi}}, \bibinfo {author} {\bibfnamefont {R.~J.}\ \bibnamefont
			{Allen}}, \bibinfo {author} {\bibfnamefont {D.}~\bibnamefont {Marenduzzo}},\
		and\ \bibinfo {author} {\bibfnamefont {M.~E.}\ \bibnamefont {Cates}},\
	}\bibfield  {title} {\bibinfo {title} {Continuum theory of phase separation
			kinetics for active brownian particles},\ }\href
	{https://doi.org/10.1103/PhysRevLett.111.145702} {\bibfield  {journal}
		{\bibinfo  {journal} {Phys. Rev. Lett.}\ }\textbf {\bibinfo {volume} {111}},\
		\bibinfo {pages} {145702} (\bibinfo {year} {2013})}\BibitemShut {NoStop}%
	\bibitem [{\citenamefont {Chakraborti}\ \emph {et~al.}(2016)\citenamefont
		{Chakraborti}, \citenamefont {Mishra},\ and\ \citenamefont
		{Pradhan}}]{chakraborti2016additivity}%
	\BibitemOpen
	\bibfield  {author} {\bibinfo {author} {\bibfnamefont {S.}~\bibnamefont
			{Chakraborti}}, \bibinfo {author} {\bibfnamefont {S.}~\bibnamefont
			{Mishra}},\ and\ \bibinfo {author} {\bibfnamefont {P.}~\bibnamefont
			{Pradhan}},\ }\bibfield  {title} {\bibinfo {title} {Additivity, density
			fluctuations, and nonequilibrium thermodynamics for active brownian
			particles},\ }\href@noop {} {\bibfield  {journal} {\bibinfo  {journal}
			{Physical Review E}\ }\textbf {\bibinfo {volume} {93}},\ \bibinfo {pages}
		{052606} (\bibinfo {year} {2016})}\BibitemShut {NoStop}%
	\bibitem [{\citenamefont {Chakraborti}\ and\ \citenamefont
		{Pradhan}(2019)}]{chakraborti2019additivity}%
	\BibitemOpen
	\bibfield  {author} {\bibinfo {author} {\bibfnamefont {S.}~\bibnamefont
			{Chakraborti}}\ and\ \bibinfo {author} {\bibfnamefont {P.}~\bibnamefont
			{Pradhan}},\ }\bibfield  {title} {\bibinfo {title} {Additivity and density
			fluctuations in vicsek-like models of self-propelled particles},\ }\href
	{https://doi.org/10.1103/PhysRevE.99.052604} {\bibfield  {journal} {\bibinfo
			{journal} {Phys. Rev. E}\ }\textbf {\bibinfo {volume} {99}},\ \bibinfo
		{pages} {052604} (\bibinfo {year} {2019})}\BibitemShut {NoStop}%
	\bibitem [{\citenamefont {{Solon A. P.}}\ \emph {et~al.}(2015)\citenamefont
		{{Solon A. P.}}, \citenamefont {{Fily Y.}}, \citenamefont {{Baskaran A.}},
		\citenamefont {{Cates M. E.}}, \citenamefont {{Kafri Y.}}, \citenamefont
		{{Kardar M.}},\ and\ \citenamefont {{Tailleur J.}}}]{KafriNature2015}%
	\BibitemOpen
	\bibfield  {author} {\bibinfo {author} {\bibnamefont {{Solon A. P.}}},
		\bibinfo {author} {\bibnamefont {{Fily Y.}}}, \bibinfo {author} {\bibnamefont
			{{Baskaran A.}}}, \bibinfo {author} {\bibnamefont {{Cates M. E.}}}, \bibinfo
		{author} {\bibnamefont {{Kafri Y.}}}, \bibinfo {author} {\bibnamefont
			{{Kardar M.}}},\ and\ \bibinfo {author} {\bibnamefont {{Tailleur J.}}},\
	}\bibfield  {title} {\bibinfo {title} {{Pressure is not a state function for
				generic active fluids}},\ }\href
	{https://doi.org/https://doi.org/10.1038/nphys3377} {\bibfield  {journal}
		{\bibinfo  {journal} {Nature Physics}\ }\textbf {\bibinfo {volume} {11}},\
		\bibinfo {pages} {673} (\bibinfo {year} {2015})}\BibitemShut {NoStop}%
	\bibitem [{\citenamefont {Soto}\ and\ \citenamefont
		{Golestanian}(2014)}]{soto2014run}%
	\BibitemOpen
	\bibfield  {author} {\bibinfo {author} {\bibfnamefont {R.}~\bibnamefont
			{Soto}}\ and\ \bibinfo {author} {\bibfnamefont {R.}~\bibnamefont
			{Golestanian}},\ }\bibfield  {title} {\bibinfo {title} {Run-and-tumble
			dynamics in a crowded environment: Persistent exclusion process for
			swimmers},\ }\href@noop {} {\bibfield  {journal} {\bibinfo  {journal}
			{Physical Review E}\ }\textbf {\bibinfo {volume} {89}},\ \bibinfo {pages}
		{012706} (\bibinfo {year} {2014})}\BibitemShut {NoStop}%
	\bibitem [{\citenamefont {Sep{\'u}lveda}\ and\ \citenamefont
		{Soto}(2016)}]{sepulveda2016coarsening}%
	\BibitemOpen
	\bibfield  {author} {\bibinfo {author} {\bibfnamefont {N.}~\bibnamefont
			{Sep{\'u}lveda}}\ and\ \bibinfo {author} {\bibfnamefont {R.}~\bibnamefont
			{Soto}},\ }\bibfield  {title} {\bibinfo {title} {Coarsening and clustering in
			run-and-tumble dynamics with short-range exclusion},\ }\href@noop {}
	{\bibfield  {journal} {\bibinfo  {journal} {Physical Review E}\ }\textbf
		{\bibinfo {volume} {94}},\ \bibinfo {pages} {022603} (\bibinfo {year}
		{2016})}\BibitemShut {NoStop}%
	\bibitem [{\citenamefont {Cates}\ and\ \citenamefont
		{Tailleur}(2013)}]{CatesEPL2013}%
	\BibitemOpen
	\bibfield  {author} {\bibinfo {author} {\bibfnamefont {M.~E.}\ \bibnamefont
			{Cates}}\ and\ \bibinfo {author} {\bibfnamefont {J.}~\bibnamefont
			{Tailleur}},\ }\bibfield  {title} {\bibinfo {title} {When are active brownian
			particles and run-and-tumble particles equivalent? consequences for
			motility-induced phase separation},\ }\href
	{https://doi.org/10.1209/0295-5075/101/20010} {\bibfield  {journal} {\bibinfo
			{journal} {{EPL} (Europhysics Letters)}\ }\textbf {\bibinfo {volume}
			{101}},\ \bibinfo {pages} {20010} (\bibinfo {year} {2013})}\BibitemShut
	{NoStop}%
	\bibitem [{\citenamefont {Dolai}\ \emph {et~al.}(2020)\citenamefont {Dolai},
		\citenamefont {Das}, \citenamefont {Kundu}, \citenamefont {Dasgupta},
		\citenamefont {Dhar},\ and\ \citenamefont {Kumar}}]{DolaiCondmat2020}%
	\BibitemOpen
	\bibfield  {author} {\bibinfo {author} {\bibfnamefont {P.}~\bibnamefont
			{Dolai}}, \bibinfo {author} {\bibfnamefont {A.}~\bibnamefont {Das}}, \bibinfo
		{author} {\bibfnamefont {A.}~\bibnamefont {Kundu}}, \bibinfo {author}
		{\bibfnamefont {C.}~\bibnamefont {Dasgupta}}, \bibinfo {author}
		{\bibfnamefont {A.}~\bibnamefont {Dhar}},\ and\ \bibinfo {author}
		{\bibfnamefont {K.~V.}\ \bibnamefont {Kumar}},\ }\href@noop {} {\bibinfo
		{title} {Universal scaling in active single-file dynamics}} (\bibinfo {year}
	{2020}),\ \Eprint {https://arxiv.org/abs/2004.01150} {arXiv:2004.01150
		[cond-mat.stat-mech]} \BibitemShut {NoStop}%
	\bibitem [{\citenamefont {Gradenigo}\ and\ \citenamefont
		{Majumdar}(2019)}]{gradenigo2019first}%
	\BibitemOpen
	\bibfield  {author} {\bibinfo {author} {\bibfnamefont {G.}~\bibnamefont
			{Gradenigo}}\ and\ \bibinfo {author} {\bibfnamefont {S.~N.}\ \bibnamefont
			{Majumdar}},\ }\bibfield  {title} {\bibinfo {title} {A first-order dynamical
			transition in the displacement distribution of a driven run-and-tumble
			particle},\ }\href@noop {} {\bibfield  {journal} {\bibinfo  {journal}
			{Journal of Statistical Mechanics: Theory and Experiment}\ }\textbf {\bibinfo
			{volume} {2019}},\ \bibinfo {pages} {053206} (\bibinfo {year}
		{2019})}\BibitemShut {NoStop}%
	\bibitem [{\citenamefont {Malakar}\ \emph {et~al.}(2018)\citenamefont
		{Malakar}, \citenamefont {Jemseena}, \citenamefont {Kundu}, \citenamefont
		{Kumar}, \citenamefont {Sabhapandit}, \citenamefont {Majumdar}, \citenamefont
		{Redner},\ and\ \citenamefont {Dhar}}]{malakar2018steady}%
	\BibitemOpen
	\bibfield  {author} {\bibinfo {author} {\bibfnamefont {K.}~\bibnamefont
			{Malakar}}, \bibinfo {author} {\bibfnamefont {V.}~\bibnamefont {Jemseena}},
		\bibinfo {author} {\bibfnamefont {A.}~\bibnamefont {Kundu}}, \bibinfo
		{author} {\bibfnamefont {K.~V.}\ \bibnamefont {Kumar}}, \bibinfo {author}
		{\bibfnamefont {S.}~\bibnamefont {Sabhapandit}}, \bibinfo {author}
		{\bibfnamefont {S.~N.}\ \bibnamefont {Majumdar}}, \bibinfo {author}
		{\bibfnamefont {S.}~\bibnamefont {Redner}},\ and\ \bibinfo {author}
		{\bibfnamefont {A.}~\bibnamefont {Dhar}},\ }\bibfield  {title} {\bibinfo
		{title} {Steady state, relaxation and first-passage properties of a
			run-and-tumble particle in one-dimension},\ }\href@noop {} {\bibfield
		{journal} {\bibinfo  {journal} {Journal of Statistical Mechanics: Theory and
				Experiment}\ }\textbf {\bibinfo {volume} {2018}},\ \bibinfo {pages} {043215}
		(\bibinfo {year} {2018})}\BibitemShut {NoStop}%
	\bibitem [{\citenamefont {Dhar}\ \emph {et~al.}(2019)\citenamefont {Dhar},
		\citenamefont {Kundu}, \citenamefont {Majumdar}, \citenamefont
		{Sabhapandit},\ and\ \citenamefont {Schehr}}]{dhar2019run}%
	\BibitemOpen
	\bibfield  {author} {\bibinfo {author} {\bibfnamefont {A.}~\bibnamefont
			{Dhar}}, \bibinfo {author} {\bibfnamefont {A.}~\bibnamefont {Kundu}},
		\bibinfo {author} {\bibfnamefont {S.~N.}\ \bibnamefont {Majumdar}}, \bibinfo
		{author} {\bibfnamefont {S.}~\bibnamefont {Sabhapandit}},\ and\ \bibinfo
		{author} {\bibfnamefont {G.}~\bibnamefont {Schehr}},\ }\bibfield  {title}
	{\bibinfo {title} {Run-and-tumble particle in one-dimensional confining
			potentials: Steady-state, relaxation, and first-passage properties},\
	}\href@noop {} {\bibfield  {journal} {\bibinfo  {journal} {Physical Review
				E}\ }\textbf {\bibinfo {volume} {99}},\ \bibinfo {pages} {032132} (\bibinfo
		{year} {2019})}\BibitemShut {NoStop}%
	\bibitem [{\citenamefont {Basu}\ \emph {et~al.}(2020)\citenamefont {Basu},
		\citenamefont {Majumdar}, \citenamefont {Rosso}, \citenamefont
		{Sabhapandit},\ and\ \citenamefont {Schehr}}]{basu2020exact}%
	\BibitemOpen
	\bibfield  {author} {\bibinfo {author} {\bibfnamefont {U.}~\bibnamefont
			{Basu}}, \bibinfo {author} {\bibfnamefont {S.~N.}\ \bibnamefont {Majumdar}},
		\bibinfo {author} {\bibfnamefont {A.}~\bibnamefont {Rosso}}, \bibinfo
		{author} {\bibfnamefont {S.}~\bibnamefont {Sabhapandit}},\ and\ \bibinfo
		{author} {\bibfnamefont {G.}~\bibnamefont {Schehr}},\ }\bibfield  {title}
	{\bibinfo {title} {Exact stationary state of a run-and-tumble particle with
			three internal states in a harmonic trap},\ }\href@noop {} {\bibfield
		{journal} {\bibinfo  {journal} {Journal of Physics A: Mathematical and
				Theoretical}\ }\textbf {\bibinfo {volume} {53}},\ \bibinfo {pages} {09LT01}
		(\bibinfo {year} {2020})}\BibitemShut {NoStop}%
	\bibitem [{\citenamefont {Basu}\ \emph {et~al.}(2018)\citenamefont {Basu},
		\citenamefont {Majumdar}, \citenamefont {Rosso},\ and\ \citenamefont
		{Schehr}}]{basu2018active}%
	\BibitemOpen
	\bibfield  {author} {\bibinfo {author} {\bibfnamefont {U.}~\bibnamefont
			{Basu}}, \bibinfo {author} {\bibfnamefont {S.~N.}\ \bibnamefont {Majumdar}},
		\bibinfo {author} {\bibfnamefont {A.}~\bibnamefont {Rosso}},\ and\ \bibinfo
		{author} {\bibfnamefont {G.}~\bibnamefont {Schehr}},\ }\bibfield  {title}
	{\bibinfo {title} {Active brownian motion in two dimensions},\ }\href@noop {}
	{\bibfield  {journal} {\bibinfo  {journal} {Physical Review E}\ }\textbf
		{\bibinfo {volume} {98}},\ \bibinfo {pages} {062121} (\bibinfo {year}
		{2018})}\BibitemShut {NoStop}%
	\bibitem [{\citenamefont {Majumdar}\ and\ \citenamefont
		{Meerson}(2020)}]{majumdar2020toward}%
	\BibitemOpen
	\bibfield  {author} {\bibinfo {author} {\bibfnamefont {S.~N.}\ \bibnamefont
			{Majumdar}}\ and\ \bibinfo {author} {\bibfnamefont {B.}~\bibnamefont
			{Meerson}},\ }\bibfield  {title} {\bibinfo {title} {Toward the full
			short-time statistics of an active brownian particle on the plane},\
	}\href@noop {} {\bibfield  {journal} {\bibinfo  {journal} {arXiv preprint
				arXiv:2004.13547}\ } (\bibinfo {year} {2020})}\BibitemShut {NoStop}%
	\bibitem [{\citenamefont {Slowman}\ \emph {et~al.}(2017)\citenamefont
		{Slowman}, \citenamefont {Evans},\ and\ \citenamefont
		{Blythe}}]{slowman2017exact}%
	\BibitemOpen
	\bibfield  {author} {\bibinfo {author} {\bibfnamefont {A.}~\bibnamefont
			{Slowman}}, \bibinfo {author} {\bibfnamefont {M.}~\bibnamefont {Evans}},\
		and\ \bibinfo {author} {\bibfnamefont {R.}~\bibnamefont {Blythe}},\
	}\bibfield  {title} {\bibinfo {title} {Exact solution of two interacting
			run-and-tumble random walkers with finite tumble duration},\ }\href@noop {}
	{\bibfield  {journal} {\bibinfo  {journal} {Journal of Physics A:
				Mathematical and Theoretical}\ }\textbf {\bibinfo {volume} {50}},\ \bibinfo
		{pages} {375601} (\bibinfo {year} {2017})}\BibitemShut {NoStop}%
	\bibitem [{\citenamefont {Das}\ \emph {et~al.}(2020)\citenamefont {Das},
		\citenamefont {Dhar},\ and\ \citenamefont {Kundu}}]{das2020gap}%
	\BibitemOpen
	\bibfield  {author} {\bibinfo {author} {\bibfnamefont {A.}~\bibnamefont
			{Das}}, \bibinfo {author} {\bibfnamefont {A.}~\bibnamefont {Dhar}},\ and\
		\bibinfo {author} {\bibfnamefont {A.}~\bibnamefont {Kundu}},\ }\bibfield
	{title} {\bibinfo {title} {Gap statistics of two interacting run and tumble
			particles in one dimension},\ }\href@noop {} {\bibfield  {journal} {\bibinfo
			{journal} {Journal of Physics A: Mathematical and Theoretical}\ }\textbf
		{\bibinfo {volume} {53}},\ \bibinfo {pages} {345003} (\bibinfo {year}
		{2020})}\BibitemShut {NoStop}%
	\bibitem [{\citenamefont {Le~Doussal}\ \emph {et~al.}(2019)\citenamefont
		{Le~Doussal}, \citenamefont {Majumdar},\ and\ \citenamefont
		{Schehr}}]{doussal2019non}%
	\BibitemOpen
	\bibfield  {author} {\bibinfo {author} {\bibfnamefont {P.}~\bibnamefont
			{Le~Doussal}}, \bibinfo {author} {\bibfnamefont {S.~N.}\ \bibnamefont
			{Majumdar}},\ and\ \bibinfo {author} {\bibfnamefont {G.}~\bibnamefont
			{Schehr}},\ }\bibfield  {title} {\bibinfo {title} {Noncrossing run-and-tumble
			particles on a line},\ }\href@noop {} {\bibfield  {journal} {\bibinfo
			{journal} {Physical Review E}\ }\textbf {\bibinfo {volume} {100}},\ \bibinfo
		{pages} {012113} (\bibinfo {year} {2019})}\BibitemShut {NoStop}%
	\bibitem [{\citenamefont {Das}\ \emph {et~al.}(2015)\citenamefont {Das},
		\citenamefont {Chatterjee}, \citenamefont {Pradhan},\ and\ \citenamefont
		{Mohanty}}]{das2015additivity}%
	\BibitemOpen
	\bibfield  {author} {\bibinfo {author} {\bibfnamefont {A.}~\bibnamefont
			{Das}}, \bibinfo {author} {\bibfnamefont {S.}~\bibnamefont {Chatterjee}},
		\bibinfo {author} {\bibfnamefont {P.}~\bibnamefont {Pradhan}},\ and\ \bibinfo
		{author} {\bibfnamefont {P.}~\bibnamefont {Mohanty}},\ }\bibfield  {title}
	{\bibinfo {title} {Additivity property and emergence of power laws in
			nonequilibrium steady states},\ }\href@noop {} {\bibfield  {journal}
		{\bibinfo  {journal} {Physical Review E}\ }\textbf {\bibinfo {volume} {92}},\
		\bibinfo {pages} {052107} (\bibinfo {year} {2015})}\BibitemShut {NoStop}%
	\bibitem [{\citenamefont {Das}\ \emph {et~al.}(2017)\citenamefont {Das},
		\citenamefont {Kundu},\ and\ \citenamefont {Pradhan}}]{das2017einstein}%
	\BibitemOpen
	\bibfield  {author} {\bibinfo {author} {\bibfnamefont {A.}~\bibnamefont
			{Das}}, \bibinfo {author} {\bibfnamefont {A.}~\bibnamefont {Kundu}},\ and\
		\bibinfo {author} {\bibfnamefont {P.}~\bibnamefont {Pradhan}},\ }\bibfield
	{title} {\bibinfo {title} {Einstein relation and hydrodynamics of
			nonequilibrium mass transport processes},\ }\href@noop {} {\bibfield
		{journal} {\bibinfo  {journal} {Physical Review E}\ }\textbf {\bibinfo
			{volume} {95}},\ \bibinfo {pages} {062128} (\bibinfo {year}
		{2017})}\BibitemShut {NoStop}%
	\bibitem [{\citenamefont {Burschka}\ \emph {et~al.}(1989)\citenamefont
		{Burschka}, \citenamefont {Doering},\ and\ \citenamefont
		{Ben-Avraham}}]{burschka1989transition}%
	\BibitemOpen
	\bibfield  {author} {\bibinfo {author} {\bibfnamefont {M.~A.}\ \bibnamefont
			{Burschka}}, \bibinfo {author} {\bibfnamefont {C.~R.}\ \bibnamefont
			{Doering}},\ and\ \bibinfo {author} {\bibfnamefont {D.}~\bibnamefont
			{Ben-Avraham}},\ }\bibfield  {title} {\bibinfo {title} {Transition in the
			relaxation dynamics of a reversible diffusion-limited reaction},\ }\href@noop
	{} {\bibfield  {journal} {\bibinfo  {journal} {Physical review letters}\
		}\textbf {\bibinfo {volume} {63}},\ \bibinfo {pages} {700} (\bibinfo {year}
		{1989})}\BibitemShut {NoStop}%
	\bibitem [{\citenamefont {Ben-Avraham}\ \emph {et~al.}(1990)\citenamefont
		{Ben-Avraham}, \citenamefont {Burschka},\ and\ \citenamefont
		{Doering}}]{ben1990statics}%
	\BibitemOpen
	\bibfield  {author} {\bibinfo {author} {\bibfnamefont {D.}~\bibnamefont
			{Ben-Avraham}}, \bibinfo {author} {\bibfnamefont {M.~A.}\ \bibnamefont
			{Burschka}},\ and\ \bibinfo {author} {\bibfnamefont {C.~R.}\ \bibnamefont
			{Doering}},\ }\bibfield  {title} {\bibinfo {title} {Statics and dynamics of a
			diffusion-limited reaction: anomalous kinetics, nonequilibrium self-ordering,
			and a dynamic transition},\ }\href@noop {} {\bibfield  {journal} {\bibinfo
			{journal} {Journal of Statistical Physics}\ }\textbf {\bibinfo {volume}
			{60}},\ \bibinfo {pages} {695} (\bibinfo {year} {1990})}\BibitemShut
	{NoStop}%
	\bibitem [{\citenamefont {Derrida}(2011)}]{derrida2011microscopic}%
	\BibitemOpen
	\bibfield  {author} {\bibinfo {author} {\bibfnamefont {B.}~\bibnamefont
			{Derrida}},\ }\bibfield  {title} {\bibinfo {title} {Microscopic versus
			macroscopic approaches to non-equilibrium systems},\ }\href@noop {}
	{\bibfield  {journal} {\bibinfo  {journal} {Journal of Statistical Mechanics:
				Theory and Experiment}\ }\textbf {\bibinfo {volume} {2011}},\ \bibinfo
		{pages} {P01030} (\bibinfo {year} {2011})}\BibitemShut {NoStop}%
	\bibitem [{\citenamefont {Chakraborti}(2018)}]{Subhadip_thesis}%
	\BibitemOpen
	\bibfield  {author} {\bibinfo {author} {\bibfnamefont {S.}~\bibnamefont
			{Chakraborti}},\ }\emph {\bibinfo {title} {Studies of fluctuations in systems
			of self-propelled particles}},\ \href@noop {} {Ph.D. thesis},\ \bibinfo
	{school} {University of Calcutta} (\bibinfo {year} {2018})\BibitemShut
	{NoStop}%
	\bibitem [{\citenamefont {Rajesh}\ and\ \citenamefont
		{Majumdar}(2000{\natexlab{a}})}]{rajesh2000conserved}%
	\BibitemOpen
	\bibfield  {author} {\bibinfo {author} {\bibfnamefont {R.}~\bibnamefont
			{Rajesh}}\ and\ \bibinfo {author} {\bibfnamefont {S.~N.}\ \bibnamefont
			{Majumdar}},\ }\bibfield  {title} {\bibinfo {title} {Conserved mass models
			and particle systems in one dimension},\ }\href@noop {} {\bibfield  {journal}
		{\bibinfo  {journal} {Journal of Statistical Physics}\ }\textbf {\bibinfo
			{volume} {99}},\ \bibinfo {pages} {943} (\bibinfo {year}
		{2000}{\natexlab{a}})}\BibitemShut {NoStop}%
	\bibitem [{\citenamefont {Rajesh}\ and\ \citenamefont
		{Majumdar}(2000{\natexlab{b}})}]{rajesh2000exact}%
	\BibitemOpen
	\bibfield  {author} {\bibinfo {author} {\bibfnamefont {R.}~\bibnamefont
			{Rajesh}}\ and\ \bibinfo {author} {\bibfnamefont {S.~N.}\ \bibnamefont
			{Majumdar}},\ }\bibfield  {title} {\bibinfo {title} {Exact calculation of the
			spatiotemporal correlations in the takayasu model and in the q model of force
			fluctuations in bead packs},\ }\href@noop {} {\bibfield  {journal} {\bibinfo
			{journal} {Physical Review E}\ }\textbf {\bibinfo {volume} {62}},\ \bibinfo
		{pages} {3186} (\bibinfo {year} {2000}{\natexlab{b}})}\BibitemShut {NoStop}%
	\bibitem [{\citenamefont {Krapivsky}\ \emph {et~al.}(2010)\citenamefont
		{Krapivsky}, \citenamefont {Redner},\ and\ \citenamefont
		{Ben-Naim}}]{krapivsky2010kinetic}%
	\BibitemOpen
	\bibfield  {author} {\bibinfo {author} {\bibfnamefont {P.~L.}\ \bibnamefont
			{Krapivsky}}, \bibinfo {author} {\bibfnamefont {S.}~\bibnamefont {Redner}},\
		and\ \bibinfo {author} {\bibfnamefont {E.}~\bibnamefont {Ben-Naim}},\
	}\href@noop {} {\emph {\bibinfo {title} {A kinetic view of statistical
				physics}}}\ (\bibinfo  {publisher} {Cambridge University Press},\ \bibinfo
	{year} {2010})\BibitemShut {NoStop}%
	\bibitem [{\citenamefont {Kang}\ and\ \citenamefont
		{Redner}(1985)}]{kang1985fluctuation}%
	\BibitemOpen
	\bibfield  {author} {\bibinfo {author} {\bibfnamefont {K.}~\bibnamefont
			{Kang}}\ and\ \bibinfo {author} {\bibfnamefont {S.}~\bibnamefont {Redner}},\
	}\bibfield  {title} {\bibinfo {title} {Fluctuation-dominated kinetics in
			diffusion-controlled reactions},\ }\href@noop {} {\bibfield  {journal}
		{\bibinfo  {journal} {Physical Review A}\ }\textbf {\bibinfo {volume} {32}},\
		\bibinfo {pages} {435} (\bibinfo {year} {1985})}\BibitemShut {NoStop}%
	\bibitem [{\citenamefont {Spouge}(1988)}]{spouge1988exact}%
	\BibitemOpen
	\bibfield  {author} {\bibinfo {author} {\bibfnamefont {J.~L.}\ \bibnamefont
			{Spouge}},\ }\bibfield  {title} {\bibinfo {title} {Exact solutions for a
			diffusion-reaction process in one dimension},\ }\href@noop {} {\bibfield
		{journal} {\bibinfo  {journal} {Physical review letters}\ }\textbf {\bibinfo
			{volume} {60}},\ \bibinfo {pages} {871} (\bibinfo {year} {1988})}\BibitemShut
	{NoStop}%
	\bibitem [{\citenamefont {Krishnamurthy}\ \emph {et~al.}(2002)\citenamefont
		{Krishnamurthy}, \citenamefont {Rajesh},\ and\ \citenamefont
		{Zaboronski}}]{krishnamurthy2002kang}%
	\BibitemOpen
	\bibfield  {author} {\bibinfo {author} {\bibfnamefont {S.}~\bibnamefont
			{Krishnamurthy}}, \bibinfo {author} {\bibfnamefont {R.}~\bibnamefont
			{Rajesh}},\ and\ \bibinfo {author} {\bibfnamefont {O.}~\bibnamefont
			{Zaboronski}},\ }\bibfield  {title} {\bibinfo {title} {Kang-redner small-mass
			anomaly in cluster-cluster aggregation},\ }\href@noop {} {\bibfield
		{journal} {\bibinfo  {journal} {Physical Review E}\ }\textbf {\bibinfo
			{volume} {66}},\ \bibinfo {pages} {066118} (\bibinfo {year}
		{2002})}\BibitemShut {NoStop}%
	\bibitem [{\citenamefont {Bertini}\ \emph {et~al.}(2001)\citenamefont
		{Bertini}, \citenamefont {De~Sole}, \citenamefont {Gabrielli}, \citenamefont
		{Jona-Lasinio},\ and\ \citenamefont {Landim}}]{BertiniPRL2001}%
	\BibitemOpen
	\bibfield  {author} {\bibinfo {author} {\bibfnamefont {L.}~\bibnamefont
			{Bertini}}, \bibinfo {author} {\bibfnamefont {A.}~\bibnamefont {De~Sole}},
		\bibinfo {author} {\bibfnamefont {D.}~\bibnamefont {Gabrielli}}, \bibinfo
		{author} {\bibfnamefont {G.}~\bibnamefont {Jona-Lasinio}},\ and\ \bibinfo
		{author} {\bibfnamefont {C.}~\bibnamefont {Landim}},\ }\bibfield  {title}
	{\bibinfo {title} {Fluctuations in stationary nonequilibrium states of
			irreversible processes},\ }\href
	{https://doi.org/10.1103/PhysRevLett.87.040601} {\bibfield  {journal}
		{\bibinfo  {journal} {Phys. Rev. Lett.}\ }\textbf {\bibinfo {volume} {87}},\
		\bibinfo {pages} {040601} (\bibinfo {year} {2001})}\BibitemShut {NoStop}%
	\bibitem [{\citenamefont {Rajesh}\ and\ \citenamefont
		{Majumdar}(2001)}]{rajesh2001exact}%
	\BibitemOpen
	\bibfield  {author} {\bibinfo {author} {\bibfnamefont {R.}~\bibnamefont
			{Rajesh}}\ and\ \bibinfo {author} {\bibfnamefont {S.~N.}\ \bibnamefont
			{Majumdar}},\ }\bibfield  {title} {\bibinfo {title} {Exact tagged particle
			correlations in the random average process},\ }\href@noop {} {\bibfield
		{journal} {\bibinfo  {journal} {Physical Review E}\ }\textbf {\bibinfo
			{volume} {64}},\ \bibinfo {pages} {036103} (\bibinfo {year}
		{2001})}\BibitemShut {NoStop}%
\end{thebibliography}

%

\end{document}